\documentclass[12pt,preprint]{aastex}

\usepackage{graphics,epsfig}
\usepackage{natbib}
\usepackage{lscape}

\def\spose#1{\hbox to 0pt{#1\hss}}
\def\lta{\mathrel{\spose{\lower 3pt\hbox{$\mathchar"218$}}
     \raise 2.0pt\hbox{$\mathchar"13C$}}}
\def\gta{\mathrel{\spose{\lower 3pt\hbox{$\mathchar"218$}}
     \raise 2.0pt\hbox{$\mathchar"13E$}}}

\def\figure#1#2 {\par{\narrower\noindent {\bf Fig. #1}
   \hskip 2mm #2\par}\bigskip\noindent}
\def\table#1#2 {\par{\narrower\noindent {\bf Tab. #1}
   \hskip 2mm #2\par}\bigskip\noindent}

\def\registered{{\ooalign{\hfil\raise .00ex\hbox{\scriptsize R}\hfil\crcr\mathhexbox20D}}}

\usepackage{amsmath}
\usepackage{accents}
\newlength{\dhatheight}

\shorttitle{Habitability in Binary Systems II}

\shortauthors{Cuntz}

\begin{document}


\title{{\it S}-Type and {\it P}-Type Habitability in Stellar Binary Systems: \\
A Comprehensive Approach \\
II. Elliptical Orbits}

\author{M. Cuntz}

\affil{Department of Physics}
\affil{University of Texas at Arlington, Arlington, TX 76019-0059;}
\email{cuntz@uta.edu}

\begin{abstract}
In the first paper of this series, a comprehensive approach has been
provided for the study of {\it S}-type and {\it P}-type habitable
regions in stellar binary systems, which was, however, restricted to
circular orbits of the stellar components.  Fortunately, a modest
modification of the method also allows for the consideration of
elliptical orbits, which of course entails a much broader range of
applicability.  This augmented method is presented here,
and numerous applications are conveyed.  In alignment with Paper~I,
the selected approach considers a variety of aspects, which comprise
the consideration of a joint constraint including orbital stability
and a habitable region for a possible system planet through the
stellar radiative energy fluxes (``radiative habitable zone"; RHZ).
The devised method is based on a combined formalism for
the assessment of both {\it S}-type and {\it P}-type habitability;
in particular, mathematical criteria are deduced for which kinds
of systems {\it S}-type and {\it P}-type habitable zones are realized.
If the RHZs are truncated by the additional constraint of
orbital stability, the notation of {\it ST}-type and {\it PT}-type
habitability applies.  In comparison to the circular case, it is
found that in systems of higher eccentricity, the range of the RHZs
is significantly reduced.  Moreover, for a considerable number
of models, the orbital stability constraint also reduces the range
of {\it S}-type and {\it P}-type habitability.  Nonetheless,
{\it S}-, {\it P}-, {\it ST}-, and {\it PT}-type habitability is
identified for a considerable set of system parameters.  The method
as presented is utilized for {\tt BinHab}, an online code available
at The University of Texas at Arlington.
\end{abstract}

\keywords{astrobiology --- binaries: general --- celestial mechanics
--- planetary systems}

%%%%%%%%%%%%%%%%%%%%%%%%%%%%%%%%%%%%%%%%%%%%%%%%%%%%%%%%%%%%%%%%%%%%%%%%

\section{INTRODUCTION}

A major task situated at the crossroad of stellar astrophysics, also
encompassing orbital stability studies and astrobiology, is the
identification of habitable zones (HZs) of stars, including binaries
and multi-stellar systems.  Recent studies for those systems
have been given by, e.g., \cite{egg12}, \cite{kal13a,kal13b},
\cite{kan13}, \cite{kan14}, \cite{cun14}, \cite{jai14}, and
\cite{cunb14}.  The fact that planets are able to exist in binary
systems has been pointed out in numerous observational investigations
and surveys, including \cite{pat02}, \cite{egg04,egg07},
\cite{rag06,rag10}, and \cite{roe12}.

Earlier work also targeted planet formation in binary systems,
including terrestrial planets, as summarized by \cite{kle12}. 
That work contains a thorough review of our current understanding
of disk--planet interactions, including prevalent processes
of planet growth within gaseous protoplanetary disks, as well as
the exchange of angular momentum between planets and the disk and
planet migration.  Important results have also been provided
by the {\it Kepler} mission, which include the detection
of planets in stellar binary systems.  Examples include Kepler-16
\citep{sla11,doy11}, and Kepler-47 \citep{oro12}.  Kepler-16b
constitutes the first circumbinary planet ({\it P}-type orbit),
which means that it orbits both binary components.  Theoretical
simulations on the habitability of possible Earth-mass planets
and moons in that system have been given by \cite{qua12}.  Kelper-47,
on the other hand, is host to at least three planets in orbit
around both stars.

A previous study focused on {\it S-}type and {\it P-}type habitable zones
in binary systems has been given by \cite{cun14}, henceforth called Paper~I.
The focus of this work was the study of habitable regions for systems
with the stellar components in circular orbits, a limitation that will
be overcome in the present work.  Observationally, it is
found that the stars residing in binary systems are almost always
in elliptical orbits, with some systems exhibiting fairly high
eccentricities; see, e.g., \cite{roe12} and references therein.  Examples
include $\gamma$~Cep~A ($e_{\rm b} = 0.41$), HD~196885~A ($e_{\rm b} = 0.42$),
and HD~126614~A ($e_{\rm b} \le 0.6$).  In these systems, the planet is
orbiting one of the stellar components ({\it S}-type orbit), with the other
stellar component acting as a perturbator.  Also, there is high diversity
regarding stellar component combinations.  For the cases discussed by
\cite{roe12} (see their Table~2), the primaries span between spectral types
F, G, and K, whereas the secondaries are mostly M-type dwarfs.

Akin to the previous work of Paper~I,
numerous aspects have been taken into account, including:
(1) the consideration of a joint constraint including orbital stability
and a habitable region for a possible system planet through the stellar
radiative energy fluxes (``radiative habitable zone"; RHZ);
(2) the treatment of conservative, general and extended zones of
habitability for the various systems as defined for the Solar System and
beyond (see Section~2.1); (3) the provision of a combined formalism for the
assessment of both {\it S}-type and {\it P}-type habitability; in particular,
mathematical criteria are presented for which kind of system {\it S}-type and
{\it P}-type habitability is realized.  Specifically, an algebraic formalism
for the assessment of both {\it S}-type and {\it P}-type habitability
is employed based on a fourth order polynomial.  Thus, an a prior
specification for the presence or absence of {\it S}-type or {\it P}-type
RHZs is neither necessary nor possible, as those are determined by
the adopted mathematical formalism.

In principle, five different cases of habitability are identified,
which are: {\it S}-type and {\it P}-type habitability provided by the
full extent of the RHZs; habitability, where the RHZs are truncated by
the additional constraint of planetary orbital stability (referred to
as {\it ST}- and {\it PT}-type, respectively); and cases of no
habitability at all.  On the other hand, Paper~I solely focused on
circular binary systems.
Fortunately, a modest modification of the method forwarded by Paper~I
allows for the consideration of elliptical orbits, which
provide a much broader range of applicability, including
detailed comparisons with observations as well as a large realm of
theoretical studies.  Emphasis is placed again on the detailed
calculations of the RHZs and the orbital stability criterion for
theoretical planets, which again will utilize the previous work by
\cite{hol99}, thereafter HW99.

Paper~II is structured as follows: In Section~2, we describe
the theoretical approach, particularly the adopted modifications to the
circular case as studied in Paper~I, which concern both the computation
of the RHZs and the planetary orbital stability limits.  In Section~3,
we provide various case studies regarding {\it S-}, {\it ST-}, {\it P-},
and {\it PT-}type habitability.  Special emphasis is placed on the
width of stellar habitable zones as function of the binary eccentricity
as well as the domains of the conservative and general zones of
habitability for different sets of system parameters.  Our summary and
conclusions are given in Section~4.

%%%%%%%%%%%%%%%%%%%%%%%%%%%%%%%%%%%%%%%%%%%%%%%%%%%%%%%%%%%%%%%%%%%%%%%%

\section{THEORETICAL APPROACH}

\subsection{Calculation of the RHZs}

In the following, we summarize the information needed for the calculations of
the RHZs in binary systems regarding both {\it S}-type and {\it P}-type orbits.
This effort deals with the requirement of providing a habitable region for a
system planet based on the radiative energy fluxes of the stellar components.
The requirement of orbital stability will be disregarded for now; it will
be discussed in Section~2.2.  The adopted approach is fairly similar to that
of Paper~I, which focused on binary systems in circular orbits.  Akin to Paper~I,
we focus on theoretical main-sequence stars (see Table~1).  A flow diagram
of the code, named {\tt BinHab}, also described by \cite{cunb14}, is given
as Figure~1.  {\tt BinHab} is an analytic code, and it is based on the same
equations and assumptions given in Paper~I and II.

The key equations of the adopted method read as follows.  For a star of
luminosity $L_i$ (in units of solar luminosity $L_\odot$), the distance $d_i$
of the habitability limit $s_\ell$ as identified for the Sun (see Paper~I and
references therein), constituting either an inner or outer limit of a stellar
habitable region (except for $\ell = 3$; see Table~2), is given as
\begin{equation}
d_i \ = \ s_\ell \sqrt{\frac{L_i}{S_{{\rm rel}, i\ell} L_\odot}} \ .
\end{equation}
Here $S_{{\rm rel}, i\ell} = S_{{\rm rel}, i\ell}(T_{\rm eff})$
describes the stellar flux in units of the solar constant, which is a function of the
stellar effective temperature $T_{\rm eff}$ \citep[e.g.,][and subsequent work]{kas93}.
The values for $S_{{\rm rel}, i\ell}$ are calculated following the formalism of
\cite{sel07}; see Paper~I for additional information.

In the solar case, it is found that the conservative habitable zone (CHZ)
extends between 0.95 and 1.37 AU (i.e., $\ell = 2$ and 4, respectively), and
the general habitable zone (GHZ) extends between 0.84 and 1.67 AU (i.e., $\ell = 1$
and 5, respectively)\footnote{Recent studies indicate that setting those
limits is a more complex process than adopted by \cite{kas93} as they depend
on, e.g., particulars of the planetary atmosphere (as manifested through
detailed 3-D modeling), geodynamics, rotation rate, and mass; see, e.g.,
\cite{blo09}, \cite{sea13}, \cite{lec13}, \cite{zso13}, \cite{kop13,kop14},
and \cite{yan14}, for details. Updated limits can thus be considered through
interpolation between the results obtained for the inner / outer limits of
the CHZ and GHZ.  Alternatively, updates in the limits for the CHZ and GHZ
can be considered through appropriate choices for $s_\ell$ (see Table~2),
as anticipated in the forthcoming version of {\tt BinHab}.}.
Following \cite{und03} and references therein, the
inner limit of the CHZ is defined by water loss, occurring when the
atmosphere is warm enough to build a stratosphere where water is gradually
lost by photodissociation and subsequent hydrogen loss to space.  The outer
limit of the CHZ is defined by first carbon dioxide condensation.  Moreover,
the GHZ is defined by the domain between the runaway greenhouse effect
(inner limit) and the maximum greenhouse effect (outer limit).
The extreme case of the extended habitable zone (EHZ), attained through
excessive planetary global greenhouse processes, is in the solar case
assumed to extend up to 2.40~AU \citep{for97,mis00}, i.e.,  $\ell = 6$.
Furthermore, $\ell = 3$ is used for identifying Earth-equivalent positions,
as done in Paper~I.

In case of a binary system with planetary distances $d_i$, limits of habitability
associated with $s_\ell$ are obtained through solving
\begin{equation}
\sum_{i=1}^2 \frac{L_i}{S_{{\rm rel}, i\ell} d_i^2} \ = \ \frac{L_\odot}{s_\ell^2} \ .
\end{equation}
It is found that
\begin{mathletters}
\begin{eqnarray}
d_1^2 \ & = & \ a^2 + z^2 + 2 a z \cos{\varphi} \\ 
d_2^2 \ & = & \ a^2 + z^2 - 2 a z \cos{\varphi} \ .
\end{eqnarray}
\end{mathletters}
Here $a \equiv a_{\rm b}$ denotes the semi-distance of the binary components (or
a modified value of it, see below, with $a_{\rm b}$ being used if the original
denotation applies), $z$ the distance of a position at the habitability
limit contour, also referred to as radiative habitable limit (RHL), and
$\varphi$ the associated angle; see Figure~2 for information on the coordinate
set-up for both {\it S}-type and {\it P}-type orbits.  It is also assumed
that $L_1 \ge L_2$ without loss of generality.  Following Paper~I, we again
introduce the concept of recast stellar luminosity $L_{i\ell}'$ defined as
\begin{equation}
L_{i\ell}' = \frac{L_i}{L_\odot S_{{\rm rel}, i\ell}} .
\end{equation}

Finally, following algebraic transformations, the equation for $z(\varphi)$
is obtained as
\begin{equation}
z^4 + A_2 z^2 + A_1 z + A_0 \ = \ 0
\end{equation}
with
\begin{mathletters}
\begin{eqnarray}
A_2 & = & 2 a^2 (1  - 2 \cos^2 \varphi) - s_\ell^2 (L_{1\ell}' + L_{2\ell}') \\
A_1 & = & 2 a s_\ell^2 \cos \varphi (L_{1\ell}'-L_{2\ell}') \\ 
A_0 & = & a^4 - a^2 s_\ell^2 (L_{1\ell}' + L_{2\ell}') \ .
\end{eqnarray}
\end{mathletters}
As discussed in Paper~I, Equation~(5) constitutes a fourth-order algebraic equation
that is known to possess four possible solutions \citep{bro97}, although some
(or all) of them may constitute unphysical solutions, i.e., $z(\varphi)$ having
a complex or imaginary value.  The solutions of this polynomial allow to define
the inner and outer limits of the RHZs based on the system parameters.  Equation~(5)
is solved using the adopted coordinate system, which in essence constitutes
a polar coordinate system, except that negative values for $z$ are permitted; in
this case the position of $z$ is found on the opposite side of angle $\varphi$.
This approach allows to identify the various cases for the existence of solutions
for $z(\varphi)$ in regard to {\it S}-type and {\it P}-type habitable regions.

It is noteworthy that heretofore the method of solution as presented is
identical to that for binary components in circular orbits.  Thus, the decisive
question arises as to how the RHZs for binary components in elliptical orbits can
adequately be addressed.  In this regard, the underlying assumption of this study
needs to be taken into account, which is that {\it any system planet deemed
potentially habitable needs to be located within the system's RHZ at all times.}
This standard needs to be applied regardless of the (time-dependent) separation
of the binary components, i.e., whether the system components are in a
periastron $a_{\rm per}$, apastron $a_{\rm ap}$, or any intermediate position.

Algebraically, it is found that for {\it P}-type orbits, the extrema
${\rm RHZ}_{\rm in}$ and ${\rm RHZ}_{\rm out}$ are given as
\begin{mathletters}
\begin{eqnarray}
{\rm RHZ}_{\rm in}  \ = \ {\rm Max}\big({\cal R}\big(z,{\varphi}{\big)}\big)\Big|_{s_{\ell,{\rm in}}} \Big|_{a=a_{\rm ap}} \\
{\rm RHZ}_{\rm out} \ = \ {\rm Min}\big({\cal R}\big(z,{\varphi}{\big)}\big)\Big|_{s_{\ell,{\rm out}}}\Big|_{a=a_{\rm ap}} ,
\end{eqnarray}
\end{mathletters}
whereas for {\it S}-type orbits, the extrema
${\rm RHZ}_{\rm in}$ and ${\rm RHZ}_{\rm out}$ are given as
\begin{mathletters}
\begin{eqnarray}
{\rm RHZ}_{\rm in}  \ = \ {\rm Max}\big({\cal R}\big(z,{\alpha}{\big)}\big)\Big|_{s_{\ell,{\rm in}}} \Big|_{a=a_{\rm per}}  \\
{\rm RHZ}_{\rm out} \ = \ {\rm Min}\big({\cal R}\big(z,{\alpha}{\big)}\big)\Big|_{s_{\ell,{\rm out}}}\Big|_{a=a_{\rm ap}}  ,
\end{eqnarray}
\end{mathletters}
respectively, noting that ${\cal R}(z,\alpha)$ and ${\cal R}(z,\varphi)$ describe
the areas bordered by the RHLs defined by ${s_{\ell,{\rm in}}}$ and ${s_{\ell,{\rm out}}}$
(see Paper~I).  The calculation of the extrema is applied to the angles $\alpha$
({\it S-}type systems) and $\varphi$ ({\it P-}type systems) for the intervals
$0 \le \alpha \le \pi$ and $0 \le \varphi \le \pi/2$, respectively.  For
{\it P-}type systems, the coordinate information is provided by Figure~2.
Note that for {\it S-}type systems, the angle's fix point is at the stellar primary S1.

The necessity for the choices $a=a_{\rm per}$ or $a=a_{\rm ap}$ is illustrated
by the following instructive
examples (conveyed in high precision for tutorial reasons), which refer to
equal-mass binary systems with $M_1 = M_2 = M_\odot$ for an eccentricity of
$e_{\rm b} = 0.30$ (see Figure~3), and with the RHLs inspected for the GHZ.  As
a first case, we consider a system assuming a major axis of $2a_{\rm b} = 1.0$~AU;
in this case, the RHZ is given as {\it P-}type.  At the periastron position for
the system components, the inner and outer limits for the RHLs are given as
1.436 and 2.567~AU, respectively.  However, at the system's apastron position, the
inner and outer limits for the RHLs are given as 1.663 and 2.508~AU, respectively.
Intermediate values for the inner and outer limits of the RHLs are obtained for
other positions of the system, as expected.  Therefore, the most stringent
inner and outer limits for the overall RHZ annulus are identified as
1.663 and 2.508~AU, which are the values for the system's apastron position.

As a second case, we study a system with a major axis of $2a_{\rm b} = 20.0$~AU;
in this case, the RHZ is given as {\it S-}type (see Figure~4).  At the system's periastron
position, the inner and outer limits for the RHLs concerning the GHZ are identified as
0.9300 and 1.8443~AU, respectively.  Furthermore, at the system's apastron position,
the inner and outer limits for the RHLs are identified as 0.9283 and 1.8359~AU,
respectively, with intermediate values for the inner and outer limits of the RHLs
obtained for other positions.  Thus, the most stringent
inner and outer limits for the overall RHZ annulus are identified as
0.9300 and 1.8359~AU.  In this case, the relevant value for the inner RHL
is given by the system's periastron position, whereas relevant value for outer RHL
is given by the system's apastron position.  Note that the same kind of behavior
is identified for other sets of system parameters, such as stellar masses (including
binary systems of unequal masses), major axes $2a_{\rm b}$, and eccentricities $e_{\rm b}$.
It also holds for the different types of HZs, i.e., CHZs, GHZs, and EHZs.

Consequently, the most stringent positioning for the inner and outer limits of
the RHZs, taking into account all possible separations of the binary components,
including the periastron and apastron positions, must be taken into account.
This implies that both for {\it S-}type and {\it P-}type habitability, for any
inner RHL, the outermost position must be selected, whereas for any outer RHL,
the innermost position must be selected to define the annuli for the RHZs.
Based on detailed analyses, exemplified by Figure~3 and 4,
as well as results given in Section~3, encompassing both equal-mass and
nonequal-mass binary systems, it is found that for {\it S}-type RHZs,
regarding inner RHLs, the periastron stellar positions need to be chosen, i.e.,
$a_{\rm b}(1-e_{\rm b})$, whereas regarding outer RHLs, the apastron stellar
positions need to be chosen, i.e., $a_{\rm b}(1+e_{\rm b})$.  However, for
{\it P}-type RHZs, both regarding the inner and outer RHLs, the apastron
positions need to be chosen throughout (see Table~2).

Thus, in terms of the derivations obtained in Paper~I, all equations still hold,
except that $a$ needs to be replaced by $a \mapsto \tilde{a}_{{\rm SP,}\ell}$,
where $\tilde{a}_{{\rm SP,}\ell}$ must be set as either $a_{\rm b}(1-e_{\rm b})$
or $a_{\rm b}(1+e_{\rm b})$, depending on the selection of $\ell$ and whether
{\it S}-type or {\it P}-type RHZs are considered (see Table~2).  Hence, the
entire previously given mathematical evaluations, including the transformations
for general binary systems represented through a fourth-order polynomial, continue
to hold as virtually all equations\footnote{The exception is Equation~(45) of
Paper~I, which however can be replaced by the more general Equation~(40) of Paper~I.}
and analyses are distinctly separate for inner and outer RHLs; this feature
applies to both equal-mass and non-equal mass binary systems.

One decisive connection between inner and outer RHLs is given by the requirement
that for a {\it P}-type habitable region, the inner RHL $s_{\ell,{\rm in}}$ must
be located completely inside the outer RHL $s_{\ell,{\rm out}}$ for the
corresponding RHZ to exist, i.e.,
\begin{equation}
{\rm Min}\Big({\cal R}\big(z,\varphi\big)\Big)\Big|_{s_{\ell,{\rm out}}} \ \ge \
{\rm Max}\Big({\cal R}\big(z,\varphi\big)\Big)\Big|_{s_{\ell,{\rm in}}} \ ,
\end{equation}
where ${\cal R}\big(z,\varphi\big)$ denotes the domain bordered by the respective
RHLs.  As noted in Paper~I, this condition is, however, violated in some models,
especially for relatively large values of $a$, as well as relatively small ratios
of ${L_{2\ell}'}/{L_{1\ell}'}$.  In this case, the RHZ for
$(s_{\ell,{\rm in}},s_{\ell,{\rm out}})$ is nullified, a behavior that may occur
for the pairings $(s_2,s_4)$, $(s_1,s_5)$, and $(s_1,s_6)$, corresponding to the
CHZ, GHZ, and EHZ, respectively.

\subsection{The Planetary Orbital Stability Constraint}

As noted in previous investigations, habitability in multi-stellar
systems (including binaries) requires---among other aspects---planetary
orbital stability.  In earlier work, \cite{dvo86} identified upper and lower
bounds of planetary orbital stability considering the orbital elements,
semimajor axis and eccentricity of the adopted binary stars.  Within the last
thirty years, a large body of additional studies has been performed encompassing
both numerical and analytical work.  Recent examples for the restricted three-body
problem, but also containing assessments about general cases, have been given
by \cite{mus14}.  Previously, HW99 derived fitting formulae of orbital
stability limits for both {\it S}-type and {\it P}-type planets in binary systems
given as
\begin{equation}
  \frac{a_{\rm cr}}{a_{\rm b}} \ = \ \sum_{i=0}^2 {\tilde A}_i \mu^i + {\cal F}_{\rm S}(\mu, e_{\rm b})
\end{equation}
and
\begin{equation}
  \frac{a_{\rm cr}}{a_{\rm b}} \ = \ \sum_{i=0}^2 {\tilde A}_i \mu^i + {\cal F}_{\rm P}(\mu, e_{\rm b}) ,
\end{equation}
respectively.

These equations communicate the critical semimajor axis $a_{\rm cr}$ in units of the
semimajor axis $a$ in the cases of {\it S}-type and {\it P}-type orbits.
For {\it S}-type orbits,
the ratio ${a_{\rm cr}}/a_{\rm b}$, see Equation~(10), conveys the {\it upper limit} of
planetary orbital stability, whereas for {\it P}-type orbits, the ratio
${a_{\rm cr}}/a_{\rm b}$, see Equation~(11), conveys the {\it lower limit} of planetary
orbital stability.  Here $\mu$ denotes the stellar mass ratio, given as
$\mu = M_2 / (M_1 + M_2)$, where $M_1$ and $M_2$ describe the two masses
of the binary components with $M_2 \le M_1$.  The above given equations also
include the parameter functions ${\cal F}_{\rm S}(\mu, e_{\rm b})$ and
${\cal F}_{\rm P}(\mu, e_{\rm b})$, which depend on the mass ratio $\mu$
and the eccentricity of the stellar binary, $e_{\rm b}$, and are non-zero
for elliptical binary systems\footnote{In Paper~I, we only included the
linear term of $\mu$ for the sake of parallelism between the treatment of
{\it S}-type and {\it P}-type orbits.}.

Following HW99, ${\cal F}_{\rm S}$ and ${\cal F}_{\rm P}$ are given as
\begin{equation}
{\cal F}_{\rm S} \ = \ e_{\rm b} \sum_{i=0}^2 {\tilde B}_i \mu^i + e_{\rm b}^2 \sum_{i=0}^2 {\tilde C}_i \mu^i
\end{equation}
and
\begin{equation}
{\cal F}_{\rm P} \ = \ e_{\rm b} \sum_{i=0}^2 {\tilde B}_i \mu^i + e_{\rm b}^2 \sum_{i=0}^2 {\tilde C}_i \mu^i ,
\end{equation}
respectively (see Table~3 for details).  Equations~(12) and (13) constitute second-order
polynomial fits on results of orbital stability simulations by HW99 and previous work.
The underlying studies explore the survival times of theoretical planets, treated
as test particles, in various binary systems.  Furthermore, based on HW99's work,
the Equations~(10) to (13) are invalid for eccentricities beyond 0.8.  Therefore,
we limit the scope of the present study to binary systems with eccentricities in
the range of $0.0 \le e_{\rm b} \le 0.80$.

\subsection{General Impact of the Binary System Eccentricity}

Next we highlight the general impact of the eccentricity of a binary
system on the different types of habitable regions through exemplifying
the effects of eccentricity on systems through two cases, which are:
a case of high eccentricity (i.e., $e_{\rm b}=0.5$) and the circular case
(i.e., $e_{\rm b}=0$).
First, we consider the impact of the binary system eccentricity
on the RHZs.  Examples are given in Figure~5, which depicts
systems of stellar masses of $M_1 = M_2 = M_\odot$, as well as
of $M_1 = 1.5~M_\odot$ and $M_2 = 0.5~M_\odot$, taking the GHZ
as test cases.  (The aspect that nonequal-mass binary systems
compared to equal-mass binary systems, with $M_1+M_2$ kept constant,
have largely reduced RHZs, albeit $L_1+L_2$ of the nonequal-mass
system is notably higher due to the mass--luminosity relationship,
has been one of the main foci of Paper~I.)  The major axis is
assumed as $2a_{\rm b}=1.0$~AU; furthermore, we consider binary
eccentricities of $e_{\rm b}=0.0$ and 0.5.  In case of
equal masses and $e_{\rm b}=0.0$ the inner and outer RHLs are
given as 1.54 and 2.54~AU, respectively.  For $e_{\rm b}=0.5$, however,
the inner and outer RHLs are given as 1.75 and 2.11~AU, respectively.
Thus, if $e_{\rm b}$ is increased, the inner RHL moves outward,
whereas the outer RHL moves inward.  Therefore, the total width
of the RHZ is reduced from 1.0~AU to 0.36~AU.

An even more dramatic reduction of the RHZ occurs for unequal
mass systems.  Taking $M_1 = 1.5~M_\odot$ and $M_2 = 0.5~M_\odot$
as an example, it is found that for $e_{\rm b}=0$, the inner and outer
RHLs are given as 2.16 and 2.66~AU, respectively.  For $e_{\rm b}=0.5$,
the inner and outer RHLs are given as 2.41 and 2.45~AU,
respectively.  In this case, the RHZ is virtually nonexistent;
its width has been reduced by 91\% relative to the circular case,
whereas in the equal-mass system previously discussed, the
reduction is ``only" 64\%.  Note that reductions in the RHZs also
occur for {\it S}-type habitable regions.  However, for those
models, the effects of increased values for $e_{\rm b}$ are very
small, or minuscule, as verified by numerous case studies (see
Section~3.3).  For the above-given examples of $M_1$ and $M_2$ with
$2a_{\rm b}=20$~AU, the RHZs are marginally reduced both for the
equal and non-equal mass system, noting that the reduction relative
to the circular case are given as 0.8\% and 0.04\%, respectively.

Another important aspect concerns the influence of $e_{\rm b}$
on the orbital stability behavior of possible planets.  Generally,
it is found that if $e_{\rm b}$ is increased, the domain of stability
is reduced both for {\it S}-type and {\it P}-type orbits.  If $e_{\rm b}$
is larger than zero, the orbital stability limit for {\it P}-type orbits
(which constitutes a lower limit) is positioned further outward,
whereas for {\it S}-type orbits, the orbital stability limit (which
constitutes an upper limit) is positioned further inward.  Evidently,
both types of behavior affect the domains of planetary orbital stability
in an adverse manner.
To showcase this type of behavior, we evaluated both ${\cal F}_{\rm P}$
and ${\cal F}_{\rm S}$ for different values of $e_{\rm b}$, given as
0.0, 0.3, and 0.5.  For $\mu = 0.25$, ${\cal F}_{\rm P}$ is identified
as 2.31, 3.35, and 3.85, respectively, corresponding to an increase by
a factor of 1.67.  For $\mu = 0.50$, ${\cal F}_{\rm P}$ is identified
as 2.39, 3.18, and 3.60, respectively, corresponding to an increase by
a factor of 1.5.  It is found that the orbital stability limit moves
outward if $e_{\rm b}$ is increased, thus reducing the width of the
RHZ (if any) compatible with planetary orbital stability.
For {\it S}-type habitability, the following results are found.
For $\mu = 0.25$, ${\cal F}_{\rm S}$ is obtained as 0.369, 0.233, and
0.152, and for $\mu = 0.50$, ${\cal F}_{\rm S}$ is obtained as 0.274,
0.177, and 0.118, respectively, for the sequence of $e_{\rm b}=0.0$,
0.3, and 0.5 (see above).  Hence, for $\mu = 0.25$ and 0.50,
${\cal F}_{\rm S}$ is reduced by a factor of about 2.4 if $e_{\rm b}$
is raised from 0.0 to 0.5.  As for {\it S}-type orbits, the planetary
orbital stability limit constitutes an upper limit, which is found to
be placed further inward for systems of higher binary eccentricity;
thus, the width of the RHZ (if any) compatible with planetary orbital
stability is reduced as well.

In summary, increased values of eccentricity in binary systems
entail an adverse impact both regarding {\it P}-type and {\it S}-type
habitability, due to their influence both on the widths of the RHZs
(if any) and on the domains of stability for possible circumbinary
or circumstellar planets.

%%%%%%%%%%%%%%%%%%%%%%%%%%%%%%%%%%%%%%%%%%%%%%%%%%%%%%%%%%%%%%%%%%%%%%%%

\section{RESULTS AND DISCUSSION}

\subsection{Examples of {\it S-}, {\it ST-}, {\it P-}, and {\it PT-}type
Habitability Solution Domains}

In the following, we study examples of {\it S-}, {\it ST-}, {\it P-},
and {\it PT-}type habitability aimed at highlighting the impact of
binary eccentricity.  Our study comprises an equal-mass system of
$M_1 = M_2 = 1.0~M_\odot$ with a luminosity of $L_1 = L_2 = 1.23~L_\odot$ 
(see Figure~6) and a nonequal-mass system of $M_1 = 1.25~M_\odot$ and
$M_2 = 0.75~M_\odot$ (see Figure~7); here the luminosities are given
as $L_1 = 2.15~L_\odot$ and $L_2 = 0.36~L_\odot$.  The habitable
regions are calculated for the GHZ.  The overall approach is to
place the two stellar components at different separation distances,
i.e., with the major axis $2a_{\rm b}$ considered the independent variable,
and to assess resulting domains of habitability.  The systems are
evaluated for fixed values of binary eccentricity, given as
$e_{\rm b} = 0.0$, 0.25, 0.5, and 0.75, respectively.

Regarding {\it P-}type orbits, it is found that increased values
of binary eccentricity affect the widths of the RHZs as well as the
maximum value of the major axis $2a_{\rm b}$, for which {\it P-}type RHZs
can exist.  This type of behavior is found for both the equal-mass and
the nonequal-mass system.  For the equal-mass binary system (see Figure~6),
the width of the RHZ at 0.05~AU (a distance where also the orbital stability
criterion is readily met) is given as 1.20, 1.16, 1.11, and 1.06~AU
for models of $e_{\rm b} = 0.0$, 0.25, 0.5, and 0.75, respectively.
For the nonequal-mass binary system (see Figure~7), the width of the RHZ
at 0.05~AU is identified as 0.91, 0.82, 0.74, and 0.67~AU, respectively.
These results show that higher values of $e_{\rm b}$ entail smaller
widths of the RHZ and that for given values of $e_{\rm b}$, the width
of the RHZ is noticeably smaller in the nonequal-mass system than
in the equal-mass system, even though the latter has a higher
combined stellar luminosity ($L_1+L_2$).

Regarding the orbital stability limit, which for {\it P-}type orbits
constitutes a lower limit (see Section~2.2), the following behavior
is found: If the major axis of the stellar components is increased,
the limit moves inward, which is found for both the equal-mass and
the nonequal-mass system (see Figure 6 and 7, respectively).  Therefore,
the stellar separation distance must be relatively small for the
full width of the RHZ to be available for {\it P-}type habitability.
For models of increased values of $e_{\rm b}$, the requirement of
stellar proximity becomes more rigorous; not that this more-stringent
requirement corresponds to a larger slope of the orbital stability
cut-off line.  In the eight examples given in Figures~6 and 7, there
are also domains of {\it PT}-type habitability, i.e., where some,
but not all, of the {\it P-}type RHZ is available for circumbinary
habitability; however, they are relatively small, if not minuscule.
Thus, in other words, irrespective of the system's eccentricity,
for small binary separation distances $2a_{\rm b}$, the habitable
region is mostly characterized as {\it P-}type instead of {\it PT}-type,
because the full width of the RHZ is available for providing habitability.

{\it S-}type RHZs are also influenced by the eccentricity of the
binary system, though the effects are relatively minor.  For binary
systems with a major axis of $2a_{\rm b}=15$~AU, the following behavior occurs.
For the equal-mass system of Figure~6, RHZ$_{\rm in}$ changes from 0.930
to 0.985~AU, and RHZ$_{\rm out}$ changes from 1.843 to 1.836~AU if the
system's eccentricity is increased from 0.0 to 0.75. Furthermore, for
the nonequal-mass system of Figure~7, RHZ$_{\rm in}$ changes from 1.209 to 1.234~AU,
and RHZ$_{\rm out}$ changes from 2.341 to 2.338~AU if the system's
eccentricity is increased from 0.0 to 0.75.  Hence, the width
of the RHZ for the equal-mass system decreases from 0.91 to 0.85~AU,
whereas the width of the RHZ for the nonequal-mass system decreases
from 1.13 to 1.10~AU between the circular case and $e_{\rm b} = 0.75$.
Thus, increased values of eccentricity reduce
the width of the RHZs for {\it S-}type systems as well.  However,
it is also found that the width of the RHZ in the nonequal-mass binary
system is consistently higher than in the equal-mass binary system owing
to the influence of $L_1$.

Regarding the orbital stability limit, which for {\it S-}type orbits
constitutes an upper limit (see Section~2.2), the following behavior is
identified: If the major axis of the stellar components is increased,
the orbital stability limit moves outward.  Therefore, the stellar
separation distance must be relatively large for the full width of
the RHZ to be available for {\it S-}type habitability.
For models of increased values of $e_{\rm b}$, the requirement of a
sufficiently large stellar separation distance is highly rigorous;
{\it S-}type habitability is only available for relatively large $2a_{\rm b}$
values.  In Figures~6 and 7, the more-stringent requirement of orbital
stability for larger values of $e_{\rm b}$ amounts to a decrease
in the slope of the orbital stability cut-off line.

The effect of the system's eccentricity on its ability to provide
{\it S-}type habitable regions is remarkable. For $e_{\rm b}$ of
0.0, 0.25, 0.5, and 0.75, {\it S-}type habitability occurs at minimum
major axes of 13.5, 19.1, 31.1, and 74.5~AU, respectively, for the
equal-mass system (see Figure~6) and at 14.8, 21.2, 35.1, and 84.9~AU,
respectively, for the nonequal-mass system (see Figure~7).
Higher values of the system's eccentricity also imply larger
domains of {\it ST-}type habitability, i.e., domains where
not the full width of the {\it S-}type RHZ constitutes a possible
region of circumstellar habitability.  If $e_{\rm b}$ is increased
from 0.0 to 0.5 or 0.75, the domains of {\it ST-}type habitability
change from 6.6 to 15.2 or 36.5~AU for the equal-mass system and
from 7.2 to 16.9 or 41.0~AU for the nonequal-mass system.  The
reason for this pattern is that in systems of high eccentricity,
the stars are in relatively close proximity at their periastron
positions, which reduces both the {\it S-}type and {\it ST-}type
habitable regions due to the orbital stability requirement.
Considering major axes of $2a_{\rm b}=15$~AU for the four equal-mass
systems as well as the four nonequal-mass systems studied here,
{\it S-}type and {\it ST-}type habitability are found for one system
each, whereas the case of no habitable region is encoutered for
two systems each.

In summary, it is found that increased values of binary eccentricity
affect the RHZs (though the change in {\it S}-type RHZs is very minor)
and pose a highly unfavorable impact on the domains of orbital stability
for possible system planet.  These findings apply to both {\it S-}type
and {\it P-}type habitability.  Moreover, higher values of binary
eccentricity also entail larger domains of $2a_{\rm b}$, where neither
{\it S}/{\it ST-}type nor {\it P}/{\it PT-}type habitable regions exist.
Additional case studies for {\it S-}type and {\it P-}type systems are
given in Section~3.3 and 3.2, respectively.  Our focus will be systems
of different combinations of $M_1$ and $M_2$ and, by implication,
$L_1$ and $L_2$, to evaluate the zones of circumstellar and
circumbinary habitability.

\subsection{Case Studies of {\it P}/{\it PT-}type Habitability}

In the following, we present case studies pertaining to
{\it P}/{\it PT-}type habitability.  Owing to the many possible sets
of parameter combinations, encompassing the stellar major axis $2a_{\rm b}$,
the binary eccentricity $e_{\rm b}$, and the stellar masses $M_1$
and $M_2$, it will not be feasible to provide a fully comprehensive
analysis.  However, it will be possible to focus on various cases,
selected for tutorial purposes, by using $2a_{\rm b}=0.5$, 0.75, and 1.0~AU.
Moreover, some intriguing aspects will be targeted such as (1) the
study of main-sequence stars of different masses, i.e., between
1.25 and 0.50~$M_\odot$, (2) the extent and position of the RHZs
in the various models, as well as the impact of the planetary orbital
stability requirement, and (3) the critical value of $e_{\rm b}$,
below which---depending on the model---{\it P}/{\it PT-}type
habitable regions are able to exist.

Akin to Paper~I, we concentrate on systems of main-sequence stars;
see Table~1 for information on the stellar parameters.
One of our goals is to inspect the values for RHZ$_{\rm in}$ and
RHZ$_{\rm out}$ at the stellar periastron and apastron position,
i.e., $a_{\rm per}$ and $a_{\rm ap}$, respectively, to obtain
additional data to justify the method adopted for calculating the
RHZs (see Section~2.1).  We also include systems of low-mass /
low-luminosity stars, corresponding to stellar types K and M,
which is motivated by their relatively high abundance (i.e.,
up to about 90\%), owing to the skewness of the Galactic initial
mass function \citep[e.g.,][]{kro01,kro02,cha03}.  Our results
are depicted in Table~4 to 8.

Table~4 and 5 convey results for equal-mass binary systems of
$M_1 = M_2 = 1.0~M_\odot$ regarding both the CHZ and GHZ, respectively, for
$2a_{\rm b}=0.5$, 0.75, and 1.0~AU for different eccentricities of
the binary components.  Data are given for both the RHZ (see
column ``Orbit") and the orbital stability limited denoted
as $a_{\rm cr}$.  It is found that for both the CHZ and GHZ,
the RHZ$_{\rm in}$ and RHZ$_{\rm out}$ significantly depend
on the system's eccentricity $e_{\rm b}$ with $2a_{\rm b}$ assumed
to be fixed.  Specifically, RHZ$_{\rm in}$ increases and RHZ$_{\rm out}$
decreases as functions of $e_{\rm b}$.  Hence, for any system with
a prescribed value for $2a_{\rm b}$, the RHZs narrow if $e_{\rm b}$
is increased.

For the CHZ (Table~4), the width of the RHZ between $e_{\rm b}=0.0$
and 0.8 changes from 0.56 to 0.41~AU in systems of $2a_{\rm b}=0.50$~AU and
from 0.48 to 0.18 in systems of $2a_{\rm b}=0.75$~AU.  For $2a_{\rm b}=1.0$~AU,
the width of the RHZ for $e_{\rm b}=0.0$ is given as 0.36~AU;
however, no RHZ is found beyond $e_{\rm b} = 0.64$, owing to the
fact that the inner and outer RHL intersect.  The planetary orbital
stability limit $a_{\rm cr}$, constituting a lower limit for
circumbinary habitability, is relatively small.  Therefore,
if the CHZ-RHZs exist, which occurs for most values of $e_{\rm b}$,
{\it P-}type habitable regions are realized.
For the GHZ (Table~5), the width of the RHZ
between $e_{\rm b}=0.0$ and 0.8 changes from 1.20 to 1.05~AU
in systems of $2a_{\rm b}=0.50$~AU, from 1.11 to 0.82~AU in systems of
$2a_{\rm b}=0.75$~AU, and from 1.00 to 0.39~AU in systems of $2a_{\rm b}=1.0$~AU.
As expected, in all of these models, the GHZ is consistently
wider than the CHZ.  Moreover, {\it P-}type habitability is
identified in all models, except beyond $e_{\rm b}=0.41$ for
$2a_{\rm b}=1.0$~AU, where {\it PT-}type habitability occurs.  For
$e_{\rm b}=0.8$ and $2a_{\rm b}=1.0$~AU, the width of the RHZ is given
as 0.55~AU, whereas the width of the {\it PT-}type HZ is given
as 0.39~AU due to truncation.

Table~6 and 7 convey results for nonequal-mass binary systems
of $M_1 = 1.25~M_\odot$ and $M_2 = 0.75~M_\odot$.  We again
focus on both the CHZ and GHZ for $2a_{\rm b}=0.5$, 0.75, and 1.0~AU
for different eccentricities of the binary components.  Again,
regarding both the CHZ and the GHZ, respectively, the RHZ$_{\rm in}$ and
RHZ$_{\rm out}$ significantly depend on $e_{\rm b}$; i.e.,
RHZ$_{\rm in}$ increases and RHZ$_{\rm out}$ decreases as a
function of $e_{\rm b}$, entailing that the RHZs narrow with
increasing values of $e_{\rm b}$.  This feature has profound
consequences for our models; particularly, for $2a_{\rm b}=0.75$~AU,
no habitable regions exist for systems of $e_{\rm b} > 0.29$
pertaining to the CHZ.  Moreover, no habitable regions exist
for systems of $2a_{\rm b}=1.0$~AU regardless of the eccentricity of
the binary components, including systems with the stars in
circular orbits.

Regarding the GHZ, a different type of picture emerges.  Here,
{\it P-}type habitability is identified for the nonequal-mass
system previously studied, irrespective of $e_{\rm b}$.
However, the {\it P-}type HZ critically narrows as a function
of $e_{\rm b}$.  The width of the RHZ between $e_{\rm b}=0.0$
and 0.8 changes from 0.91 to 0.66~AU in systems of $2a_{\rm b}=0.50$~AU,
from 0.74 to 0.38~AU in systems of $2a_{\rm b}=0.75$~AU, and from
0.60 to 0.09~AU in systems of $2a_{\rm b}=1.0$~AU.  Furthermore, owing
to the orbital stability limit, which constitutes a lower limit,
the kind of habitability identified is {\it P-}type rather than
{\it PT-}type.  Comparing our findings for the nonequal-mass
binary system to the results obtained for the equal-mass binary
system indicates that circumbinary habitable regions (if existing)
are much narrower in nonequal-mass compared to equal-mass systems,
a result in alignment with the discussion of Section~2.3.

Another approach to demonstrate the impact of the binary eccentricity
$e_{\rm b}$ on the domains of {\it P}/{\it PT-}type habitability is
to evaluate the quantities {\%}RHZ and {\%}HZ for the various models.
{\%}RHZ denotes the relative size of the RHZ for a given set of
models with the special case of $e_{\rm b} = 0$ set as 100 (see
Tables~4 to 7).  Moreover, {\%}HZ denotes relative size of the
HZ.  In the absence of restrictions due to the orbital stability
constrait, {\%}RHZ and {\%}HZ are identical; otherwise, {\%}HZ is
smaller than {\%}RHZ.  For a distinct type of model, defined
by a given value of $2a_{\rm b}$, the choice of CHZ or GHZ (with
the EHZ disregarded in the following), and the stellar masses
$M_1$ and $M_2$, it is found that both {\%}RHZ and {\%}HZ decrease
with increasing values for $e_{\rm b}$.

The results provided by the Tables~4 to 7 allow us to assess
for which type of models the decrease in {\%}RHZ and {\%}HZ for
increasing values of $e_{\rm b}$ is most severe.  In conclusion,
it is found that the decrease is most severe for relatively large
values of $2a_{\rm b}$, if the CHZ considered instead of the GHZ,
and for unequal-mass systems compared to equal-mass systems if
$(M_1 + M_2)$ remained unaltered.  With $e_{\rm b} = 0.5$ taken as
an instructive example, the results of our models read as follows:
For the CHZ and regarding $2a_{\rm b}=0.5$, 0.75, and 1.0~AU,
{\%}RHZ is given as 85, 64, and 24, respectively.  Furthermore,
there is also no difference between {\%}RHZ and {\%}HZ, as for
those models the habitable regions are consistently identified
as {\it P-}type.  Regarding the GHZ, {\%}RHZ is identified as
93, 84, and 73, respectively.  However, the model for
$2a_{\rm b}=1.0$~AU is identified as {\it PT-}type; hence,
{\%}HZ is found as 68, which is smaller than {\%}RHZ.

Additionally, we focus on the results for nonequal-mass binary
systems of $M_1 = 1.25~M_\odot$ and $M_2 = 0.75~M_\odot$.  With
$e_{\rm b} = 0.5$ again used as an example, we obtained the following
results.  For the CHZ and for $2a_{\rm b}=0.5$~AU, {\%}RHZ is given
as 45.  However, there are no solutions for either $2a_{\rm b}=0.75$
or 1.0~AU.  For the latter case, there is even no solution for circular
systems defined through $e_{\rm b} = 0.0$.  For the GHZ, the results
read as follows:  Regarding $2a_{\rm b}=0.5$, 0.75, and 1.0~AU,
{\%}RHZ is given as 82, 70, and 49, respectively.  {\%}RHZ follows
the same trend previously identified.  Regarding the GHZ, there
is also no difference between the values for the {\%}RHZ and the
{\%}HZ, as the habitable regions are identified as {\it P-}type.

Table~8 lists the results for $e_{\rm b}$ (crit.), defined as
the maximum value of $e_{\rm b}$ for a given system, classified by
$2a_{\rm b}$, $M_1$, and $M_2$, for which a habitable region can
still be identified.  Note that $e_{\rm b}$ (crit.) has been
evaluated in increments of 0.01, except when it was identified
as below 0.01.  In this case, increments of 0.001 were used.
Regarding {\it P}/{\it PT-}type systems,
the absence of habitable regions is either due to
the intersect between RHL$_{\rm in}$ and RHL$_{\rm out}$
(see Equation~9), or because the planetary orbital stability
requirement is violated.  Table~8 shows that for most systems
of $2a_{\rm b}=0.5$~AU studied, the choice of $e_{\rm b}$ presents
little hindrance for the occurrence of circumbinary HZs.
Exceptions include, however, some cases of low-mass stellar
systems, especially systems of $M_2=0.50~M_\odot$.
If the GHZ is assumed instead of the CHZ, the
number of cases without {\it P}/{\it PT-}type habitability
is further reduced.  For the majority of models studied
with major axes of 0.5~AU, it is found that
$0.8 \le e_{\rm b} {\rm (crit.)} < 1$.  Furthermore, for
models with major axes of 1.0~AU, $e_{\rm b}$ (crit.)
is often identified as relatively small, or, alternatively,
circumbinary habitability is absent altogether.
Especially if the CHZ is considered rather than the GHZ,
circumbinary habitability is highly restricted.  The
only exception is the case of $M_1=M_2=1.25~M_\odot$,
where in the framework of our model no restriction for
$e_{\rm b}$ has been identified.

Figure~8 provides a summary of {\it P}/{\it PT-}type
habitability for several equal-mass and nonequal-mass
binary systems by displaying the widths of the respective
HZs.  It is found that for $2a_{\rm b}=0.5$~AU, the impact of
$e_{\rm b}$ on the size of the circumbinary HZs is not
very large for most systems.  However, $2a_{\rm b}=1.0$~AU, the
situation is notably different, also taking into account
that for many cases circumbinary HZs can be found, especially
for cases of intermediate or high values of $e_{\rm b}$.
Moreover, Figure~8 also displays the widths of the EHZ
for a selected number of cases.  Here it is found that
in many cases, the extent of the EHZ exceeds that for the
GHZ by about a factor of 2, irrespective of the
value adopted for $e_{\rm b}$.

\subsection{Case Studies of {\it S}/{\it ST-}type Habitability}

In the following, we describe results pertaining to {\it S-}type
and {\it ST-}type habitability.  Akin to Section~3.2, we again
consider systems of main-sequence stars; see Table~1
for information on the stellar parameters.  We also inspect
the values for RHZ$_{\rm in}$ and RHZ$_{\rm out}$ at the stellar
periastron and apastron position, denoted as $a_{\rm per}$ and
$a_{\rm ap}$, respectively, to provide additional verification
for the method of calculating the RHZs for $e_{\rm b} > 0$ (see
Section~2.1).  Particularly, we include systems of low-mass /
low-luminosity stars owing to their relatively high abundance;
see Table~9 to 11 for a depiction of our results.  Additional
studies, including systems with binary masses of $0.65~M_\odot$,
were given by \cite{cunb14}.  These studies are based on the
same rigor and limitations as considered in Paper~I and II.

Table~9 conveys results for equal-mass binary systems of
$M_1 = M_2 = 1.0~M_\odot$ regarding both the CHZ and GHZ
for $2a_{\rm b}=20$~AU for different eccentricities of the binary
components.  Information is given regarding both the RHZ
(see column ``Orbit") and the orbital stability limited
denoted as $a_{\rm cr}$.  It is found that regarding both
the CHZ and GHZ, the RHZ$_{\rm in}$ and RHZ$_{\rm out}$ are
only very weakly dependent on $e_{\rm b}$ except for cases
beyond $e_{\rm b} \simeq 0.65$, which are however of little
interest because those cases lack habitability
due to the orbital stability requirement.  Thus, generally,
the RHZs extend from 1.05 to 1.50~AU for the CHZ and from
0.93 to 1.84~AU for the GHZ.  The orbital stability limit,
which constitutes an upper limit, moves inward as a
function of the binary eccentricity, starting at 2.74~AU
for the circular case and taking values of 1.77 and 0.90~AU
at $e_{\rm b} = 0.30$ and 0.60, respectively.  Therefore,
for the CHZ, {\it S}-type habitability is obtained at
eccentricities below 0.39, and {\it ST}-type habitability
is obtained at eccentricities below 0.54; the latter value
is again referred to as critical value of $e_{\rm b}$
(see also Table~11).  Moreover, for the GHZ, {\it S}-type
and {\it ST}-type habitability occurs below eccentricities
of 0.28 and 0.59, respectively.

Table~10 conveys results for nonequal-mass binary systems of
$M_1 = 1.25~M_\odot$ and $M_2 = 0.75~M_\odot$.  It is again
found that for both the CHZ and GHZ, the RHZ$_{\rm in}$ and
RHZ$_{\rm out}$ are barely dependent on $e_{\rm b}$ except
for cases beyond $e_{\rm b} \simeq 0.65$, which again are
of little interest because they lack habitability due to
the orbital stability requirement.  It is found that the
RHZs extend from 1.37 to 1.90~AU for the CHZ and from
1.21 to 2.34~AU for the GHZ.  The orbital stability limit,
which constitutes an upper limit, moves inward as a
function of the binary eccentricity, starting at 3.22~AU
for the circular case and taking values of 2.02 and 1.01~AU
at $e_{\rm b} = 0.30$ and 0.60, respectively.  Compared
to the previous case of the equal-mass binary system,
both the RHZ-CHZ and the RHZ-GHZ are broader and are at
a larger stellar distance due to the higher value of $L_1$,
i.e., 2.15~$L_\odot$ versus 1.23~$L_\odot$.  For the CHZ
of the nonequal-mass system, {\it S}-type habitability is
identified at eccentricities below 0.33, and {\it ST}-type
habitability is obtained at eccentricities up to 0.49.
Furthermore, for the GHZ, {\it S}-type and {\it ST}-type
habitability exists up to eccentricities of 0.21 and 0.54,
respectively.

Figure~9 depicts the widths of {\it S}/{\it ST}-type habitable
zones for the GHZ of various binary systems for $2a_{\rm b}=15.0$~AU
and $2a_{\rm b}=20.0$~AU for various types of systems between masses
of 1.25~$M_\odot$ and 0.50~$M_\odot$.  For some of the systems,
information is also given regarding the CHZ and EHZ.  The
finding for the various models is that between the circular
case of $e_{\rm b} = 0$ up to a distinct value of $e_{\rm b}$,
the widths of the HZs are constant ({\it S}-type habitability);
thereafter, they decrease linearly as a function of $e_{\rm b}$
({\it ST}-type habitability).  Finally, for each model, the
critical value of $e_{\rm b}$ is found.  Beyond that value,
no habitable region exists.
Figure~9 and Table~11 allow for the identification of general properties
of systems with different combinations of masses and, by
implication, luminosities of the stellar components.  The
findings include that in models of low eccentricity, i.e.,
$e_{\rm b} \simeq 0$, the widths of the HZs are largely
determined by the luminosity of the stellar primary, which
means that the higher its luminosity the larger is the
resulting HZ.  This result is valid for the CHZ, GHZ, as
well as the EHZ.  In the absence of truncation imposed by the
orbital stability limit, the widths for the EHZs are largest
and the widths for the GHZs and CHZs are respectively smaller,
as expected. 

The width of the {\it S}-type habitable region is largest
for systems where the stellar primary possesses the greatest
luminosity.  However, these systems are also characterized by
relatively low critical values of $e_{\rm b}$ owing to highly
efficient truncation of the RHZ due to the orbital stability
limit because of the primary's relatively high mass.  Conversely,
for systems of low luminosity, relatively high numbers for the
critical value of $e_{\rm b}$ are obtained.  Even though the
{\it S}/{\it ST}-type habitable zones have small widths in most
of those systems, habitability is still possible for systems of
high eccentricity.  For example, for systems of $M_1 = 1.25~M_\odot$
and $M_2 = 0.75~M_\odot$ with $2a_{\rm b}=20$~AU, the critical values for
the eccentricity of the binary components are given as 0.49
and 0.54 for the CHZ and GHZ, respectively.  However, for
systems of $M_1 = M_2 = 0.75~M_\odot$, the critical values of
$e_{\rm b}$ are given as 0.72 and 0.74, respectively.  Moreover,
for equal-mass systems of 0.50~$M_\odot$, the critical values of
$e_{\rm b}$ are larger than 0.80, which is the upper limit for
$e_{\rm b}$ considered in our study; see Table~11 and the
work of \cite{cunb14} for additional results.

For $2a_{\rm b}=20$~AU, we again studied the domains of
the quantities {\%}RHZ and {\%}HZ for the various models.
{\%}RHZ denotes the relative size of the RHZ for a given set of
models with the special case of $e_{\rm b} = 0$ set as 100 (see
Tables~9 to 10).  Furthermore, {\%}HZ denotes the relative size
of the HZ, which contrary to the {\%}RHZ also takes into account
the orbital stability constraint; see Section~3.2 for results on
the {\it P}/{\it PT-}type habitable regions.  According to the
model calculations-as-pursued, it was found that the decrease in
the {\%}RHZ for increasing values of $e_{\rm b}$ is very small.
Regarding the 18 models for the equal-mass binary of $M_1 = M_2
= 1.0~M_\odot$, obtained with respect to the CHZ and GHZ, it is
found that for 10 models, the reduction in the RHZ is less than
1\%, and for 14 models, it is less than 2\%.  Nevertheless, for
relatively high values of $e_{\rm b}$, the reduction in the {\%}HZ 
is highly notable due to the impact of the orbital stability
constraint.

Regarding the 18 models for the nonequal-mass binary system
given as $M_1 = 1.25~M_\odot$ and $M_2 = 0.75~M_\odot$, the
following behavior was identified.  Here the reduction of the
RHZ is even less significant than in the equal-mass system
previously discussed.  In fact, regarding the 18 models
considered, the RHZ is reduced by 1\% or less in 15 models,
and by 2\% or less in 17 models.  This result is due to the
the lower luminosity of the secondary star compared to
the previous system.  However, the impact of the orbital
stability constraint in the nonequal-mass system is much more
severe than in the equal-mass system previously discussed,
as communicated by the numerical values of {\%}HZ.  Regarding
the CHZ and GHZ, habitable regions are rendered impossible
for eccentricities beyond 0.49 and 0.54, respectively (see
Table~11).

Finally, we also compared systems of different major axes,
specifically $2a_{\rm b}=15$~AU and $2a_{\rm b}=20$~AU.  It was found that
the widths of the HZs, especially for models of low eccentricity,
were largely unaffected by the value of binary major axis; however,
the critical values of $e_{\rm b}$ were still influenced by the
choice of the major axis.  These results are also consistent with
the analysis of Figures~6 and 7 (see Section~3.1), which consider
systems of different major axes $2a_{\rm b}$.  In those systems, the
{\it S-}type RHZs were found to be largely independent of the values
of the major axes; however, in systems of high eccentricity, habitability
has been highly restricted or nullified due to the planetary orbital
stability requirement.

%%%%%%%%%%%%%%%%%%%%%%%%%%%%%%%%%%%%%%%%%%%%%%%%%%%%%%%%%%%%%%%%%%%%%%%%

\section{SUMMARY AND CONCLUSIONS}

The method as described allows the calculation of {\it S}-type and
{\it P}-type habitable zones in stellar binary systems.  {\it P}-type
orbits occur when the planet orbits both binary components, whereas in
the case of {\it S}-type orbits, the planet orbits only one of the binary
components with the second component considered a perturbator.   
The selected approach considers a variety of aspects including:
(1) besides simple cases, the treatment of nonequal-mass systems and
systems in elliptical orbits; this latter aspect is a significant
augmentation of the methodology given in Paper~I; (2) the
consideration of a joint constraint, including orbital stability
and a habitable region for a possible system planet through the stellar
radiative energy fluxes; (3) the provision of a combined
formalism for the assessment of both {\it S}-type and {\it P}-type
habitability; in particular, through the solution of a fourth-order
polynomial, mathematical criteria are employed for the kind of system
in which {\it S}-type and {\it P}-type habitability is realized, an
approach in alignment with that of Paper~I.  

If the RHZs are truncated by the additional constraint of
orbital stability, the notation of {\it ST}-type and {\it PT}-type
habitability is used.  In comparison to the circular case, it is
found that in systems of higher eccentricity, the domains of the RHZs
are significantly reduced.  Furthermore, the orbital stability
constraint also impacts {\it S}-type and {\it P}-type habitability,
in an unfavorable manner; this latter aspect is particularly
relevant to the overall context of astrobiology.
Nonetheless, {\it S}-, {\it P}-, {\it ST}-, and {\it PT}-type
habitability is identified for a considerable set of system parameters.
Compared to Paper~I, which was focused on the circular case, a modest
modification of the method allows for the consideration of elliptical
orbits, entailing a significantly augmented range of applicability for
the proposed method.

It is beyond the scope of this work to present a comprehensive parameter
study.  Besides the choice of the appropriate type of HZ, the system
parameters for standard main-sequence stars (see Table~1) pertaining
to our model include the major axis $2a_{\rm b}$, the eccentricity of
the binary components $e_{\rm b}$, and the stellar luminosities $L_1$
and $L_2$, or alternatively, the stellar masses $M_1$, and $M_2$.  In the
case of general stars, include subgiants and giants, the set of parameters
would be further enlarged due to the lack of a universal mass--luminosity
relationship.  Selected results for low-mass main-sequence star binary
systems have previously been given by \cite{cunb14}.

It is found that an increased value for the eccentricity in stellar
binary systems results in an adverse influence both regarding the RHZs
and the orbital stability of possible system planets.  For the latter,
it is found that for {\it P}-type habitability, the limits of orbital
stability move inward, and for {\it S}-type habitability, the limits
of orbital stability move outward, thus potentially reducing the size
of habitable regions for, respectively, circumbinary and circumstellar
planets.  Restrictions due to the orbital stability criterion
are especially significant for {\it S}-type habitability for systems
in highly elliptical orbits, since at the periastron position large segments
of the RHZs are nullified.  However, restrictions due to limited orbital
stability in systems of highly elliptical orbits are less relevant 
regarding {\it P}-type habitability; in this case the most influential
factor is given by the distribution of luminosity between the stellar
components.  As discussed in Paper~I, highly unequal distributions of
luminosity often entail small or nonexisting RHZs.

Our studies also show that increased eccentricities for the stellar
binary components lead to smaller widths of the HZs, as well as the
lack of existence of HZs for relatively large values of $e_{\rm b}$.
In order to nonetheless offer some degree of habitability for systems
in notably eccentric orbits, the finding places extra weight on previous
studies, which show that under distinct conditions (e.g., relatively
thick planetary atmospheres, distinct types of planetary atmospheric
compositions) exoplanets are able to leave the RHZs without forfeiting
habitability.  For example, \cite{wil02} argued---based on 3-D
general-circulation climate models and 1-D energy balance models aimed
at examining Earth-type planets on extremely elliptical orbits near the
HZ---exoplanets, as well as exomoons hosted by Jupiter-type planets
\citep[e.g.,][]{hel12,hel14}, will be able to harbor life
even when undergoing excursions beyond the HZ.  The reason is that
in the framework of their models, long-term climate stability depends
primarily on the average stellar flux received over an entire orbit,
rather than the length of time spent within the HZ.

As previously discussed, a second criterion---besides the existence
of the RHZs---for facilitating habitability consists in the orbital
stability requirement.  This criterion has heightened relevance compared
to the radiative criterion because evidently ``there is no way of
escaping gravity" as due to system conditions, gravity-imposed planetary orbital
instabilities may occur.  Our present study employs the stability limits
given by \cite{hol99}.  They are based on numerical simulations for test
particles able to survive $10^4$ binary orbits.  Considering the need for
detailed habitability studies based on timescales of up to several billions
of years, as implied by terrestrial biology, it appears appropriate to
update the previous work of \citeauthor{hol99} based on significantly
elongated timescales.  An expected outcome of this type of investigation
will be updated stability limits for both {\it P}-type and {\it S}-type
orbits, which will likely be more stringent than currently attained. 

Based on existing studies and tools given in the literature,
future studies should include: (1) how limits of habitability
(i.e., $s_\ell$; see Section~2.1) depend on the properties of the
planet itself, such as planetary atmosphere, geodynamics, rotation rate,
and mass; see, e.g., \cite{blo09}, \cite{sea13}, \cite{lec13},
\cite{zso13}, \cite{kop13,kop14}, and \cite{yan14}, (2) models
of binary systems where the component undergoes significant
time-dependent evolution, and (3) models where the orbit of
the terrestrial planet is misaligned with the orbital plane of
the stellar binary components.  This latter aspect is motivated
by ongoing discoveries, including the detection of misaligned
protoplanetary disks in the young binary system HK Tauri
\citep{jen14}.  In this case, it is found that the disks are
misaligned by 60$^\circ$ to 68$^\circ$, such that one or both
of the disks are significantly inclined to the binary orbital
plane.  Any future system planet will thus most likely result
in a highly eccentric as well as  inclined orbit.  Such outcomes
will pose a serious challenge to the calculation of both RHZs
and orbital stability limits needed for future investigation
of {\it S-} and {\it P-}type habitability.

\acknowledgments
This work has been supported in part
by NASA's Goddard Space Flight Center.  The author
acknowledges comments by Robert Bruntz and Billy Quarles as well as
assistance by Satoko Sato and Zhaopeng Wang with computer graphics.
He also wishes to draw the reader's attention to the online code
{\tt BinHab}, hosted at The University of Texas at Arlington (UTA),
which allows the calculation of habitable regions in binary systems
based on the developed method.

%%%%%%%%%%%%%%%%%%%%%%%%%%%%%%%%%%%%%%%%%%%%%%%%%%%%%%%%%%%%%%%%%%%%%%%%

%%%%%%%%%%%%%%%%%%%%%%%%%%%%%%%%%%%%%%%%%%%%%%%%%%%%%%%%%%%%%%%%%%%%%%%%

\clearpage

%+++++++++++++++++++++++++++++++++++++++++++++++++++++++++++++++++++++++++

%%% *** Fig.1
%%%%%%%%%%%%%%%%%%%%%%%%%%%%%%%%%%%%%%%%%%%%%%%%%%%%%%%%%%%%%%%%%
\begin{figure*} 
\centering
\begin{tabular}{c}
\epsfig{file=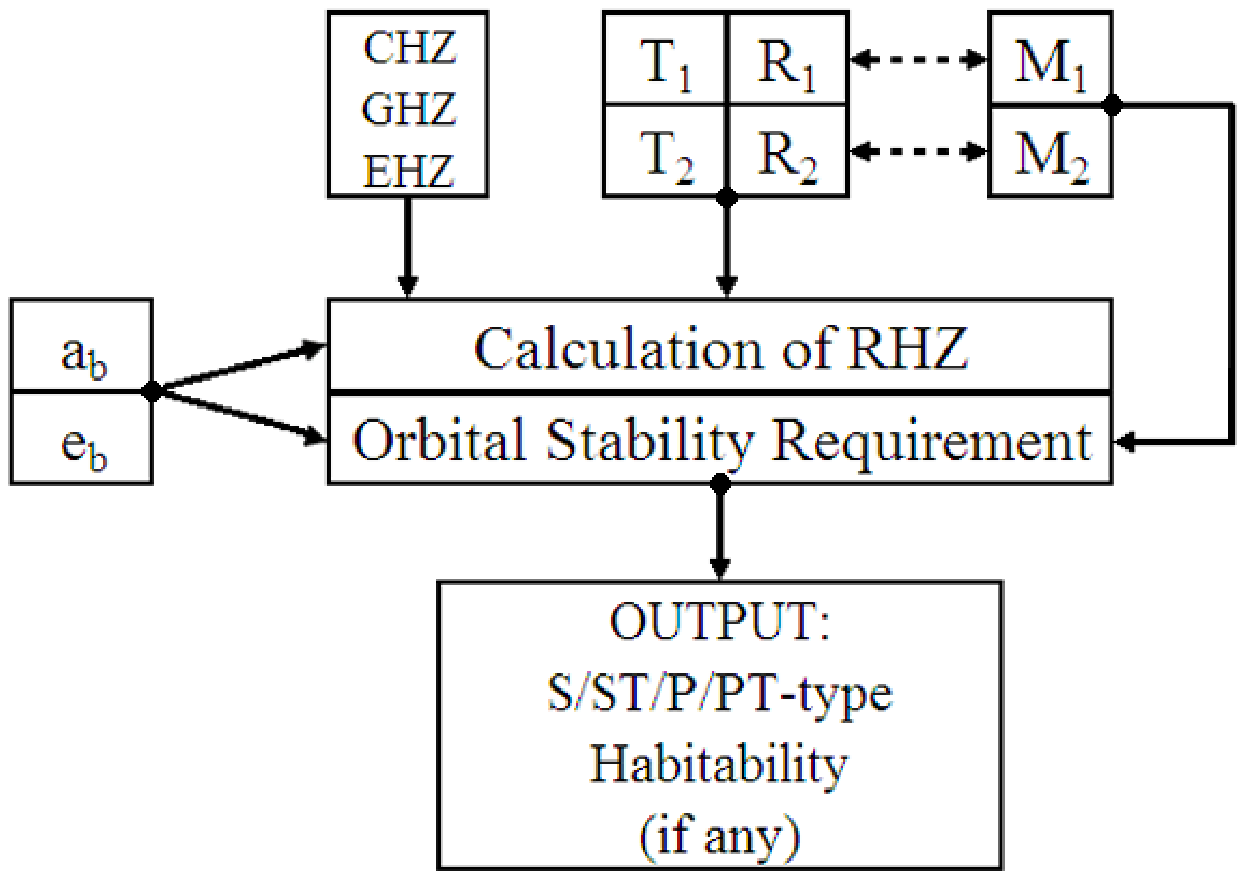,width=0.75\linewidth} \\
\end{tabular}
\caption{Flow diagram of {\tt BinHab} indicating the adopted method of solution.
Feed-ins and the generation of the output are indicated by arrowed solid lines.
For theoretical main-sequence stars, effective temperatures and stellar radii
($T_i$, $R_i$, with $i=1,2$), on the one hand, or stellar masses ($M_i$), on the
other hand, may serve as input parameters, as indicated by double-arrowed dashed
lines.  See \cite{cunb14} for additional information.
}
\end{figure*}

\clearpage

%+++++++++++++++++++++++++++++++++++++++++++++++++++++++++++++++++++++++++

%%% *** Fig.2
%%%%%%%%%%%%%%%%%%%%%%%%%%%%%%%%%%%%%%%%%%%%%%%%%%%%%%%%%%%%%%%%%
\begin{figure*} 
\centering
\begin{tabular}{c}
\epsfig{file=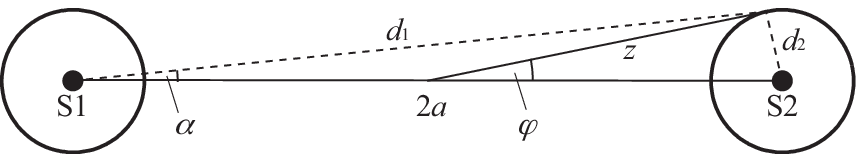,width=0.75\linewidth} \\
\noalign{\bigskip}
\noalign{\bigskip}
\epsfig{file=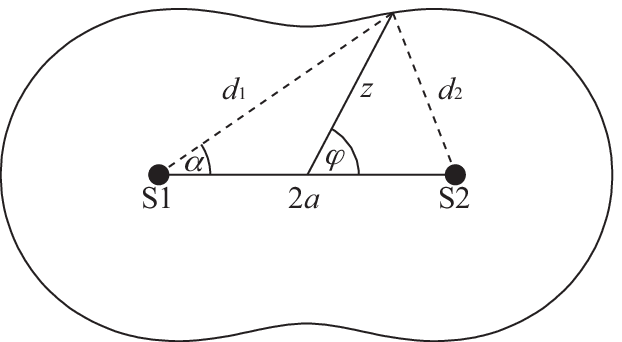,width=0.50\linewidth} \\
\end{tabular}
\caption{Set-up for the mathematical treatment of {\it S}-type (top)
and {\it P}-type (bottom) habitable zones of binary systems as given
by the stellar radiative fluxes with $a \equiv a_{\rm b}$.  Note that
the stars S1 and S2 have been depicted as identical for convenience.
(Adopted from Paper~I.)
}
\end{figure*}

\clearpage

%+++++++++++++++++++++++++++++++++++++++++++++++++++++++++++++++++++++++++

%%% *** Fig.3
%%%%%%%%%%%%%%%%%%%%%%%%%%%%%%%%%%%%%%%%%%%%%%%%%%%%%%%%%%%%%%%%%
\begin{figure*} 
\centering
\begin{tabular}{c}
\epsfig{file=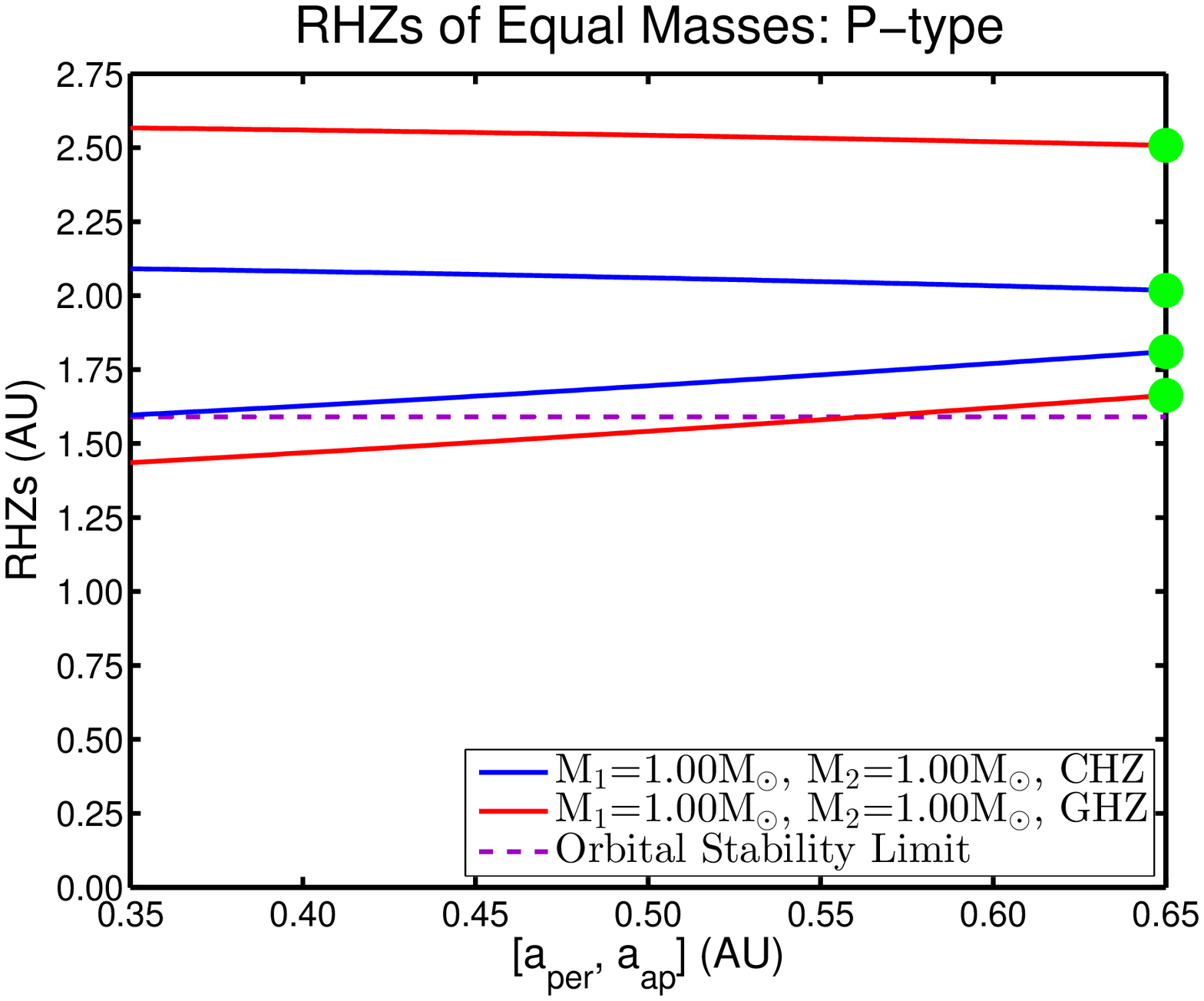,width=0.60\linewidth} \\
\epsfig{file=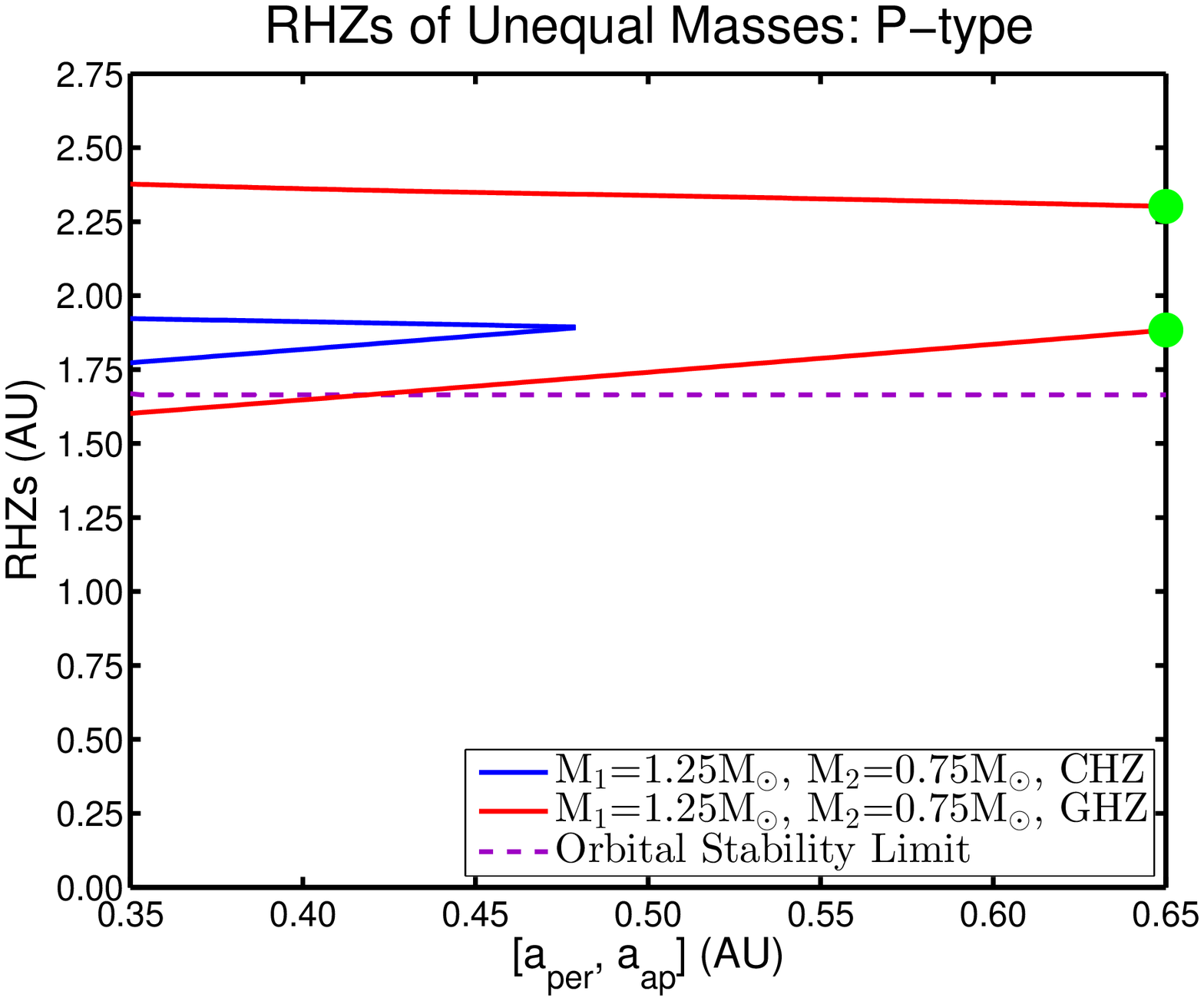,width=0.60\linewidth} \\
\end{tabular}
\caption{
RHLs for {\it P}-type habitability for various
models in the situations when the binary components are in periastron
($a_{\rm per}$), apastron ($a_{\rm ap}$), or intermediate position.  
The top panel depicts the case of $M_1 = M_2 = M_\odot$, whereas
the bottom panel depicts the case of $M_1 = 1.25~M_\odot$ and 
$M_2 = 0.75~M_\odot$.  The models are for $2a_{\rm b}=1.0$~AU and
$e_{\rm b}=0.30$; i.e., $a_{\rm per} = 0.35$~AU and $a_{\rm ap} = 0.65$~AU.
The blue and red lines indicate the RHLs for the CHZs and
GHZs, respectively.  The planetary orbital stability limits are given
as dashed purple lines for comparison.  Note that for defining the
stringent inner and outer limits of the RHZs, the RHLs at the apastron
positions are decisive; they are indicated by green dots.  Also, in case
of the CHZ for unequal masses, no orbit-wide RHZ exists. 
}
\end{figure*}

\clearpage

%+++++++++++++++++++++++++++++++++++++++++++++++++++++++++++++++++++++++++

%%% *** Fig.4
%%%%%%%%%%%%%%%%%%%%%%%%%%%%%%%%%%%%%%%%%%%%%%%%%%%%%%%%%%%%%%%%%
\begin{figure*} 
\centering
\begin{tabular}{c}
\epsfig{file=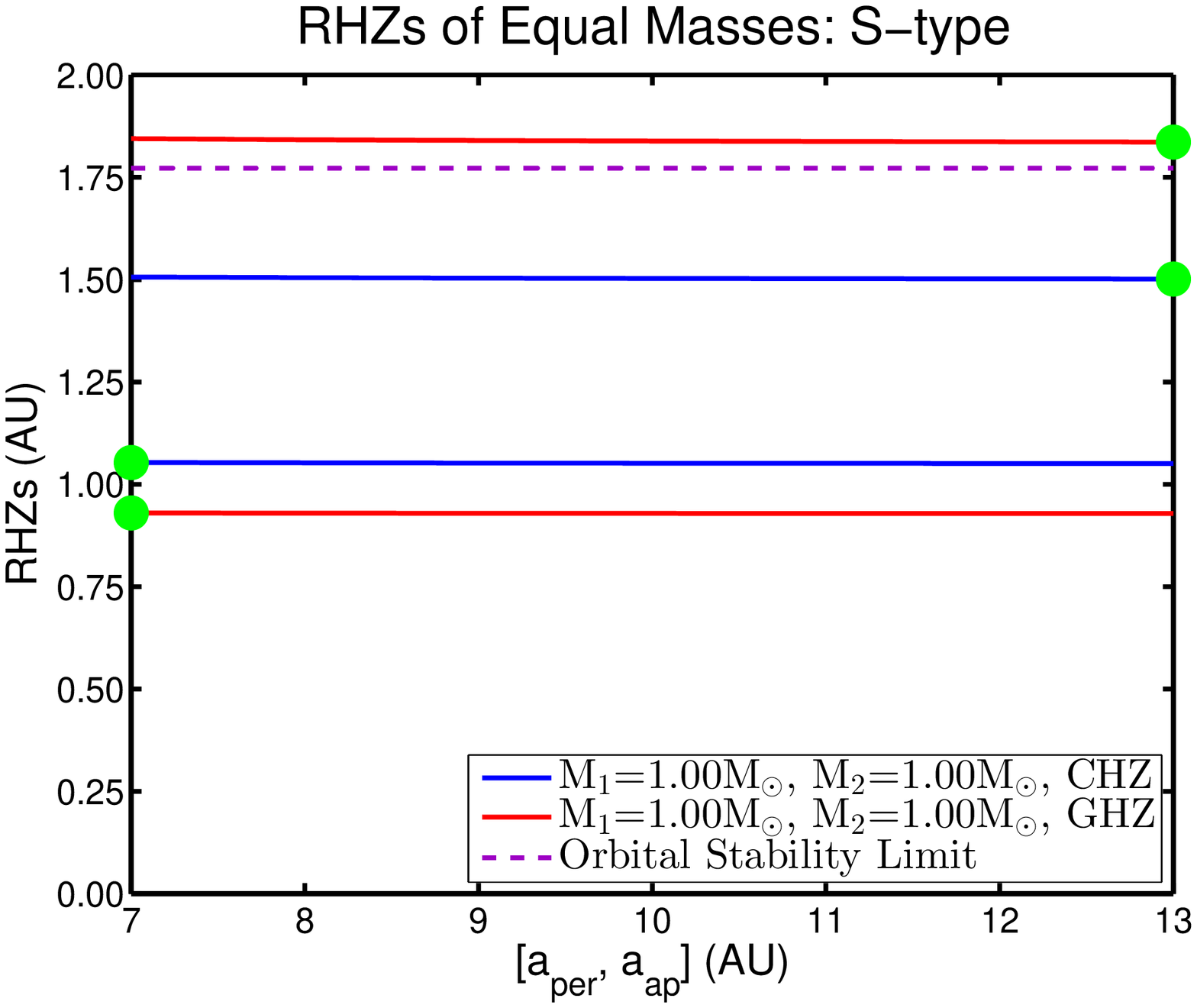,width=0.60\linewidth} \\
\epsfig{file=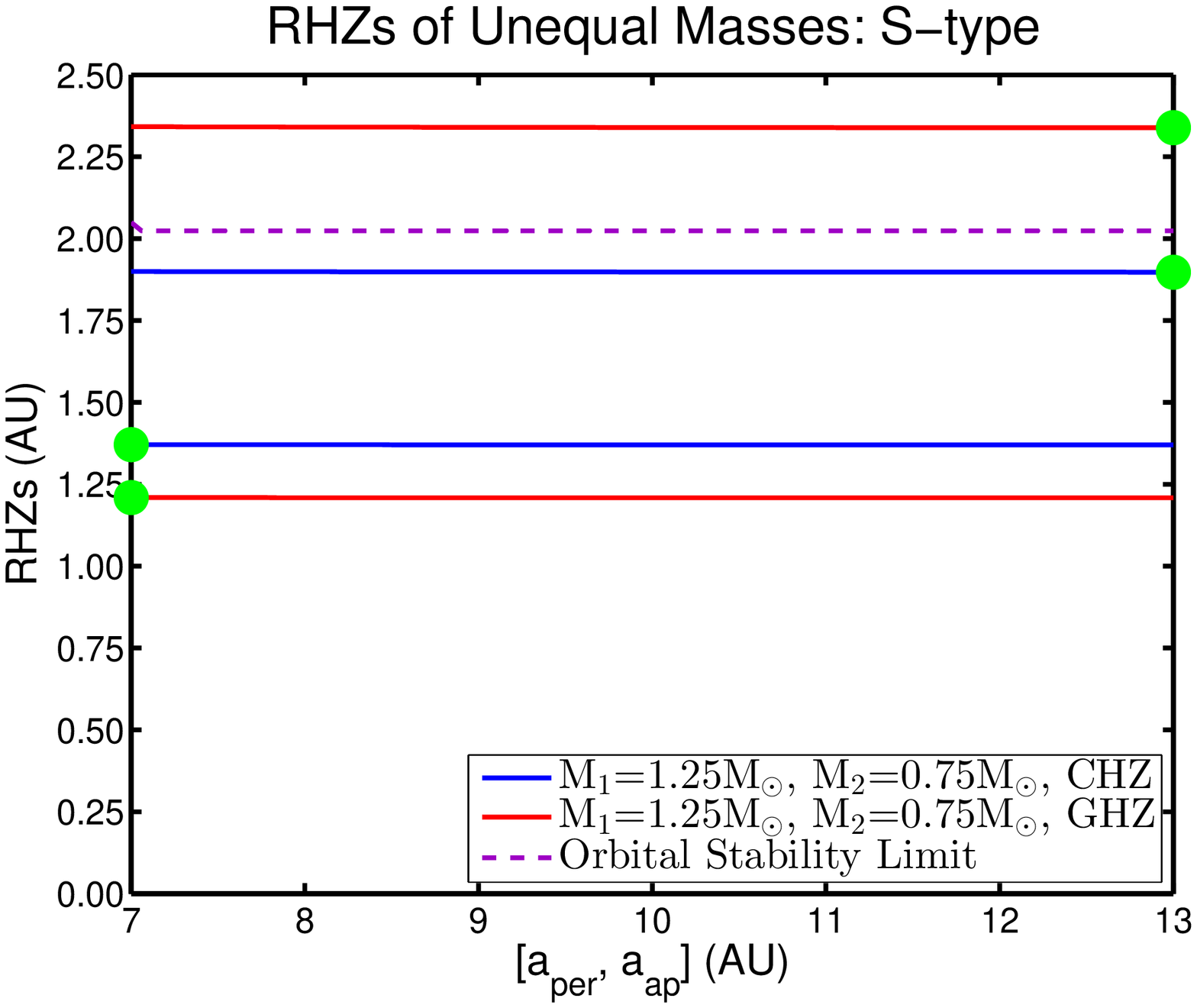,width=0.60\linewidth} \\
\end{tabular}
\caption{
Same as Figure~3, but now regarding the RHLs for {\it S}-type
habitability.  The models are for $2a_{\rm b}=20.0$~AU and $e_{\rm b}=0.30$;
i.e., $a_{\rm per} = 7.0$~AU and $a_{\rm ap} = 13.0$~AU.
The blue and red lines indicate the RHLs for the CHZs and
GHZs, respectively.  The planetary orbital stability limits are given
as dashed purple lines for comparison.  Note that in case of {\it S}-type
habitable regions, the stringent inner limits of the RHZs are set by the RHLs
at the periastron positions, whereas the stringent outer limits of the RHZs
are set by the RHLs at the apastron positions (see green dots).  However, the
differences in the RHLs between the $a_{\rm per}$ and $a_{\rm ap}$ positions
are very minor.
}
\end{figure*}

\clearpage

%+++++++++++++++++++++++++++++++++++++++++++++++++++++++++++++++++++++++++

%%% *** Fig.5
%%%%%%%%%%%%%%%%%%%%%%%%%%%%%%%%%%%%%%%%%%%%%%%%%%%%%%%%%%%%%%%%%
\begin{figure*} 
\centering
\begin{tabular}{cc}
\epsfig{file=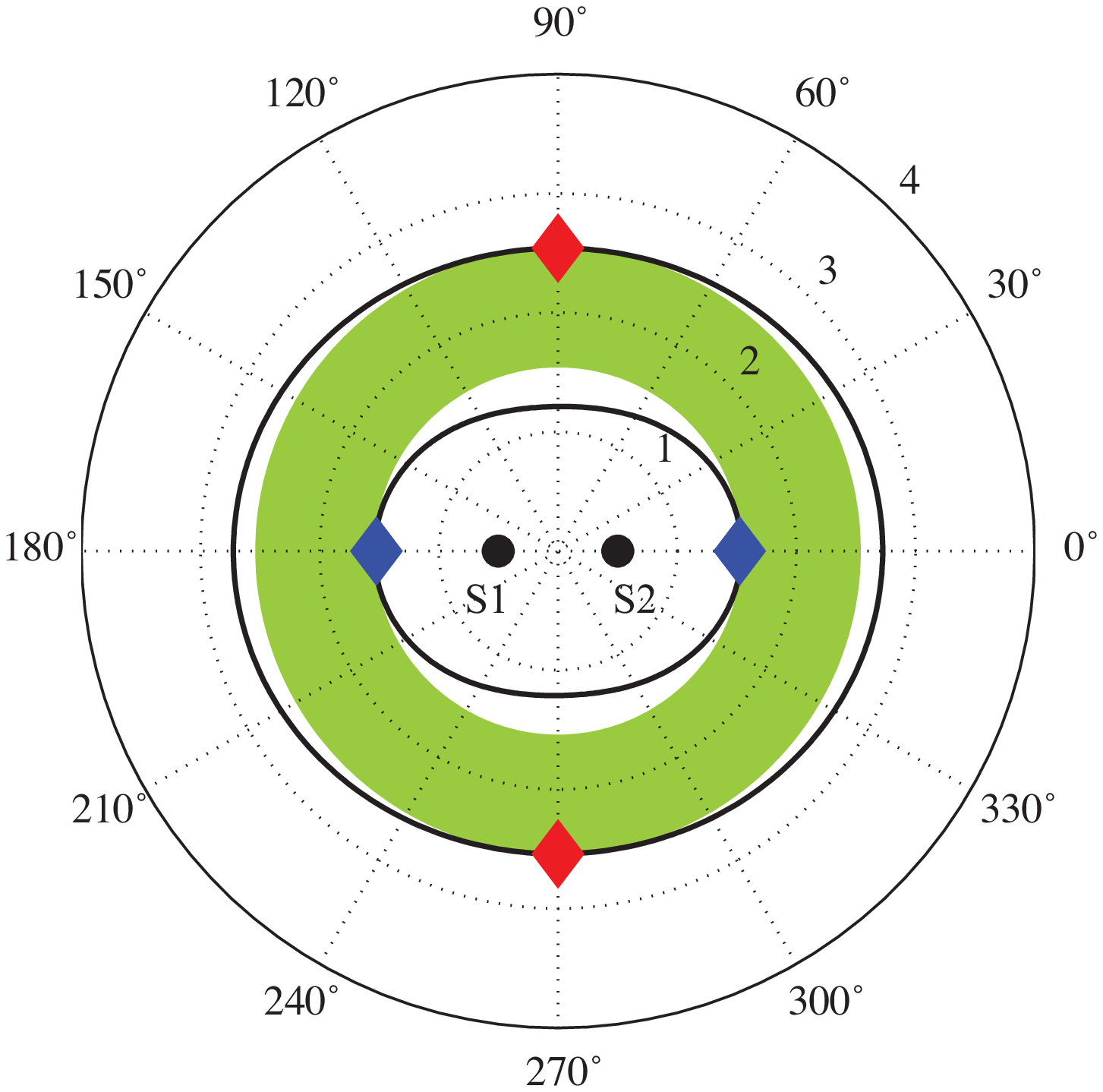,width=0.45\linewidth} &
\epsfig{file=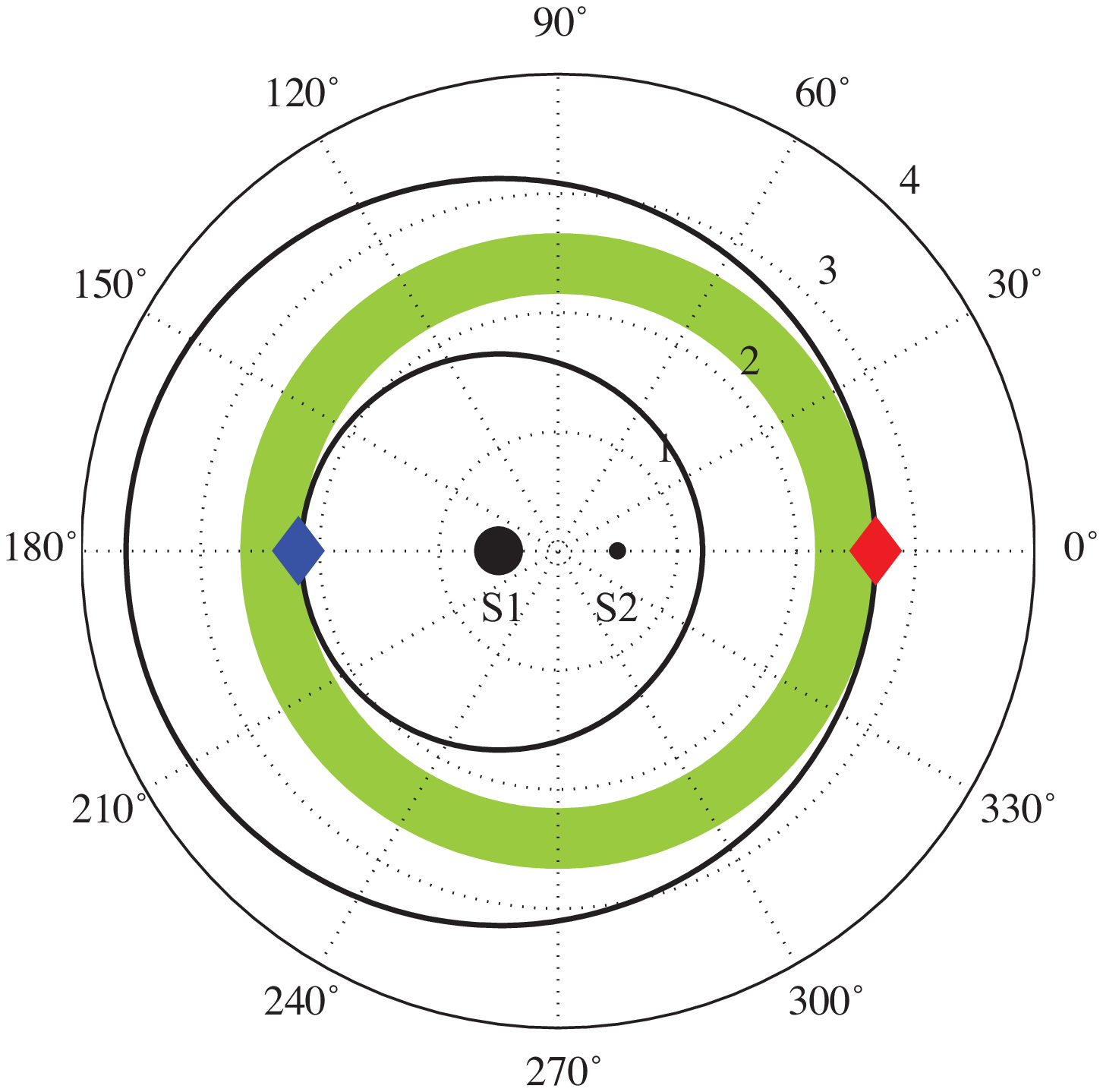,width=0.45\linewidth} \\
\epsfig{file=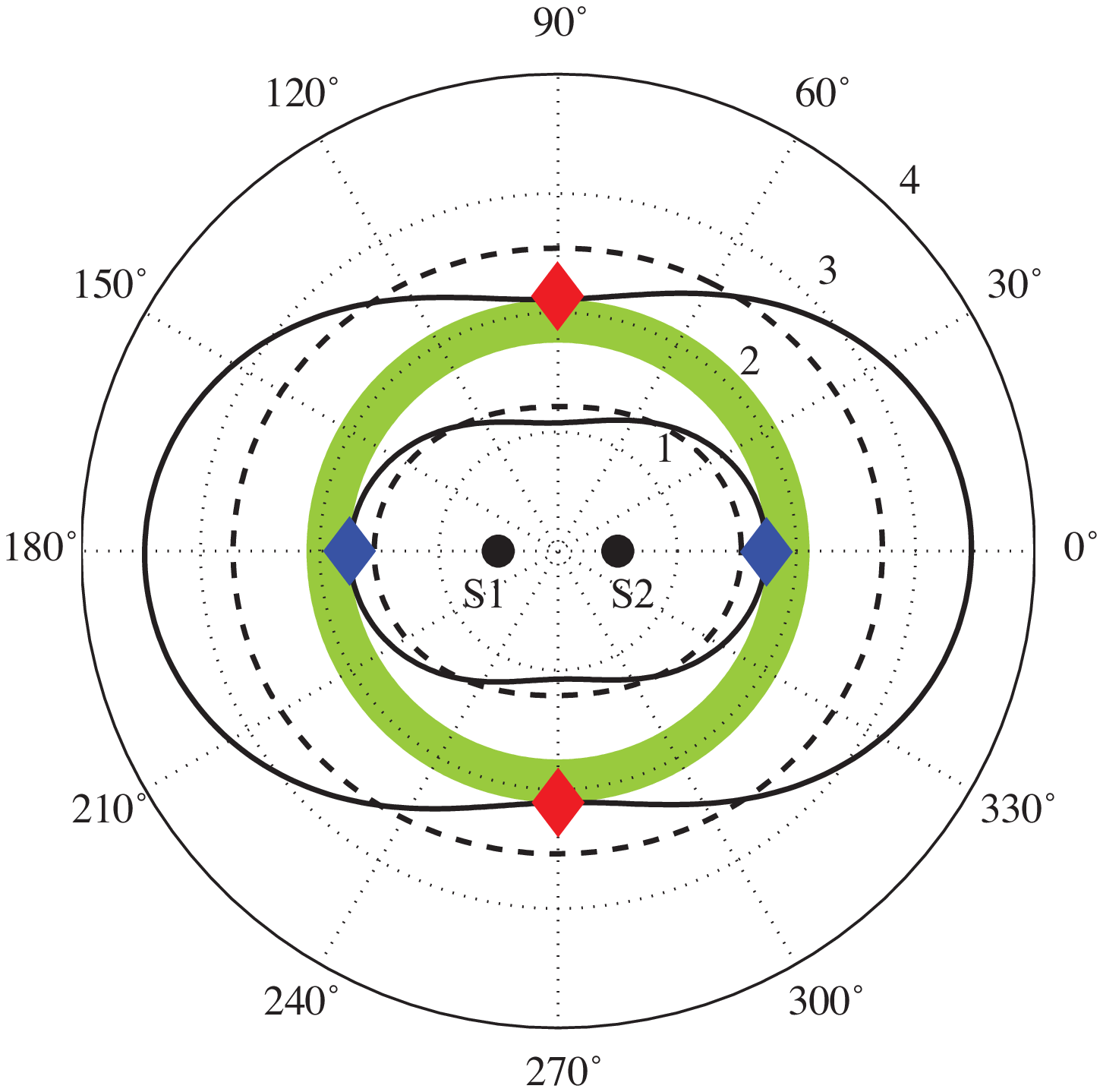,width=0.45\linewidth} &
\epsfig{file=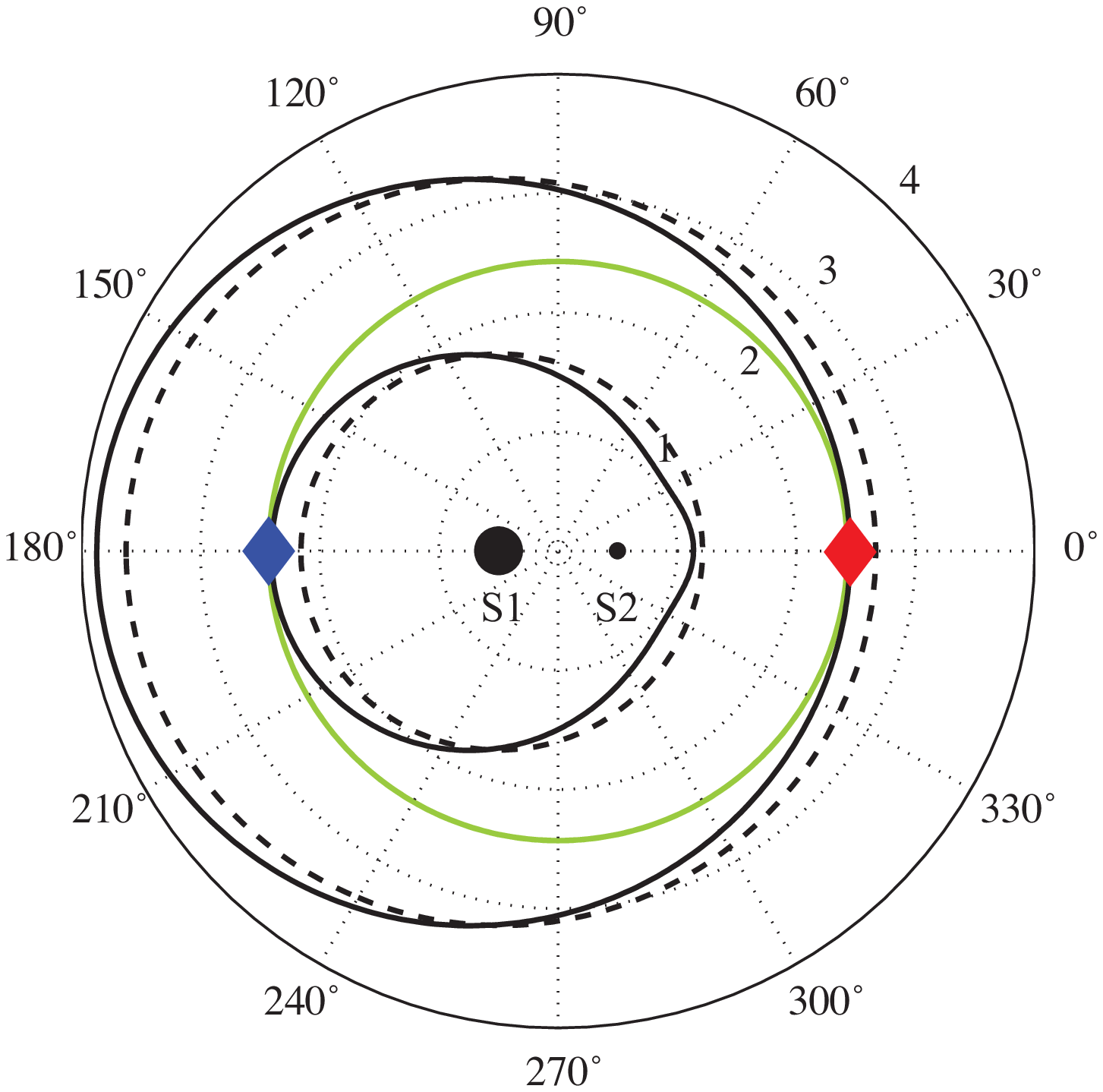,width=0.45\linewidth} \\
\end{tabular}
\caption{
Examples of {\it P}-type RHZs for different systems.  The solutions are given
in polar coordinates, with the radial coordinate depicted in units of AU.
The thick solid lines indicate the RHLs corresponding to the inner and
outer limit of habitability, based on $s_\ell = 0.84$ and 1.67~AU, respectively.
The top row displays systems with $e_{\rm b} = 0.0$, whereas the bottom row
displays systems with eccentricity $e_{\rm b} = 0.5$.  Furthermore,
there are different cases of stellar mass, and by implication
stellar luminosity, noting that the left column features systems with
$M_1 = M_2 = M_\odot$, whereas the right column features systems with
$M_1 = 1.5~M_\odot$ and $M_2 = 0.5~M_\odot$.  The green areas indicate the
appropriate circular regions (annuli), referred to as RHZs, for each case.
The touching points between the RHZs and the inner and outer RHLs (utilized
for the definition of the RHZs) are depicted as blue and red diamonds,
respectively.  For the elliptical systems (bottom row), the RHLs of
the corresponding circular systems (top row) are given for comparison
as dashed lines.  Note that the extent of the
RHZ is significantly reduced in systems of unequal stellar luminosities
or nonzero eccentricity of the stellar components, with the most extreme
reductions occurring for combined conditions.
}
\end{figure*}

\clearpage

%+++++++++++++++++++++++++++++++++++++++++++++++++++++++++++++++++++++++++

%%% *** Fig.6
%%%%%%%%%%%%%%%%%%%%%%%%%%%%%%%%%%%%%%%%%%%%%%%%%%%%%%%%%%%%%%%%%
\begin{figure*} 
\centering
\begin{tabular}{cc}
\epsfig{file=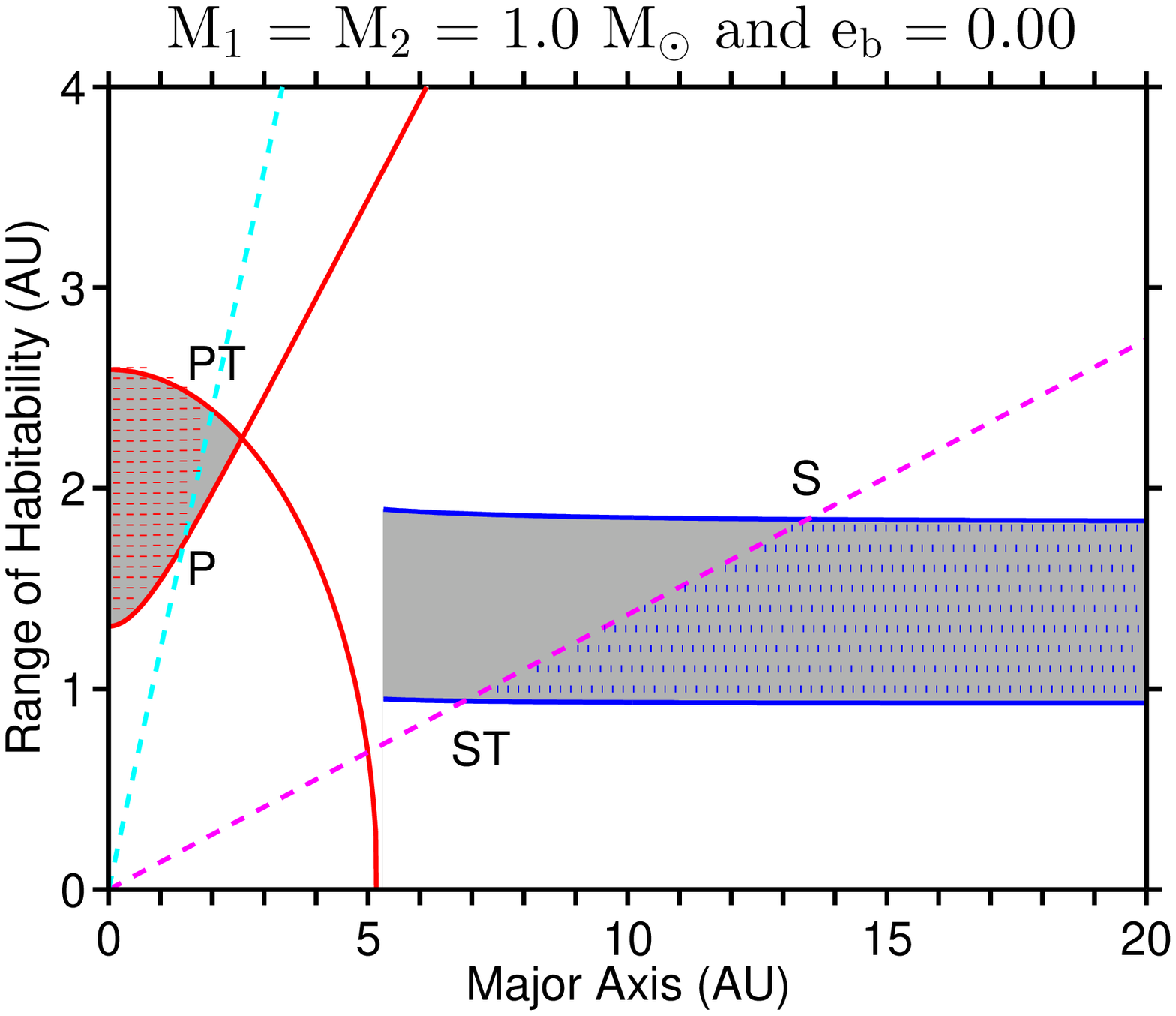,width=0.45\linewidth} &
\epsfig{file=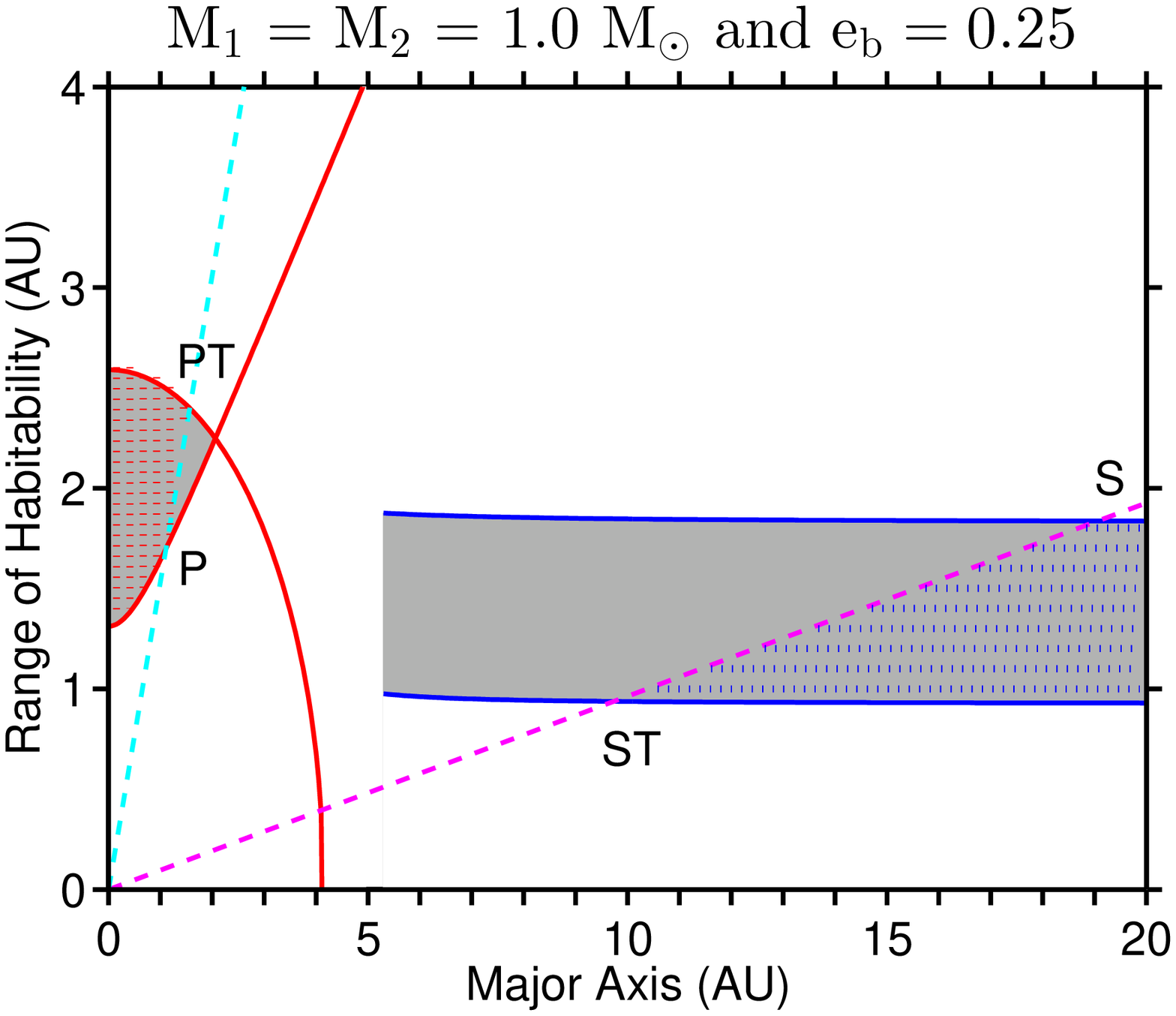,width=0.45\linewidth} \\
\epsfig{file=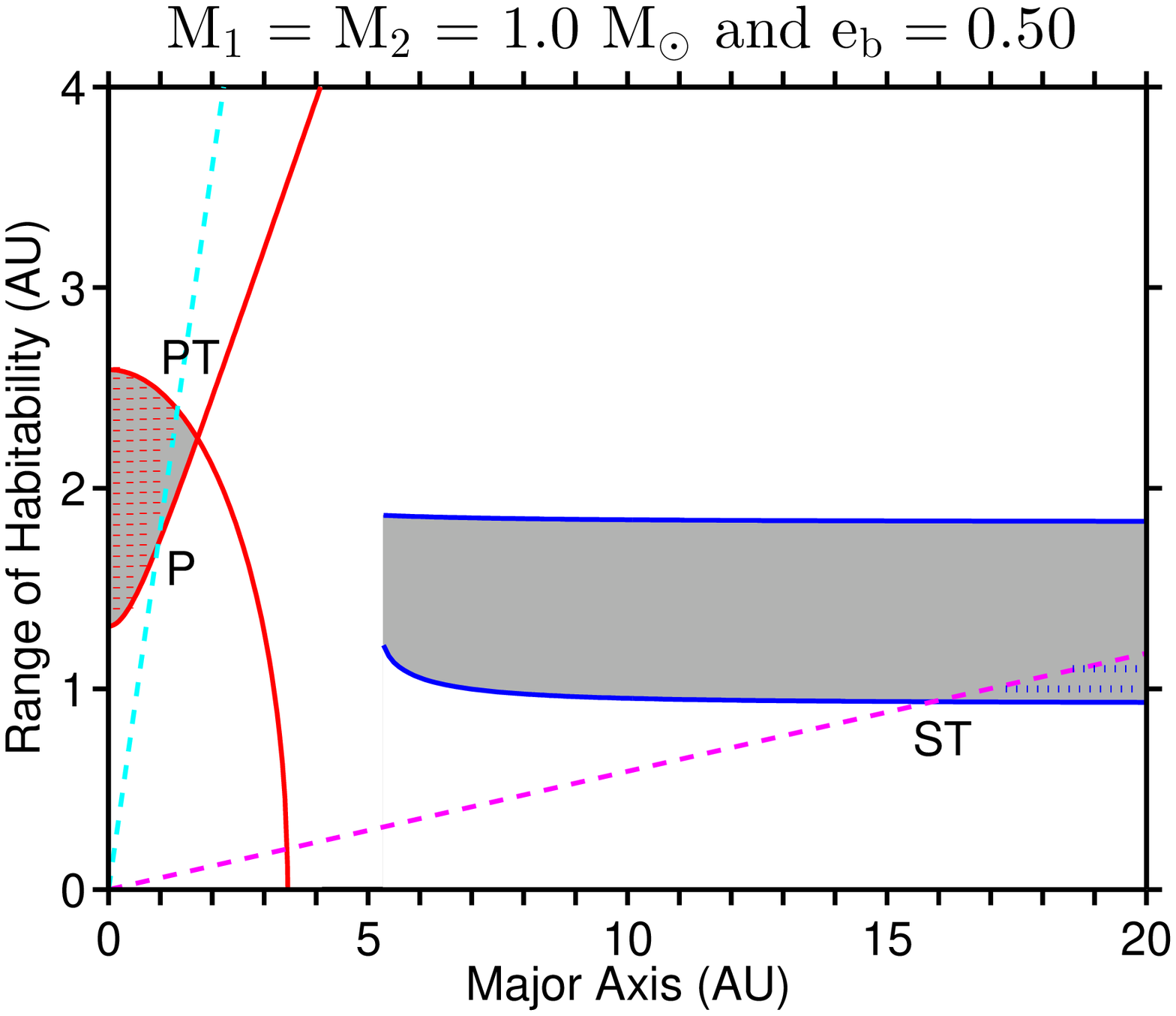,width=0.45\linewidth} &
\epsfig{file=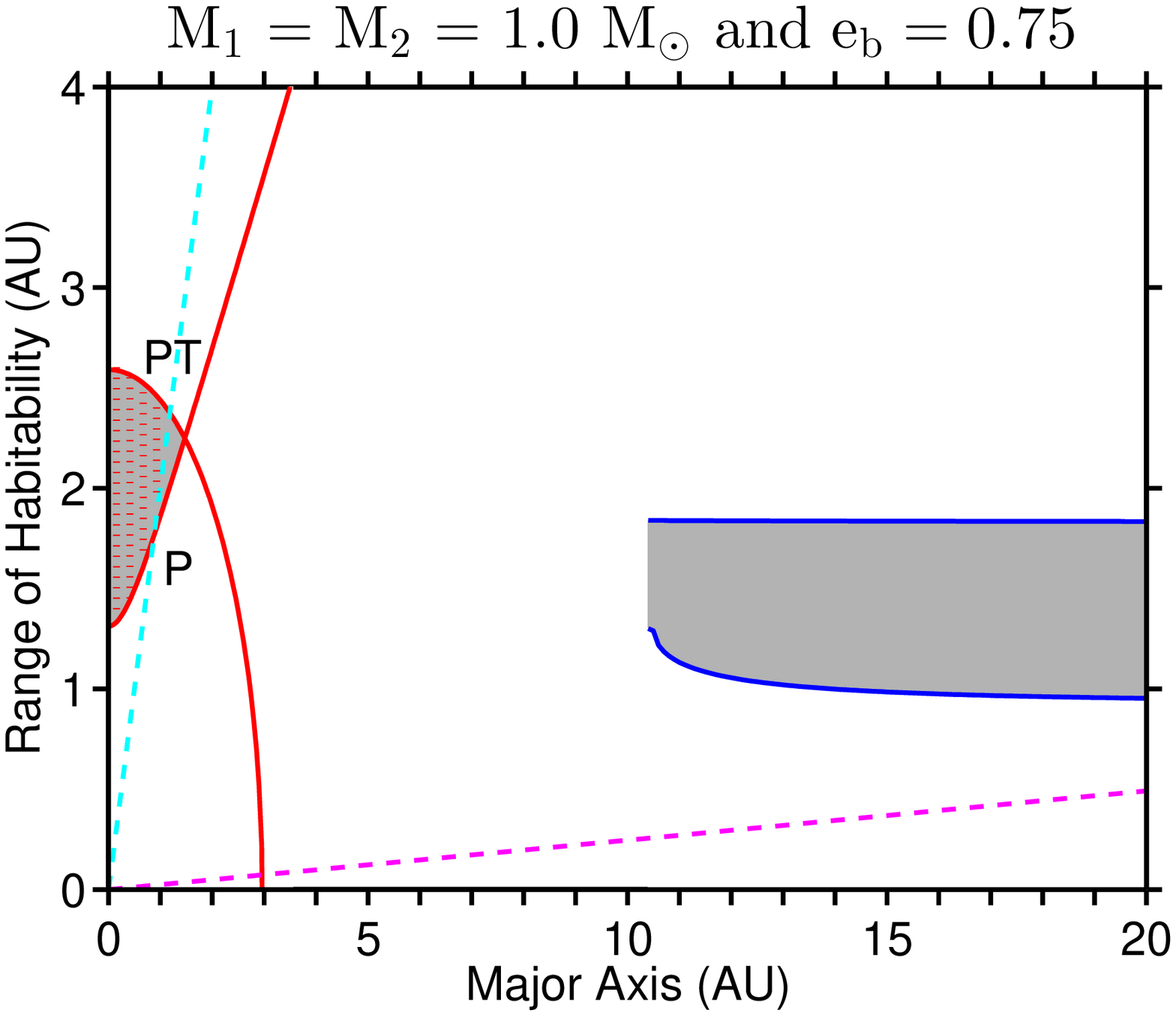,width=0.45\linewidth} \\
\end{tabular}
\caption{Range of habitability in equal-star binary systems with
$M_1 = M_2 = M_\odot$ for binary eccentricities of $e_{\rm b} =
0.00$, 0.25, 0.50, and 0.75, respectively.  Results are obtained as
a function of the binary major axis $2a_{\rm b}$ pertaining to the GHZ.
The two red lines indicate the limits of the {\it P}-type RHZ (i.e.,
RHZ$_{\rm in}$ and RHZ$_{\rm out}$), whereas the two blue lines indicate
the limits of the {\it S}-type RHZ.  The available {\it S-} and
{\it P}-type RHZs are depicted as grayish areas.  The cyan dashed line
indicates the {\it P}-type orbital stability limit, whereas the violet
dashed line indicates the {\it S}-type orbital stability limit.  Note
that the {\it P}-type orbital stability limit constitutes a lower limit,
whereas the {\it S}-type orbital stability limit constitutes an upper limit;
thus, the available ranges of habitability within the RHZs are indicated as
red-hatched and blue-hatched areas, respectively.  Hence, {\it P}-type
habitability is attained in the range beneath the P intersection point,
{\it PT}-type habitability between the intersection points P and PT,
{\it ST}-type habitability between the intersection points ST and S, and
{\it S}-type habitability beyond the S intersection point.  No habitability
is found between the intersection points PT and ST.  For $e_{\rm b}=0.50$,
the intersects of ST and S are given as 15.9 and 31.1~AU, whereas for
$e_{\rm b}=0.75$, they are given as 38.0 and 74.5~AU, respectively. 
}
\end{figure*}

\clearpage

%+++++++++++++++++++++++++++++++++++++++++++++++++++++++++++++++++++++++++

%%% *** Fig.7
%%%%%%%%%%%%%%%%%%%%%%%%%%%%%%%%%%%%%%%%%%%%%%%%%%%%%%%%%%%%%%%%%
\begin{figure*} 
\centering
\begin{tabular}{cc}
\epsfig{file=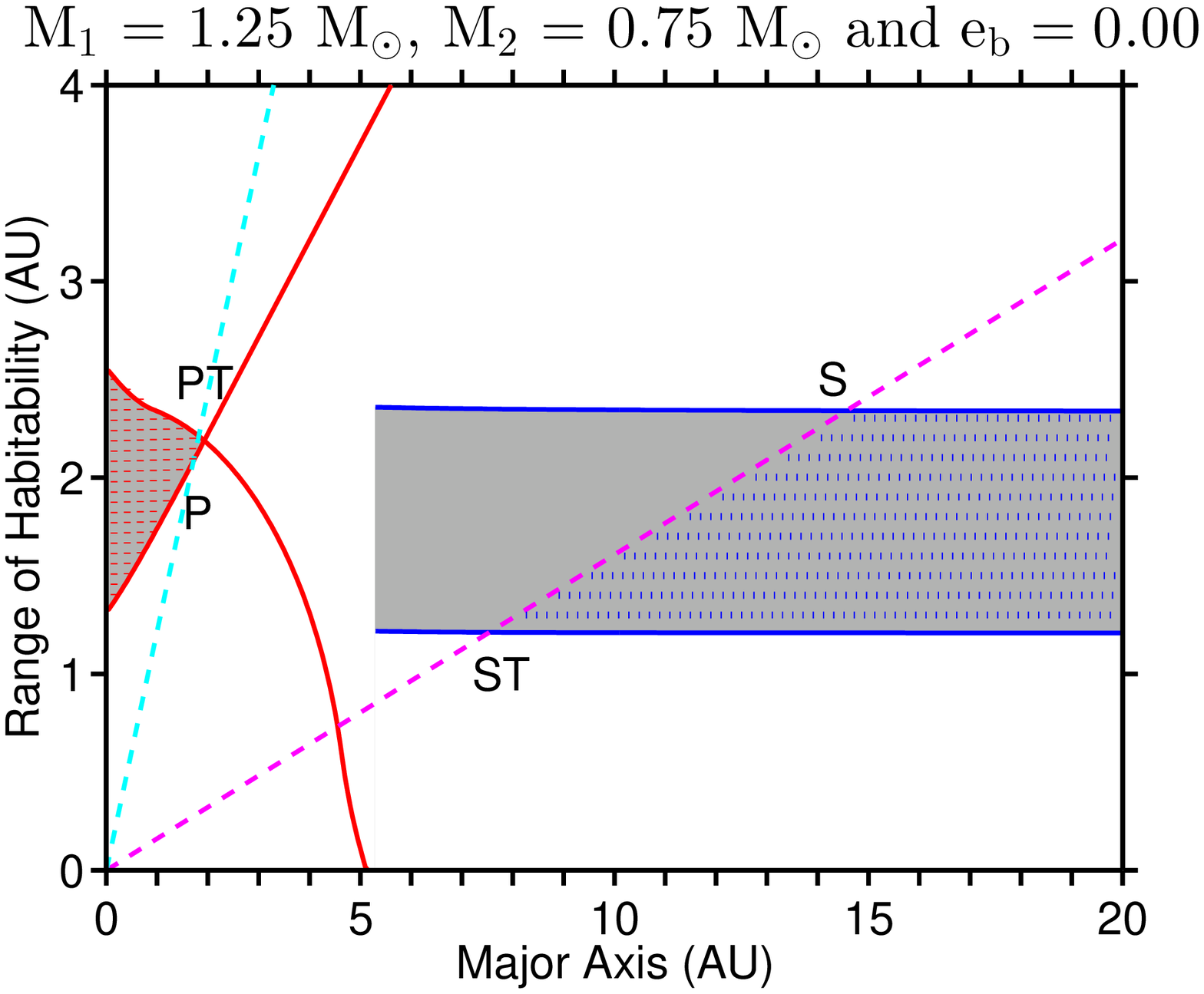,width=0.45\linewidth} &
\epsfig{file=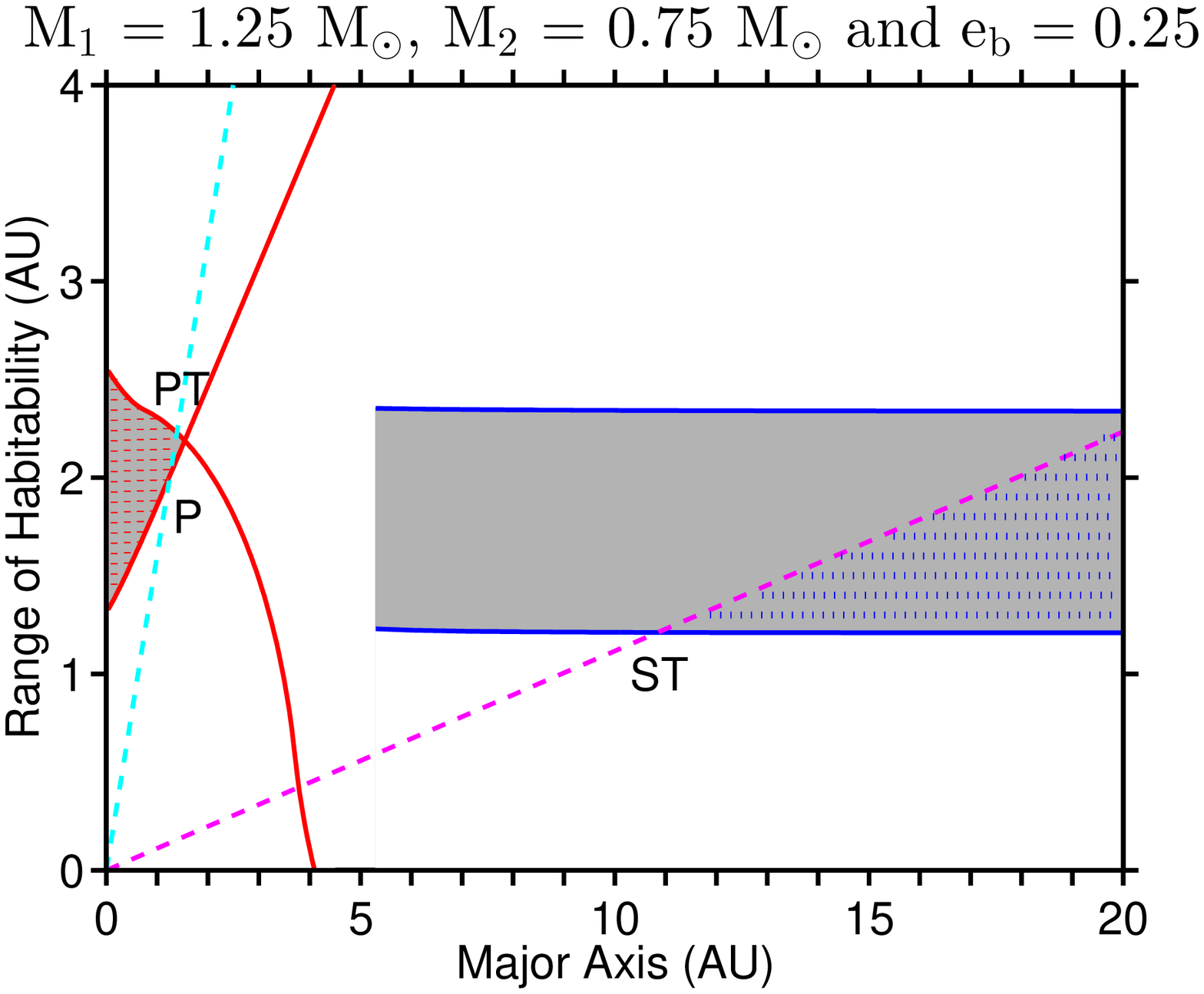,width=0.45\linewidth} \\
\epsfig{file=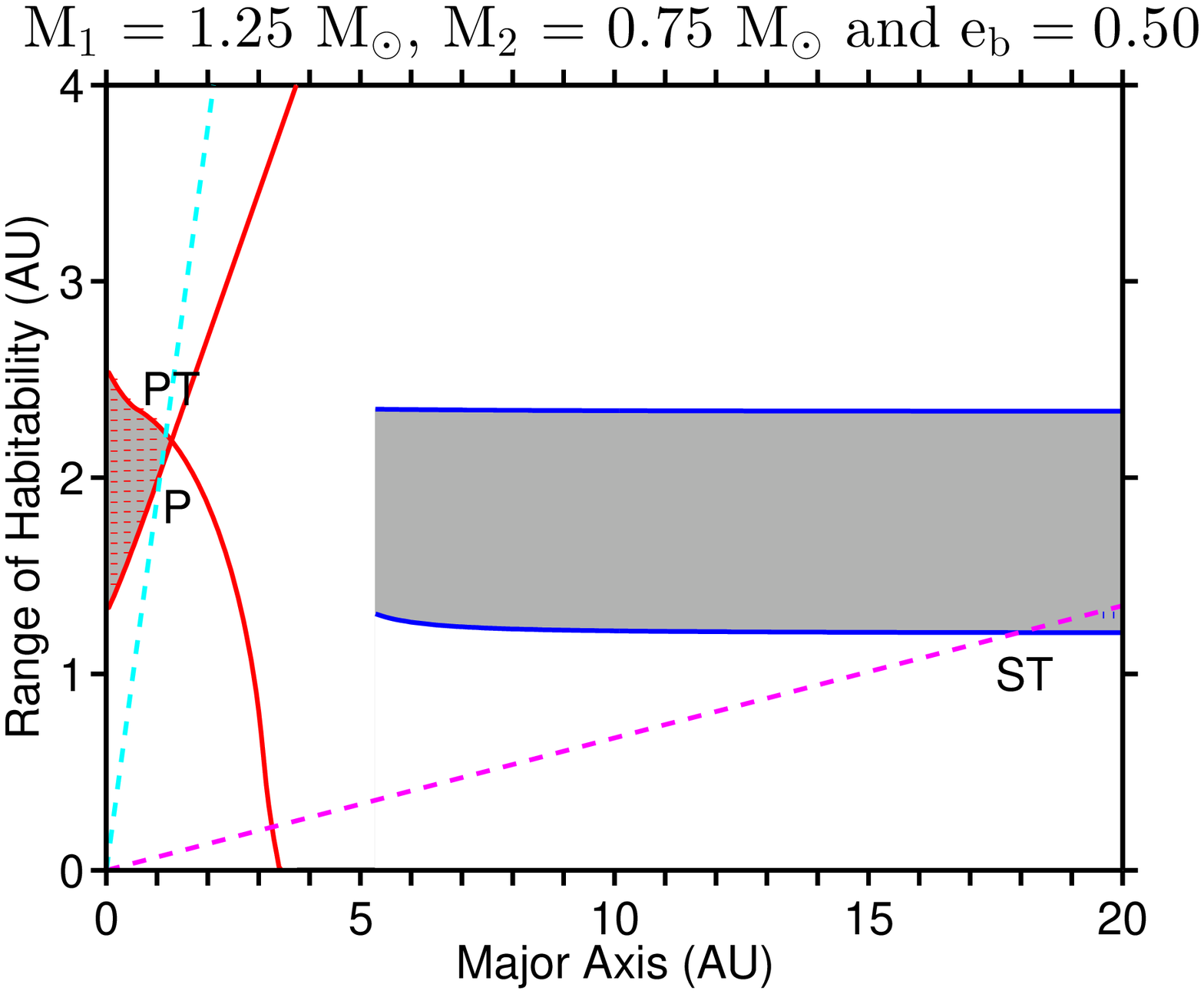,width=0.45\linewidth} &
\epsfig{file=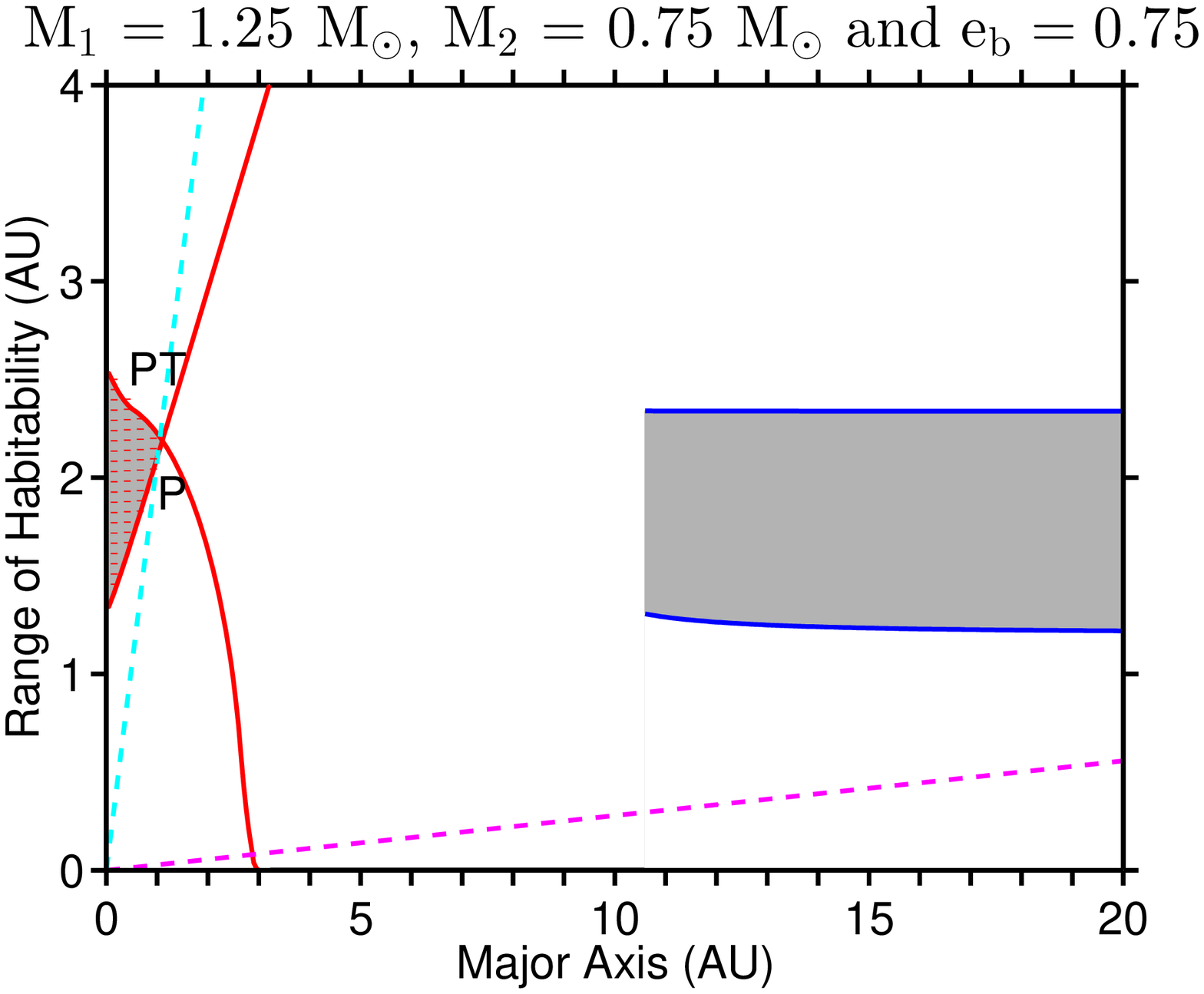,width=0.45\linewidth} \\
\end{tabular}
\caption{Same as Figure~6, but now for an unequal-star binary system with
$M_1 = 1.25~M_\odot$ and $M_2 = 0.75~M_\odot$.  For $e_{\rm b}=0.50$, the
intersects of ST and S are given as 18.2 and 35.1~AU, denoting the domain
of {\it ST-}type habitability, whereas for $e_{\rm b}=0.75$, they are given
as 43.9 and 84.9~AU, respectively.
}
\end{figure*}

\clearpage

%+++++++++++++++++++++++++++++++++++++++++++++++++++++++++++++++++++++++++

%%% *** Fig.8
%%%%%%%%%%%%%%%%%%%%%%%%%%%%%%%%%%%%%%%%%%%%%%%%%%%%%%%%%%%%%%%%%
\begin{figure*} 
\centering
\begin{tabular}{cc}
\epsfig{file=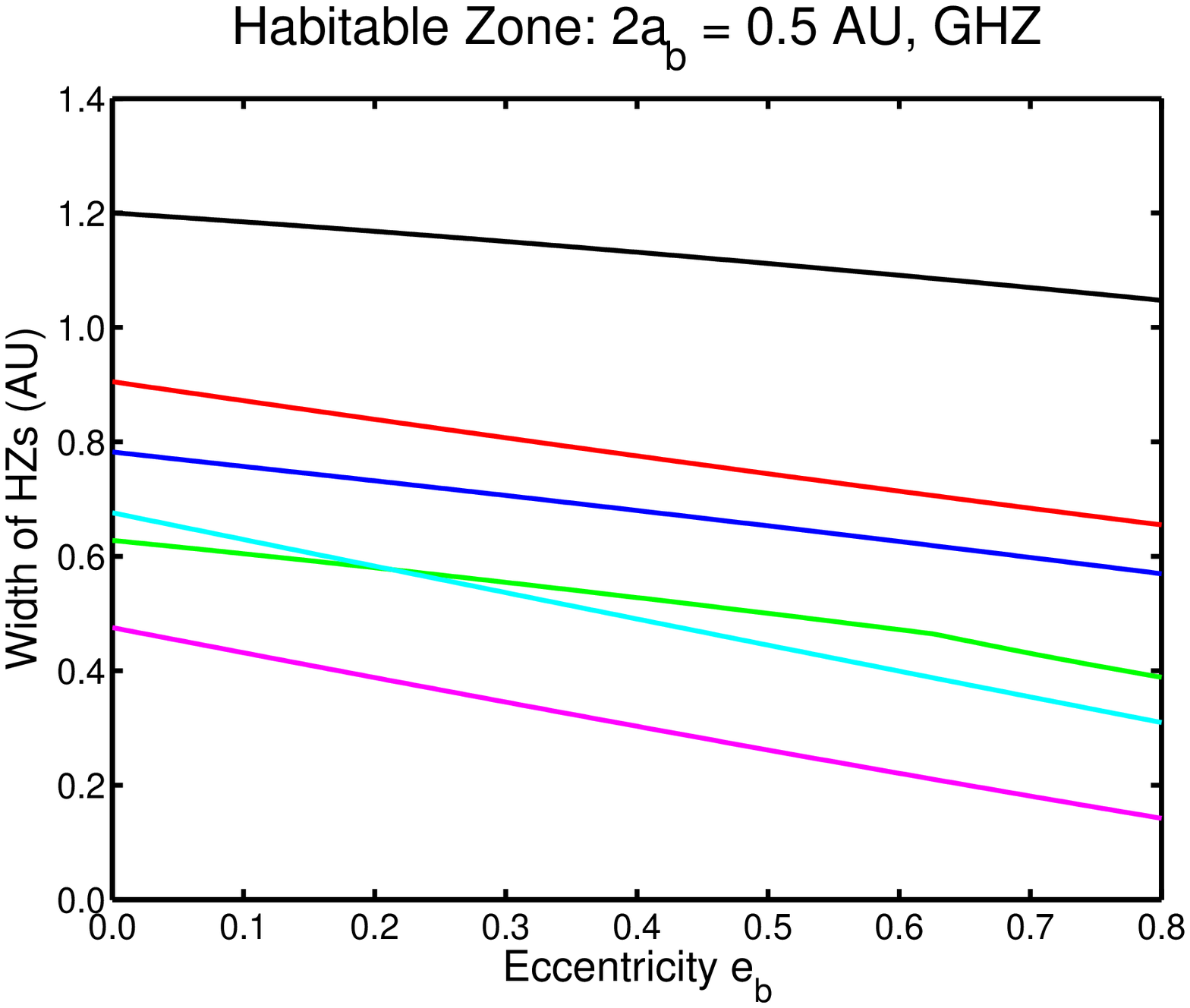,width=0.45\linewidth} &
\epsfig{file=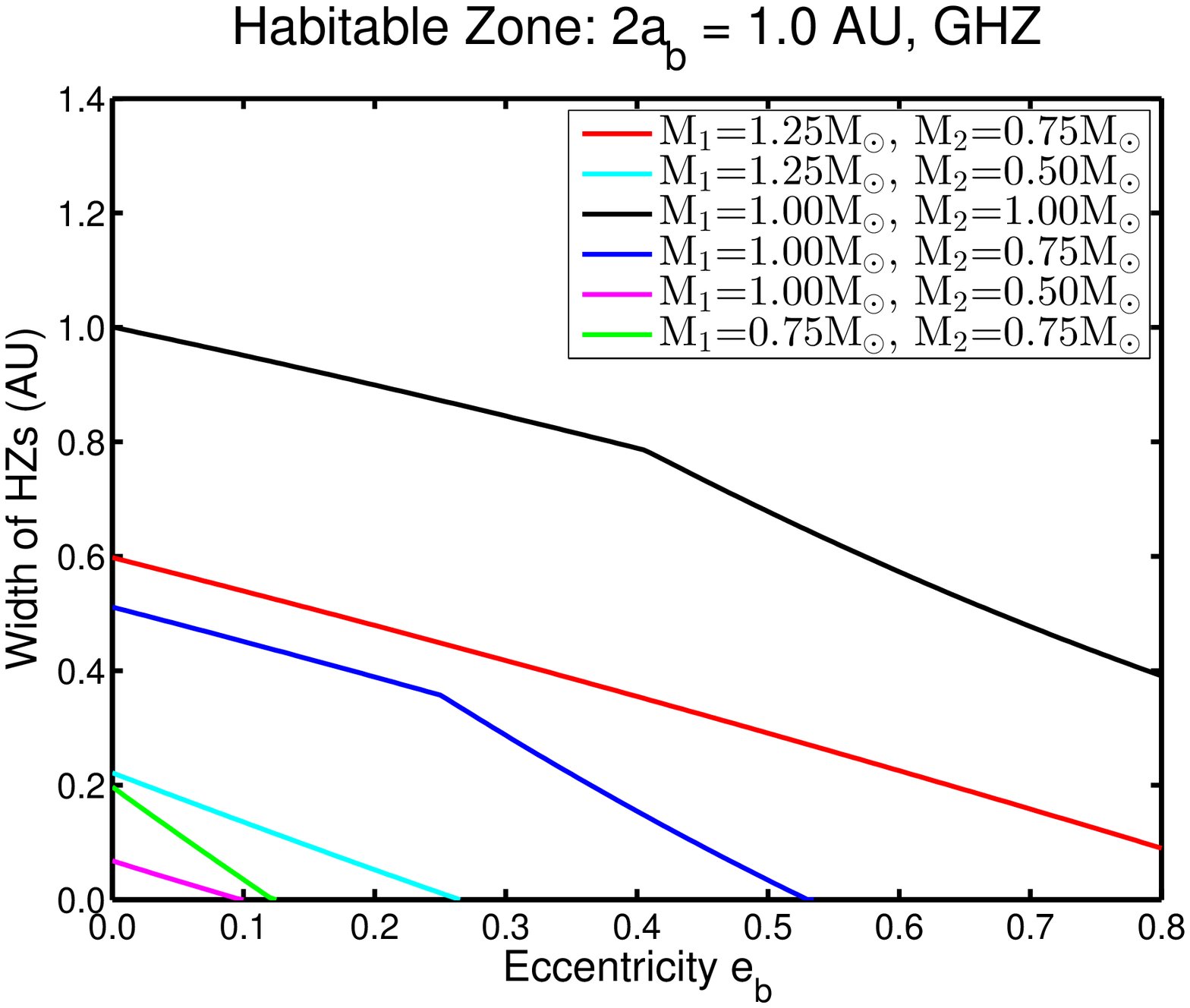,width=0.45\linewidth} \\
\epsfig{file=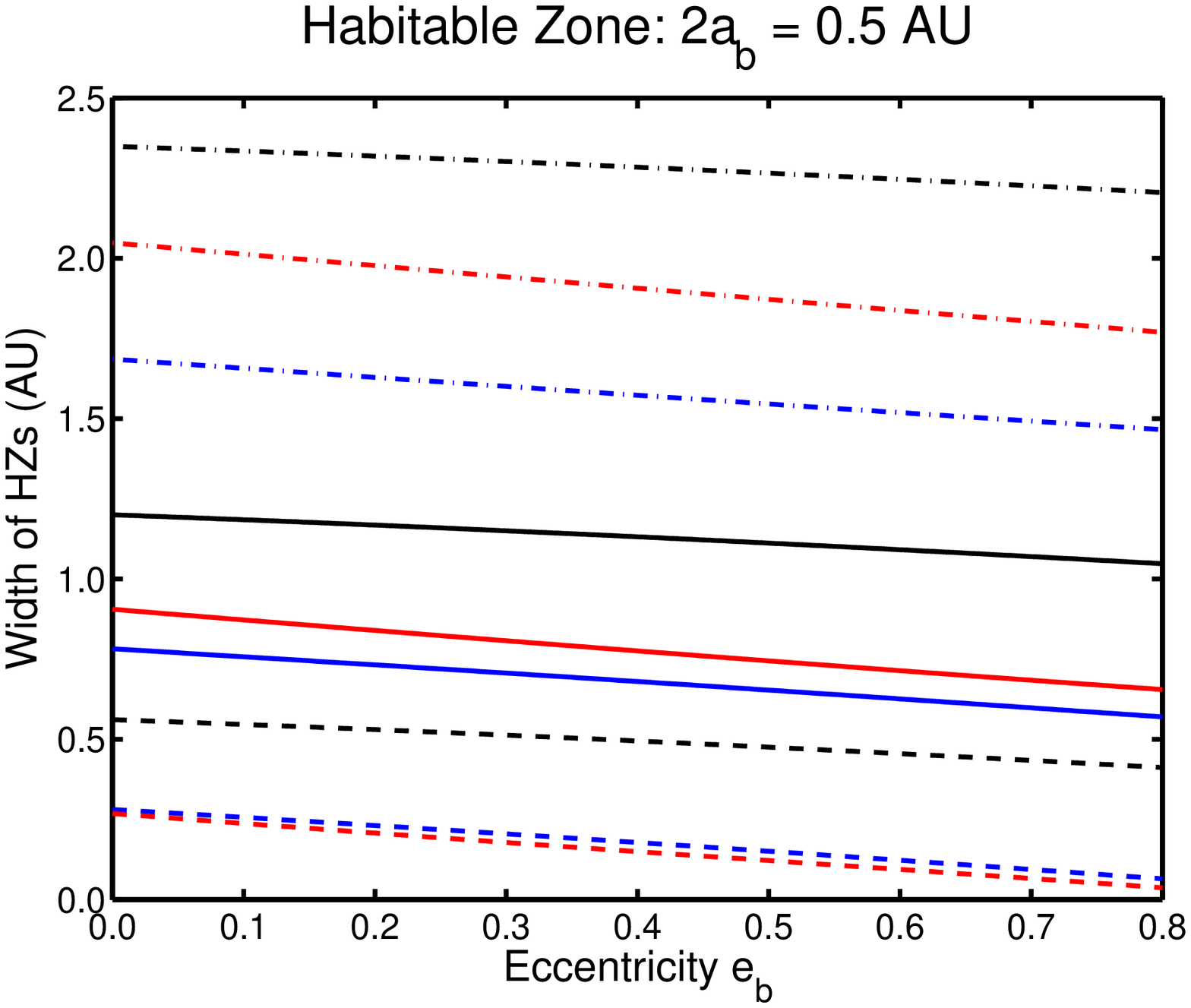,width=0.45\linewidth} &
\epsfig{file=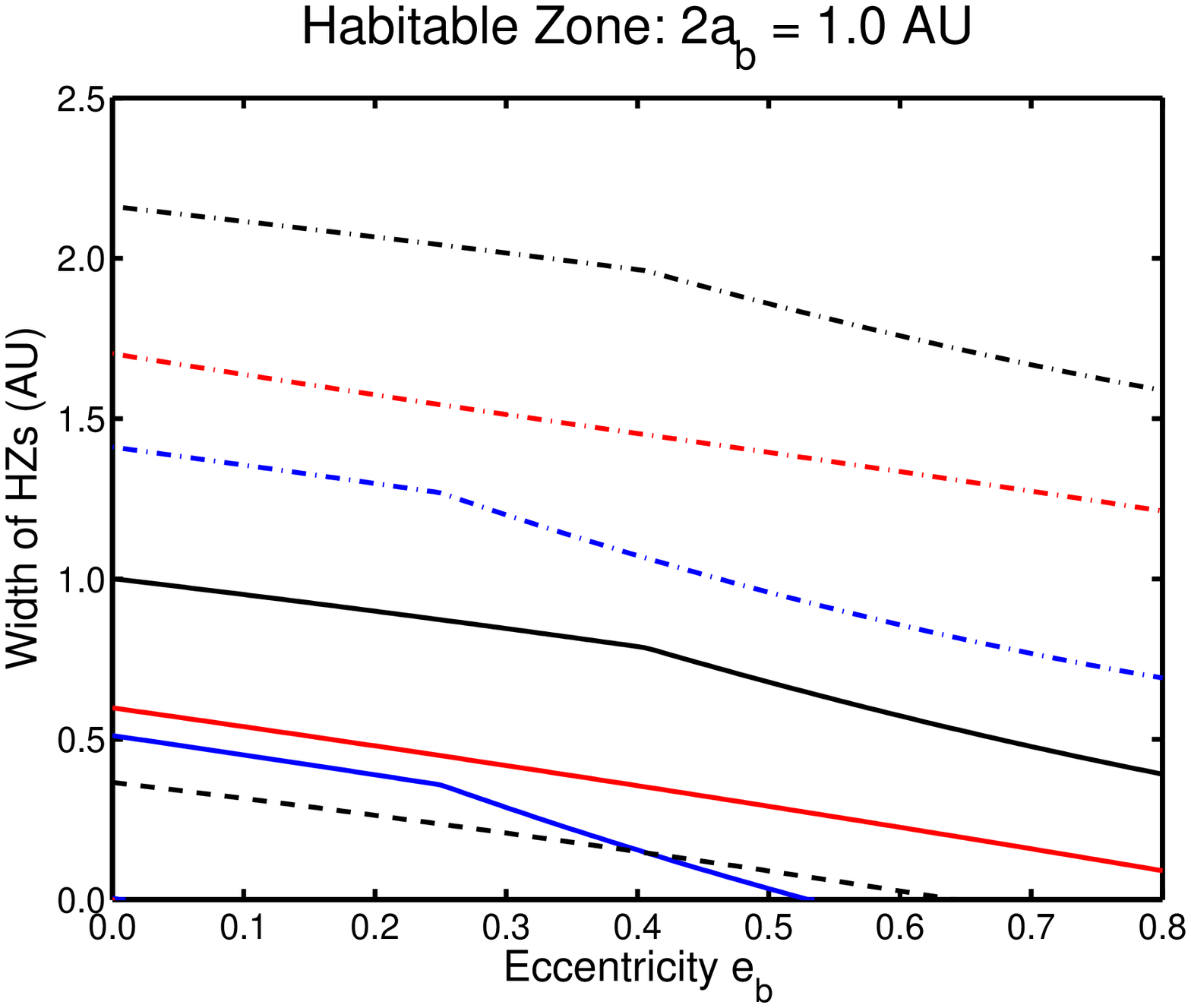,width=0.45\linewidth} \\
\end{tabular}
\caption{
Widths of {\it P}/{\it PT}-type habitable zones of various binary systems
for $2a_{\rm b}=0.5$~AU (left column) and $2a_{\rm b}=1.0$~AU (right column).
The top row depicts results for the GHZ for six systems, whereas the bottom row
depicts results for the EHZ (dash-dotted), GHZ (solid), and CHZ (dashed)
for a selection of three systems (see top right subfigure for color code).
The line explanations for the bottom row is given as part of Fig.~9 for
stylistic reasons.
}
\end{figure*}

\clearpage

%+++++++++++++++++++++++++++++++++++++++++++++++++++++++++++++++++++++++++

%%% *** Fig.9
%%%%%%%%%%%%%%%%%%%%%%%%%%%%%%%%%%%%%%%%%%%%%%%%%%%%%%%%%%%%%%%%%
\begin{figure*} 
\centering
\begin{tabular}{cc}
\epsfig{file=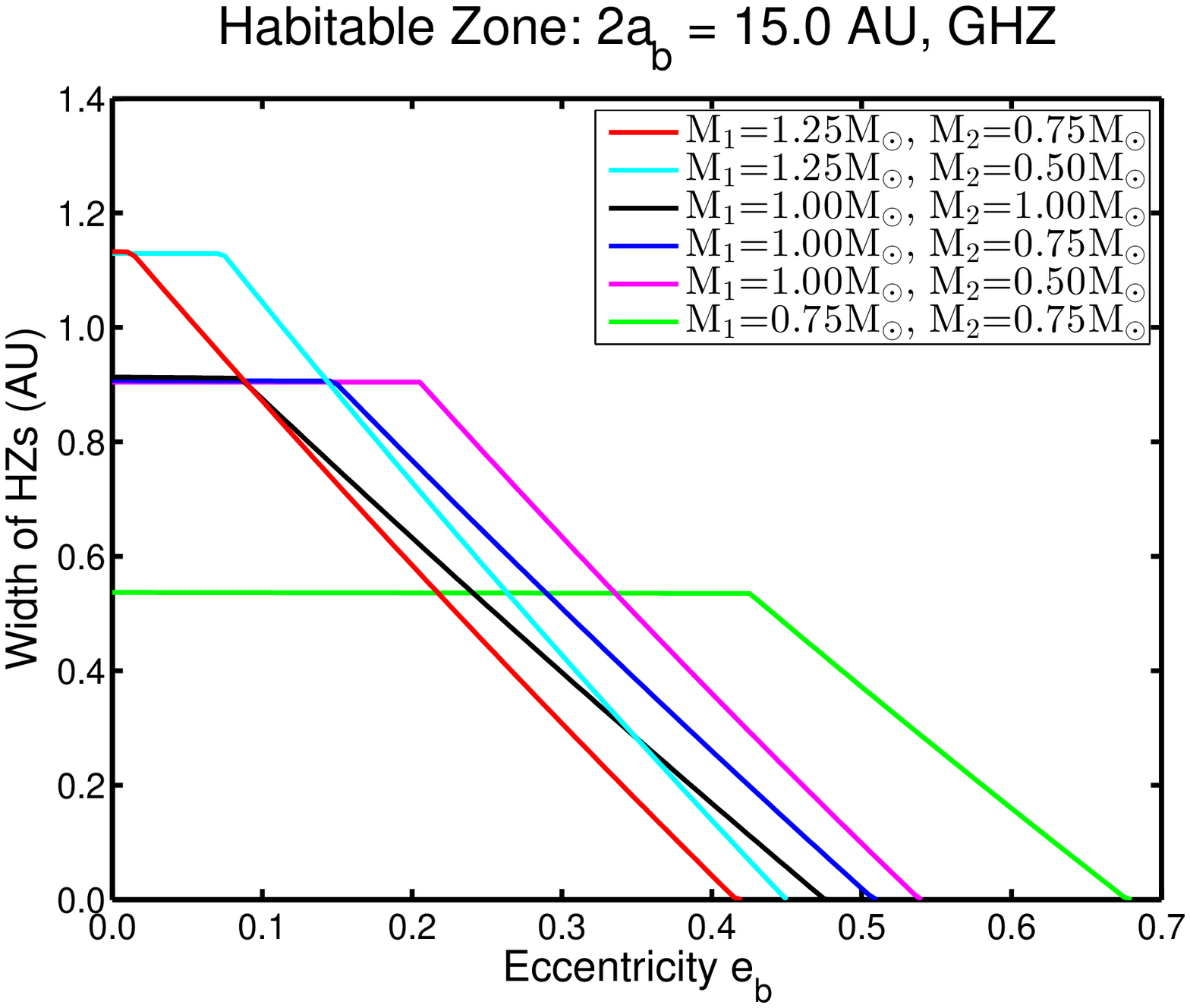,width=0.45\linewidth} &
\epsfig{file=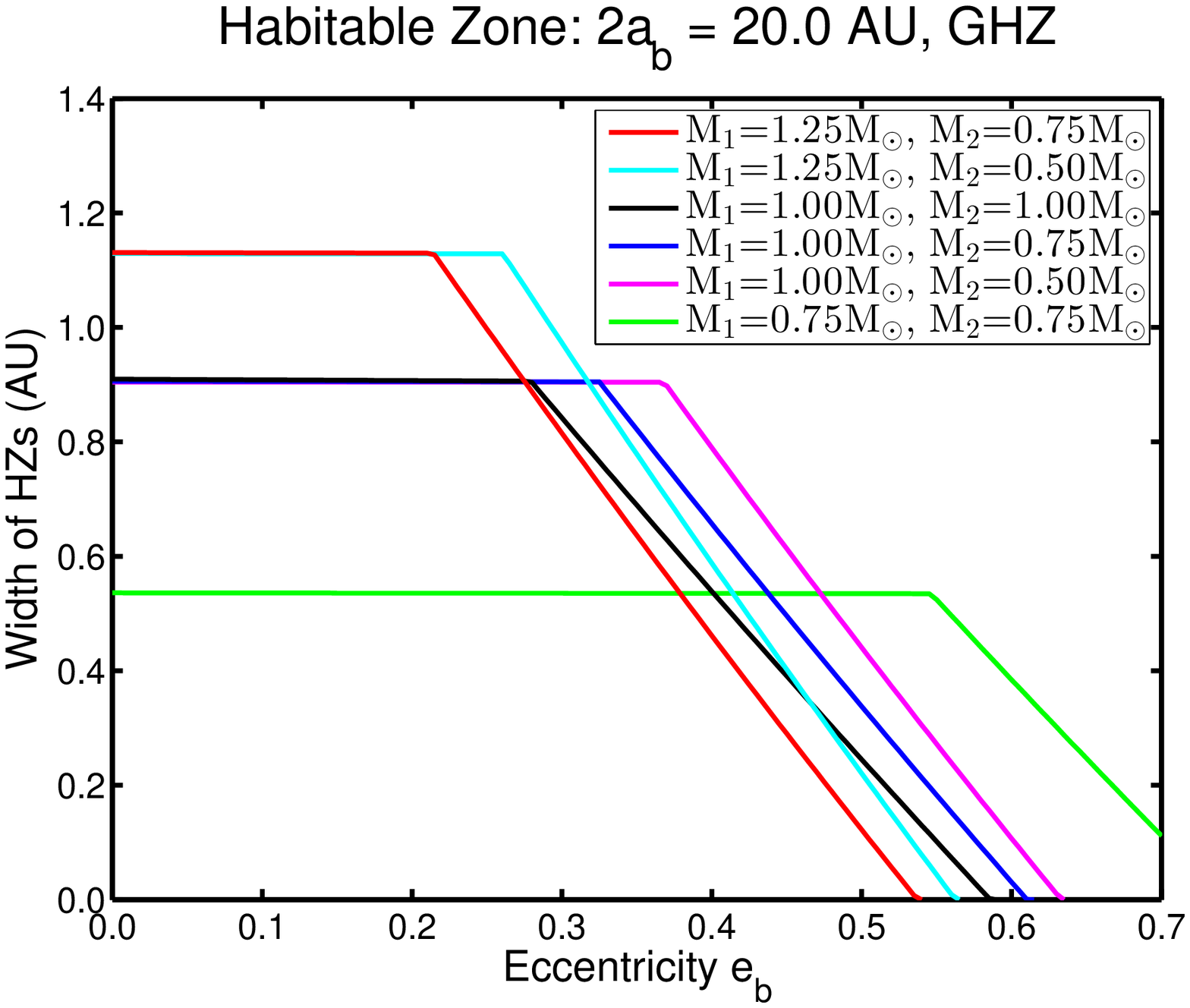,width=0.45\linewidth} \\
\epsfig{file=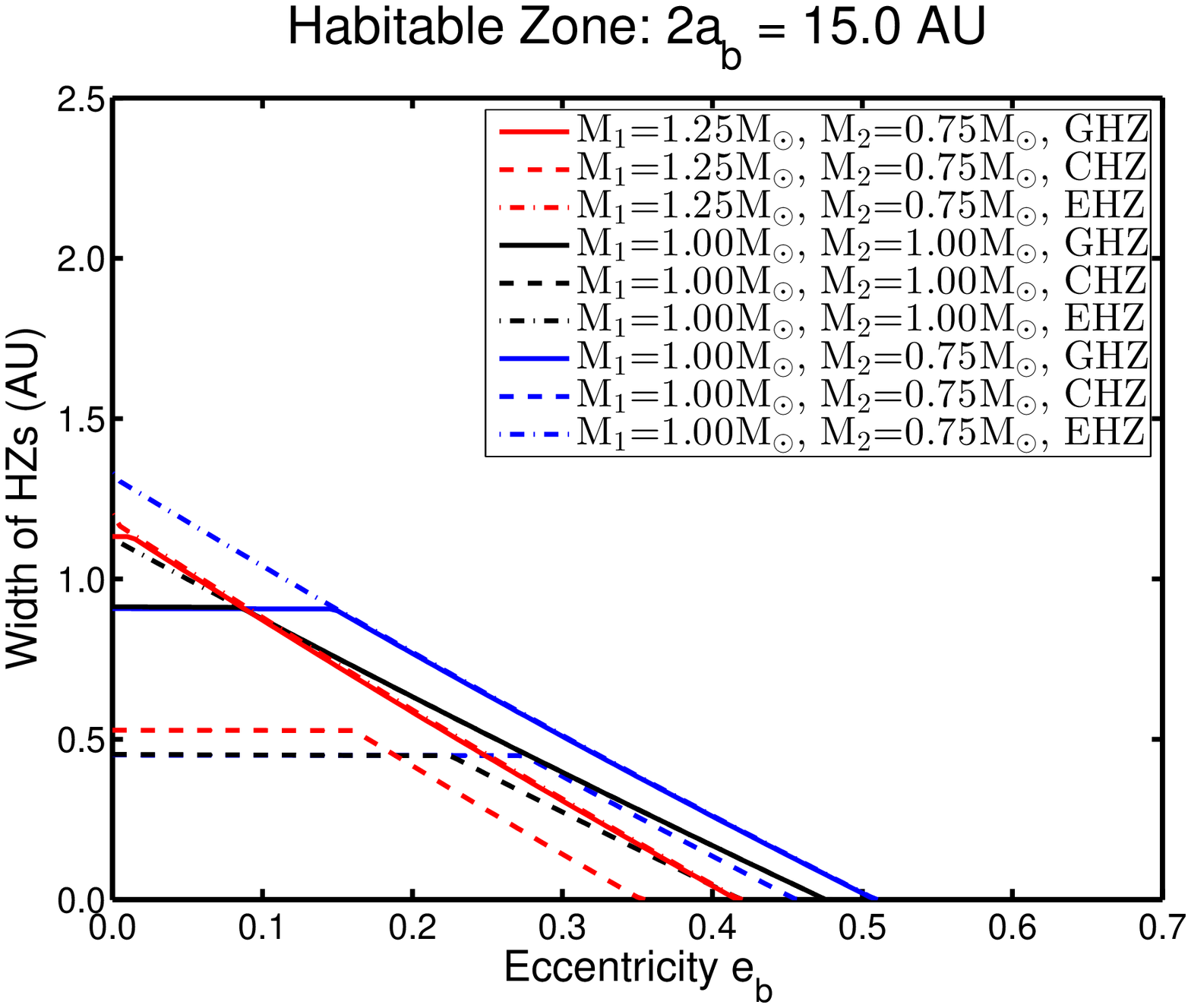,width=0.45\linewidth} &
\epsfig{file=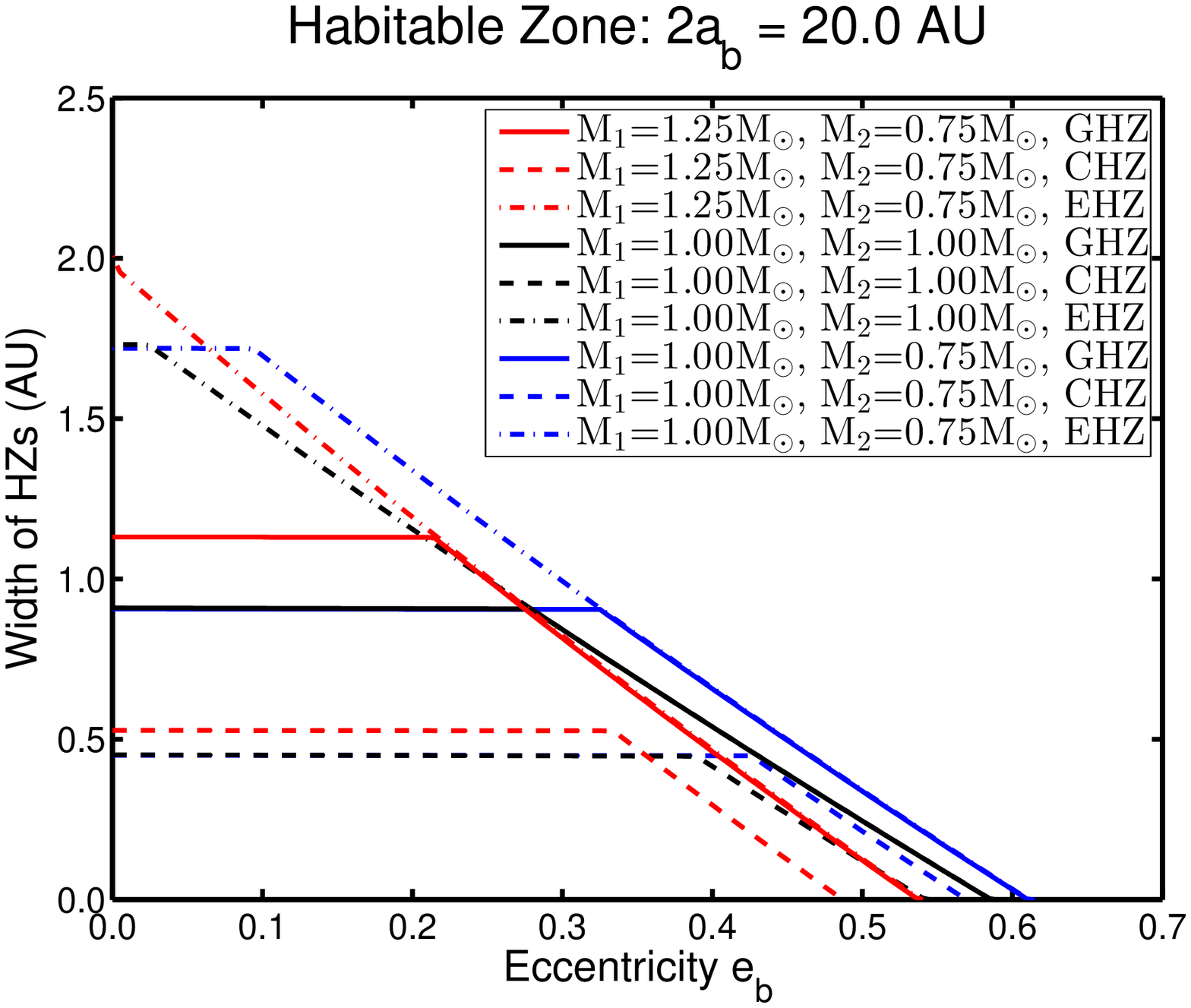,width=0.45\linewidth} \\
\end{tabular}
\caption{
Widths of {\it S}/{\it ST}-type habitable zones of various binary systems
for $2a_{\rm b}=15.0$~AU (left column) and $2a_{\rm b}=20.0$~AU (right column).
The top row depicts results for the GHZ for six systems, whereas the bottom row
depicts results for the EHZ (dash-dotted), GHZ (solid), and CHZ (dashed)
for a selection of three systems.
}
\end{figure*}

\clearpage

%+++++++++++++++++++++++++++++++++++++++++++++++++++++++++++++++++++++++++

%%% *** Table 1
%%%%%%%%%%%%%%%%%%%%%%%%%%%%%%%%%%%%%%%%%%%%%%%%%%%%%%%%%%%%%%%%%%%%%%%%%
\begin{deluxetable}{lccccc}
\tablecaption{Stellar Parameters}
\tablewidth{0pt}
\tablehead{
Sp. Type & $T_{\rm eff}$ & $R_\ast$    & $L_\ast$    & $S_{{\rm rel},\ast}$ & $M_\ast$ \\
       ... &  (K)        & ($R_\odot$) & ($L_\odot$) & ...                  & ($M_\odot$)
}
\startdata
    F0  &  7178  &  1.62  &  6.255  &  1.145  &  1.60 \\
    F2  &  6909  &  1.48  &  4.481  &  1.113  &  1.52 \\
    F5  &  6528  &  1.40  &  3.196  &  1.072  &  1.40 \\
    F8  &  6160  &  1.20  &  1.862  &  1.037  &  1.19 \\
    G0  &  5943  &  1.12  &  1.405  &  1.019  &  1.05 \\
    G2  &  5811  &  1.08  &  1.194  &  1.009  &  0.99 \\
    G5  &  5657  &  0.95  &  0.830  &  0.997  &  0.91 \\
    G8  &  5486  &  0.91  &  0.673  &  0.985  &  0.84 \\
    K0  &  5282  &  0.83  &  0.481  &  0.971  &  0.79 \\
    K2  &  5055  &  0.75  &  0.330  &  0.957  &  0.74 \\
    K5  &  4487  &  0.64  &  0.149  &  0.926  &  0.67 \\
    K8  &  4006  &  0.53  &  0.066  &  0.905  &  0.58 \\
    M0  &  3850  &  0.48  &  0.045  &  0.900  &  0.51 \\
\enddata
\tablecomments{
Adopted from Paper~I.
}
\end{deluxetable}

\clearpage

%+++++++++++++++++++++++++++++++++++++++++++++++++++++++++++++++++++++++++

%%% *** Table 2
%%%%%%%%%%%%%%%%%%%%%%%%%%%%%%%%%%%%%%%%%%%%%%%%%%%%%%%%%%%%%%%%%%%%%%%%%
\begin{deluxetable}{lcccc}
\tablecaption{Definition of $\tilde{a}_{{\rm SP,}\ell}$}
\tablewidth{0pt}
\tablehead{
$\ell$ & $s_\ell$ & \multicolumn{2}{c}{$\tilde{a}_{{\rm SP,}\ell}$} & HZ Limit \\
\noalign{\smallskip}
\hline
\noalign{\smallskip}
...    & ...      & {\it S}-Type & {\it P}-Type & ...      \\
...    & (AU)     & (AU)         & (AU)         & ...      
}
\startdata
 1 &  0.84  & $a_{\rm b} (1-e_{\rm b})$  & $a_{\rm b} (1+e_{\rm b})$  & GHZ / EHZ  \\
 2 &  0.95  & $a_{\rm b} (1-e_{\rm b})$  & $a_{\rm b} (1+e_{\rm b})$  & CHZ        \\
 3 &  1.00  & $a_{\rm b}$                & $a_{\rm b}$                & ...        \\
 4 &  1.37  & $a_{\rm b} (1+e_{\rm b})$  & $a_{\rm b} (1+e_{\rm b})$  & CHZ        \\
 5 &  1.67  & $a_{\rm b} (1+e_{\rm b})$  & $a_{\rm b} (1+e_{\rm b})$  & GHZ        \\
 6 &  2.40  & $a_{\rm b} (1+e_{\rm b})$  & $a_{\rm b} (1+e_{\rm b})$  & EHZ        \\
\enddata
\tablecomments{
The case of $\ell  = 3$ depicts an Earth-equivalent position denoting
a planetary reference distance without the consideration of eccentricity.
}
\end{deluxetable}

\clearpage

%+++++++++++++++++++++++++++++++++++++++++++++++++++++++++++++++++++++++++

%%% *** Table 3
%%%%%%%%%%%%%%%%%%%%%%%%%%%%%%%%%%%%%%%%%%%%%%%%%%%%%%%%%%%%%%%%%%%%%%%%%
\begin{deluxetable}{lrr}
\tablecaption{Orbital Stability Parameters Following HW99}
\tablewidth{0pt}
\tablehead{
Parameter	& {\it S}-Type & {\it P}-Type
}
\startdata
${\tilde A}_0$    &  $ 0.464$    &  $ 1.60$   \\
${\tilde A}_1$    &  $-0.380$    &  $ 4.12$   \\
${\tilde A}_2$    &  $ 0.0  $    &  $-5.09$   \\
${\tilde B}_0$    &  $-0.631$    &  $ 5.10$   \\
${\tilde B}_1$    &  $ 0.586$    &  $-4.27$   \\
${\tilde B}_2$    &  $ 0.0  $    &  $ 0.0 $   \\
${\tilde C}_0$    &  $ 0.150$    &  $-2.22$   \\
${\tilde C}_1$    &  $-0.198$    &  $ 0.0 $   \\
${\tilde C}_2$    &  $ 0.0  $    &  $ 4.61$   \\
\enddata
\end{deluxetable}

\clearpage

%+++++++++++++++++++++++++++++++++++++++++++++++++++++++++++++++++++++++++

%%% *** Table 4
%%%%%%%%%%%%%%%%%%%%%%%%%%%%%%%%%%%%%%%%%%%%%%%%%%%%%%%%%%%%%%%%%%%%%%%%%
\begin{deluxetable}{lcccccccccc}
\tabletypesize{\scriptsize}
% \tabletypesize{\scriptsize}
\tablecaption{RHZs of $P$-Type Orbits for $M_1 = M_2 = 1.0~M_\odot$, CHZ}
\tablewidth{0pt}
\tablehead{
$e_{\rm b}$ & \multicolumn{2}{c}{$a_{\rm per}$} & \multicolumn{2}{c}{$a_{\rm ap}$} & \multicolumn{2}{c}{Orbit} & $a_{\rm cr}$ & {\%}RHZ & {\%}HZ & Type \\
\noalign{\smallskip}
\hline
\noalign{\smallskip}
...    & RHZ$_{\rm in}$ & RHZ$_{\rm out}$ & RHZ$_{\rm in}$ & RHZ$_{\rm out}$ & RHZ$_{\rm in}$ & RHZ$_{\rm out}$ & ...  & ... & ... & ... \\
...    & (AU)           & (AU)            & (AU)           & (AU)            & (AU)           & (AU)            & (AU) & ... & ... & ...
}
\startdata
\multicolumn{9}{l}{Model:~~$2a_{\rm b}= 0.50$~AU} \\
\noalign{\smallskip}
\hline
\noalign{\smallskip}
 0.00  &   1.544   &   2.105   &   1.544   &   2.105    &   1.54  &  2.11  &  0.60   & 100 & 100 & {\it P}   \\        
 0.10  &   1.533   &   2.108   &   1.556   &   2.102    &   1.56  &  2.10  &  0.67   &  97 &  97 & {\it P}   \\       
 0.20  &   1.523   &   2.111   &   1.569   &   2.099    &   1.57  &  2.10  &  0.73   &  94 &  94 & {\it P}   \\    
 0.30  &   1.515   &   2.113   &   1.582   &   2.095    &   1.58  &  2.09  &  0.80   &  91 &  91 & {\it P}   \\      
 0.40  &   1.507   &   2.115   &   1.596   &   2.091    &   1.60  &  2.09  &  0.85   &  88 &  88 & {\it P}   \\      
 0.50  &   1.500   &   2.116   &   1.611   &   2.087    &   1.61  &  2.09  &  0.90   &  85 &  85 & {\it P}   \\    
 0.60  &   1.494   &   2.118   &   1.627   &   2.082    &   1.63  &  2.08  &  0.95   &  81 &  81 & {\it P}   \\   
 0.70  &   1.490   &   2.119   &   1.643   &   2.077    &   1.64  &  2.08  &  0.98   &  77 &  77 & {\it P}   \\   
 0.80  &   1.487   &   2.119   &   1.660   &   2.072    &   1.66  &  2.07  &  1.02   &  73 &  73 & {\it P}   \\   
\noalign{\smallskip}
\hline
\noalign{\smallskip}
\multicolumn{9}{l}{Model:~~$2a_{\rm b}= 0.75$~AU} \\
\noalign{\smallskip}
\hline
\noalign{\smallskip}
 0.00  &   1.611   &   2.087   &   1.611   &   2.087    &   1.61  &  2.09  &  0.90  & 100 & 100 & {\it P}   \\
 0.10  &   1.589   &   2.093   &   1.635   &   2.080    &   1.64  &  2.08  &  1.00  &  94 &  94 & {\it P}   \\
 0.20  &   1.569   &   2.099   &   1.660   &   2.072    &   1.66  &  2.07  &  1.10  &  87 &  87 & {\it P}   \\
 0.30  &   1.550   &   2.104   &   1.686   &   2.063    &   1.69  &  2.06  &  1.19  &  79 &  79 & {\it P}   \\
 0.40  &   1.533   &   2.108   &   1.713   &   2.054    &   1.71  &  2.05  &  1.28  &  72 &  72 & {\it P}   \\
 0.50  &   1.519   &   2.112   &   1.742   &   2.044    &   1.74  &  2.04  &  1.35  &  64 &  64 & {\it P}   \\
 0.60  &   1.507   &   2.115   &   1.771   &   2.033    &   1.77  &  2.03  &  1.42  &  55 &  55 & {\it P}   \\
 0.70  &   1.497   &   2.117   &   1.800   &   2.022    &   1.80  &  2.02  &  1.48  &  47 &  47 & {\it P}   \\
 0.80  &   1.490   &   2.119   &   1.831   &   2.010    &   1.83  &  2.01  &  1.53  &  38 &  38 & {\it P}   \\
\noalign{\smallskip}
\hline
\noalign{\smallskip}
\multicolumn{9}{l}{Model:~~$2a_{\rm b}= 1.0$~AU} \\
\noalign{\smallskip}
\hline
\noalign{\smallskip}
 0.00  &   1.695   &   2.060   &   1.695   &   2.060    &  1.70  &  2.06  &  1.19  & 100 & 100 & {\it P}    \\
 0.10  &   1.660   &   2.072   &   1.732   &   2.047    &  1.73  &  2.05  &  1.34  &  86 &  86 & {\it P}    \\
 0.20  &   1.627   &   2.082   &   1.771   &   2.033    &  1.77  &  2.03  &  1.47  &  72 &  72 & {\it P}    \\
 0.30  &   1.596   &   2.091   &   1.811   &   2.018    &  1.81  &  2.02  &  1.59  &  57 &  57 & {\it P}    \\
 0.40  &   1.569   &   2.099   &   1.852   &   2.001    &  1.85  &  2.00  &  1.70  &  41 &  41 & {\it P}    \\
 0.50  &   1.544   &   2.105   &   1.894   &   1.983    &  1.89  &  1.98  &  1.80  &  24 &  24 & {\it P}    \\
 0.60  &   1.523   &   2.111   &   1.937   &   1.963    &  1.94  &  1.96  &  1.89  &   7 &   7 & {\it P}    \\
 0.70  &   1.507   &   2.115   &   ...     &   ...      &  ...   &  ...   &  1.97  &   0 &   0 & ...        \\
 0.80  &   1.494   &   2.118   &   ...     &   ...      &  ...   &  ...   &  2.04  &   0 &   0 & ...        \\
\enddata
\tablecomments{
Here as well as in the subsequent Tables 5, 6, 7, 9, and 10, the data for $a_{\rm per}$
and $a_{\rm ap}$ are given in high precision for tutorial reasons.
}
\end{deluxetable}

\clearpage

%+++++++++++++++++++++++++++++++++++++++++++++++++++++++++++++++++++++++++

%%% *** Table 5
%%%%%%%%%%%%%%%%%%%%%%%%%%%%%%%%%%%%%%%%%%%%%%%%%%%%%%%%%%%%%%%%%%%%%%%%%
\begin{deluxetable}{lcccccccccc}
\tabletypesize{\scriptsize}
% \tabletypesize{\scriptsize}
\tablecaption{RHZs of $P$-Type Orbits for $M_1 = M_2 = 1.0~M_\odot$, GHZ}
\tablewidth{0pt}
\tablehead{
$e_{\rm b}$ & \multicolumn{2}{c}{$a_{\rm per}$} & \multicolumn{2}{c}{$a_{\rm ap}$} & \multicolumn{2}{c}{Orbit} & $a_{\rm cr}$ & {\%}RHZ & {\%}HZ & Type \\
\noalign{\smallskip}
\hline
\noalign{\smallskip}
...    & RHZ$_{\rm in}$ & RHZ$_{\rm out}$ & RHZ$_{\rm in}$ & RHZ$_{\rm out}$ & RHZ$_{\rm in}$ & RHZ$_{\rm out}$ & ...  & ... & ... & ... \\
...    & (AU)           & (AU)            & (AU)           & (AU)            & (AU)           & (AU)            & (AU) & ... & ... & ...
}
\startdata
\multicolumn{9}{l}{Model:~~$2a_{\rm b}= 0.50$~AU} \\
\noalign{\smallskip}
\hline
\noalign{\smallskip}
 0.00  &   1.379   &   2.579   &   1.379   &   2.579    &   1.38   &  2.58  &  0.60  & 100 & 100 & {\it P}    \\
 0.10  &   1.367   &   2.581   &   1.392   &   2.576    &   1.39   &  2.58  &  0.67  &  99 &  99 & {\it P}    \\
 0.20  &   1.356   &   2.583   &   1.405   &   2.573    &   1.41   &  2.57  &  0.73  &  97 &  97 & {\it P}    \\  
 0.30  &   1.346   &   2.585   &   1.420   &   2.570    &   1.42   &  2.57  &  0.80  &  96 &  96 & {\it P}    \\
 0.40  &   1.337   &   2.586   &   1.436   &   2.567    &   1.44   &  2.57  &  0.85  &  94 &  94 & {\it P}    \\
 0.50  &   1.329   &   2.588   &   1.452   &   2.563    &   1.45   &  2.56  &  0.90  &  93 &  93 & {\it P}    \\
 0.60  &   1.323   &   2.589   &   1.469   &   2.560    &   1.47   &  2.56  &  0.95  &  91 &  91 & {\it P}    \\ 
 0.70  &   1.318   &   2.590   &   1.486   &   2.556    &   1.49   &  2.56  &  0.98  &  89 &  89 & {\it P}    \\
 0.80  &   1.315   &   2.590   &   1.504   &   2.551    &   1.50   &  2.55  &  1.02  &  87 &  87 & {\it P}    \\
\noalign{\smallskip}
\hline
\noalign{\smallskip}
\multicolumn{9}{l}{Model:~~$2a_{\rm b}= 0.75$~AU} \\
\noalign{\smallskip}
\hline
\noalign{\smallskip}
 0.00  &   1.452   &   2.563   &   1.452   &   2.563    &   1.45   &  2.56  &  0.90  & 100 & 100 & {\it P}    \\
 0.10  &   1.428   &   2.569   &   1.477   &   2.558    &   1.48   &  2.56  &  1.00  &  97 &  97 & {\it P}    \\
 0.20  &   1.405   &   2.573   &   1.504   &   2.551    &   1.50   &  2.55  &  1.10  &  94 &  94 & {\it P}    \\
 0.30  &   1.385   &   2.577   &   1.532   &   2.544    &   1.53   &  2.54  &  1.19  &  91 &  91 & {\it P}    \\
 0.40  &   1.367   &   2.581   &   1.561   &   2.537    &   1.56   &  2.54  &  1.28  &  88 &  88 & {\it P}    \\
 0.50  &   1.350   &   2.584   &   1.590   &   2.529    &   1.59   &  2.53  &  1.35  &  84 &  84 & {\it P}    \\
 0.60  &   1.337   &   2.586   &   1.621   &   2.520    &   1.62   &  2.52  &  1.42  &  81 &  81 & {\it P}    \\
 0.70  &   1.326   &   2.588   &   1.652   &   2.511    &   1.65   &  2.51  &  1.48  &  77 &  77 & {\it P}    \\
 0.80  &   1.318   &   2.590   &   1.684   &   2.501    &   1.68   &  2.50  &  1.53  &  74 &  74 & {\it P}    \\
 \noalign{\smallskip}
\hline
\noalign{\smallskip}
\multicolumn{9}{l}{Model:~~$2a_{\rm b}= 1.0$~AU} \\
\noalign{\smallskip}
\hline
\noalign{\smallskip}
 0.00  &   1.541   &   2.542   &   1.541   &   2.542    &   1.54   &  2.54  &  1.19  & 100 & 100 & {\it P}   \\
 0.10  &   1.504   &   2.551   &   1.580   &   2.532    &   1.58   &  2.53  &  1.34  &  95 &  95 & {\it P}   \\
 0.20  &   1.469   &   2.560   &   1.621   &   2.520    &   1.62   &  2.52  &  1.47  &  90 &  90 & {\it P}   \\
 0.30  &   1.436   &   2.567   &   1.663   &   2.508    &   1.66   &  2.51  &  1.59  &  84 &  84 & {\it P}   \\
 0.40  &   1.405   &   2.573   &   1.705   &   2.494    &   1.71   &  2.49  &  1.70  &  79 &  79 & {\it P}   \\
 0.50  &   1.379   &   2.579   &   1.749   &   2.480    &   1.75   &  2.48  &  1.80  &  73 &  68 & {\it PT}  \\
 0.60  &   1.356   &   2.583   &   1.793   &   2.464    &   1.79   &  2.46  &  1.89  &  67 &  57 & {\it PT}  \\
 0.70  &   1.337   &   2.586   &   1.838   &   2.447    &   1.84   &  2.45  &  1.97  &  61 &  48 & {\it PT}  \\
 0.80  &   1.323   &   2.589   &   1.884   &   2.429    &   1.88   &  2.43  &  2.04  &  55 &  39 & {\it PT}  \\
\enddata
\end{deluxetable}

\clearpage

%+++++++++++++++++++++++++++++++++++++++++++++++++++++++++++++++++++++++++

%%% *** Table 6
%%%%%%%%%%%%%%%%%%%%%%%%%%%%%%%%%%%%%%%%%%%%%%%%%%%%%%%%%%%%%%%%%%%%%%%%%
\begin{deluxetable}{lcccccccccc}
\tabletypesize{\scriptsize}
\tablecaption{RHZs of $P$-Type Orbits for $M_1 = 1.25~M_\odot$, $M_2 = 0.75~M_\odot$, CHZ}
\tablewidth{0pt}
\tablehead{
$e_{\rm b}$ & \multicolumn{2}{c}{$a_{\rm per}$} & \multicolumn{2}{c}{$a_{\rm ap}$} & \multicolumn{2}{c}{Orbit} & $a_{\rm cr}$ & {\%}RHZ & {\%}HZ & Type \\
\noalign{\smallskip}
\hline
\noalign{\smallskip}
...    & RHZ$_{\rm in}$ & RHZ$_{\rm out}$ & RHZ$_{\rm in}$ & RHZ$_{\rm out}$ & RHZ$_{\rm in}$ & RHZ$_{\rm out}$ & ...  & ... & ... & ... \\
...    & (AU)           & (AU)            & (AU)           & (AU)            & (AU)           & (AU)            & (AU) & ... & ... & ...
}
\startdata
\multicolumn{9}{l}{Model:~~$2a_{\rm b}= 0.50$~AU} \\
\noalign{\smallskip}
\hline
\noalign{\smallskip}
 0.00  &   1.685   &    1.953   &   1.685   &   1.953   &   1.68  &  1.95  &  0.61   & 100 & 100 & {\it P}  \\
 0.10  &   1.663   &    1.963   &   1.706   &   1.944   &   1.71  &  1.94  &  0.69   &  89 &  89 & {\it P}  \\
 0.20  &   1.642   &    1.973   &   1.728   &   1.936   &   1.73  &  1.94  &  0.76   &  77 &  77 & {\it P}  \\ 
 0.30  &   1.621   &    1.985   &   1.751   &   1.929   &   1.75  &  1.93  &  0.83   &  66 &  66 & {\it P}  \\
 0.40  &   1.600   &    1.997   &   1.773   &   1.922   &   1.77  &  1.92  &  0.89   &  56 &  56 & {\it P}  \\
 0.50  &   1.580   &    2.010   &   1.796   &   1.917   &   1.80  &  1.92  &  0.94   &  45 &  45 & {\it P}  \\
 0.60  &   1.560   &    2.024   &   1.818   &   1.912   &   1.82  &  1.91  &  0.99   &  35 &  35 & {\it P}  \\
 0.70  &   1.541   &    2.038   &   1.841   &   1.907   &   1.84  &  1.91  &  1.02   &  24 &  24 & {\it P}  \\
 0.80  &   1.522   &    2.053   &   1.864   &   1.901   &   1.86  &  1.90  &  1.05   &  14 &  14 & {\it P}  \\
\noalign{\smallskip}
\hline
\noalign{\smallskip}
\multicolumn{9}{l}{Model:~~$2a_{\rm b}= 0.75$~AU} \\
\noalign{\smallskip}
\hline
\noalign{\smallskip}
 0.00  &   1.796   &   1.917   &   1.796   &   1.917    &   1.80   &  1.92  &  0.91  & 100 & 100 & {\it P}  \\
 0.10  &   1.762   &   1.925   &   1.830   &   1.909    &   1.83   &  1.91  &  1.04  &  66 &  66 & {\it P}  \\
 0.20  &   1.728   &   1.936   &   1.864   &   1.901    &   1.86   &  1.90  &  1.15  &  30 &  30 & {\it P}  \\
 0.30  &   1.695   &   1.948   &   ...     &   ...      &   ...    &  ...   &  1.25  &   0 &   0 & ...      \\
 0.40  &   1.663   &   1.963   &   ...     &   ...      &   ...    &  ...   &  1.34  &   0 &   0 & ...      \\
 0.50  &   1.631   &   1.979   &   ...     &   ...      &   ...    &  ...   &  1.41  &   0 &   0 & ...      \\
 0.60  &   1.600   &   1.997   &   ...     &   ...      &   ...    &  ...   &  1.48  &   0 &   0 & ...      \\
 0.70  &   1.570   &   2.017   &   ...     &   ...      &   ...    &  ...   &  1.54  &   0 &   0 & ...      \\
 0.80  &   1.541   &   2.038   &   ...     &   ...      &   ...    &  ...   &  1.58  &   0 &   0 & ...      \\
\enddata
\tablecomments{
Regarding the model $2a_{\rm b}=1.0$~AU, the values for RHZ$_{\rm in}$ and RHZ$_{\rm out}$ are
undefined as no HZs are identified for any of the models.
}
\end{deluxetable}

\clearpage

%+++++++++++++++++++++++++++++++++++++++++++++++++++++++++++++++++++++++++

%%% *** Table 7
%%%%%%%%%%%%%%%%%%%%%%%%%%%%%%%%%%%%%%%%%%%%%%%%%%%%%%%%%%%%%%%%%%%%%%%%%
\begin{deluxetable}{lcccccccccc}
\tabletypesize{\scriptsize}
\tablecaption{RHZs of $P$-Type Orbits for $M_1 = 1.25~M_\odot$, $M_2 = 0.75~M_\odot$, GHZ}
\tablewidth{0pt}
\tablehead{
$e_{\rm b}$ & \multicolumn{2}{c}{$a_{\rm per}$} & \multicolumn{2}{c}{$a_{\rm ap}$} & \multicolumn{2}{c}{Orbit} & $a_{\rm cr}$ & {\%}RHZ & {\%}HZ & Type \\
\noalign{\smallskip}
\hline
\noalign{\smallskip}
...    & RHZ$_{\rm in}$ & RHZ$_{\rm out}$ & RHZ$_{\rm in}$ & RHZ$_{\rm out}$ & RHZ$_{\rm in}$ & RHZ$_{\rm out}$ & ...  & ... & ... & ... \\
...    & (AU)           & (AU)            & (AU)           & (AU)            & (AU)           & (AU)            & (AU) & ... & ... & ...
}
\startdata
\multicolumn{9}{l}{Model:~~$2a_{\rm b}= 0.50$~AU} \\
\noalign{\smallskip}
\hline
\noalign{\smallskip}
 0.00  &   1.512   &   2.417   &   1.512   &   2.417    &   1.51   &  2.42  &  0.61  & 100 & 100 & {\it P}   \\
 0.10  &   1.490   &   2.429   &   1.534   &   2.406    &   1.53   &  2.41  &  0.69  &  96 &  96 & {\it P}   \\
 0.20  &   1.469   &   2.441   &   1.557   &   2.396    &   1.56   &  2.40  &  0.76  &  93 &  93 & {\it P}   \\
 0.30  &   1.448   &   2.454   &   1.579   &   2.386    &   1.58   &  2.39  &  0.83  &  89 &  89 & {\it P}   \\
 0.40  &   1.427   &   2.468   &   1.602   &   2.377    &   1.60   &  2.38  &  0.89  &  86 &  86 & {\it P}   \\
 0.50  &   1.406   &   2.482   &   1.625   &   2.369    &   1.62   &  2.37  &  0.94  &  82 &  82 & {\it P}   \\
 0.60  &   1.386   &   2.497   &   1.648   &   2.361    &   1.65   &  2.36  &  0.99  &  79 &  79 & {\it P}   \\
 0.70  &   1.367   &   2.512   &   1.671   &   2.355    &   1.67   &  2.35  &  1.02  &  76 &  76 & {\it P}   \\
 0.80  &   1.348   &   2.527   &   1.694   &   2.349    &   1.69   &  2.35  &  1.05  &  72 &  72 & {\it P}   \\
\noalign{\smallskip}
\hline
\noalign{\smallskip}
\multicolumn{9}{l}{Model:~~$2a_{\rm b}= 0.75$~AU} \\
\noalign{\smallskip}
\hline
\noalign{\smallskip}
 0.00  &   1.625   &   2.369   &   1.625   &   2.369    &   1.62   &  2.37  &  0.91  & 100 & 100 & {\it P}   \\
 0.10  &   1.590   &   2.382   &   1.659   &   2.358    &   1.66   &  2.36  &  1.04  &  94 &  94 & {\it P}   \\
 0.20  &   1.557   &   2.396   &   1.694   &   2.349    &   1.69   &  2.35  &  1.15  &  88 &  88 & {\it P}   \\
 0.30  &   1.523   &   2.412   &   1.729   &   2.341    &   1.73   &  2.34  &  1.25  &  82 &  82 & {\it P}   \\
 0.40  &   1.490   &   2.429   &   1.764   &   2.333    &   1.76   &  2.33  &  1.34  &  76 &  76 & {\it P}   \\
 0.50  &   1.458   &   2.448   &   1.800   &   2.324    &   1.80   &  2.32  &  1.42  &  70 &  70 & {\it P}   \\
 0.60  &   1.427   &   2.468   &   1.836   &   2.315    &   1.84   &  2.31  &  1.48  &  64 &  64 & {\it P}   \\
 0.70  &   1.396   &   2.489   &   1.872   &   2.305    &   1.87   &  2.30  &  1.54  &  58 &  58 & {\it P}   \\
 0.80  &   1.367   &   2.512   &   1.908   &   2.294    &   1.91   &  2.29  &  1.58  &  52 &  52 & {\it P}   \\
\noalign{\smallskip}
\hline
\noalign{\smallskip}
\multicolumn{9}{l}{Model:~~$2a_{\rm b}= 1.0$~AU} \\
\noalign{\smallskip}
\hline
\noalign{\smallskip}
 0.00  &   1.741   &   2.338   &   1.741   &   2.338    &   1.74   &  2.34  &  1.21  & 100 & 100 & {\it P}   \\
 0.10  &   1.694   &   2.349   &   1.788   &   2.327    &   1.79   &  2.33  &  1.38  &  90 &  90 & {\it P}   \\
 0.20  &   1.648   &   2.361   &   1.836   &   2.315    &   1.84   &  2.31  &  1.53  &  80 &  80 & {\it P}   \\
 0.30  &   1.602   &   2.377   &   1.883   &   2.301    &   1.88   &  2.30  &  1.66  &  70 &  70 & {\it P}   \\
 0.40  &   1.557   &   2.396   &   1.932   &   2.287    &   1.93   &  2.29  &  1.78  &  59 &  59 & {\it P}   \\
 0.50  &   1.512   &   2.417   &   1.980   &   2.271    &   1.98   &  2.27  &  1.89  &  49 &  49 & {\it P}   \\
 0.60  &   1.469   &   2.441   &   2.028   &   2.254    &   2.03   &  2.25  &  1.98  &  38 &  38 & {\it P}   \\
 0.70  &   1.427   &   2.468   &   2.077   &   2.235    &   2.08   &  2.24  &  2.05  &  26 &  26 & {\it P}   \\
 0.80  &   1.386   &   2.497   &   2.126   &   2.216    &   2.13   &  2.22  &  2.11  &  15 &  15 & {\it P}   \\
\enddata
\end{deluxetable}

\clearpage

%+++++++++++++++++++++++++++++++++++++++++++++++++++++++++++++++++++++++++

%%% *** Table 8
%%%%%%%%%%%%%%%%%%%%%%%%%%%%%%%%%%%%%%%%%%%%%%%%%%%%%%%%%%%%%%%%%%%%%%%%%
\begin{deluxetable}{lcccc}
\tablecaption{Critical Values of $e_{\rm b}$ for Models of {\it P}/{\it PT}-Type Habitability}
\tablewidth{0pt}
\tablehead{
Binary Major Axis $(2a_{\rm b})$ &  \multicolumn{2}{c}{0.5~AU} & \multicolumn{2}{c}{1.0~AU} \\
\noalign{\smallskip}
\hline
\noalign{\smallskip}
Model  & CHZ    &  GHZ   & CHZ    & GHZ
}
\startdata
$M_1=1.25~M_\odot$, $M_2=1.25~M_\odot$ & $\dag$ & $\dag$  &  $\dag$ & $\dag$ \\
$M_1=1.25~M_\odot$, $M_2=1.00~M_\odot$ & $\dag$ & $\dag$  &    0.64 & $\dag$ \\
$M_1=1.25~M_\odot$, $M_2=0.75~M_\odot$ & $\dag$ & $\dag$  &    ...  & $\dag$ \\
$M_1=1.25~M_\odot$, $M_2=0.50~M_\odot$ &   0.16 & $\dag$  &    ...  &  0.26  \\
$M_1=1.00~M_\odot$, $M_2=1.00~M_\odot$ & $\dag$ & $\dag$  &    0.64 & $\dag$ \\
$M_1=1.00~M_\odot$, $M_2=0.75~M_\odot$ & $\dag$ & $\dag$  &   0.006 &  0.53  \\
$M_1=1.00~M_\odot$, $M_2=0.50~M_\odot$ &   0.05 & $\dag$  &    ...  &  0.10  \\
$M_1=0.75~M_\odot$, $M_2=0.75~M_\odot$ & $\dag$ & $\dag$  &    ...  &  0.12  \\
$M_1=0.75~M_\odot$, $M_2=0.50~M_\odot$ &   ...  &  0.53   &    ...  &  ...   \\
$M_1=0.50~M_\odot$, $M_2=0.50~M_\odot$ &   ...  &  ...    &    ...  &  ...   \\
\enddata
\tablecomments{
($\dag$) means that the critical value of $e_{\rm b}$ is larger than 0.80; it
could not be determined owing to the limitations of the work by HW99.  Dots
indicate that there is no solution.
}
\end{deluxetable}

\clearpage

%+++++++++++++++++++++++++++++++++++++++++++++++++++++++++++++++++++++++++

%%% *** Table 9
%%%%%%%%%%%%%%%%%%%%%%%%%%%%%%%%%%%%%%%%%%%%%%%%%%%%%%%%%%%%%%%%%%%%%%%%%
\begin{deluxetable}{lcccccccccc}
\tabletypesize{\scriptsize}
\tablecaption{RHZs of $S$-Type Orbits for $M_1 = M_2 = 1.0~M_\odot$}
\tablewidth{0pt}
\tablehead{
$e_{\rm b}$ & \multicolumn{2}{c}{$a_{\rm per}$} & \multicolumn{2}{c}{$a_{\rm ap}$} & \multicolumn{2}{c}{Orbit} & $a_{\rm cr}$ & {\%}RHZ & {\%}HZ & Type \\
\noalign{\smallskip}
\hline
\noalign{\smallskip}
...    & RHZ$_{\rm in}$ & RHZ$_{\rm out}$ & RHZ$_{\rm in}$ & RHZ$_{\rm out}$ & RHZ$_{\rm in}$ & RHZ$_{\rm out}$ & ...  & ... & ... & ... \\
...    & (AU)           & (AU)            & (AU)           & (AU)            & (AU)           & (AU)            & (AU) & ... & ... & ...
}
\startdata
% \noalign{\smallskip}
% \hline
% \noalign{\smallskip}
\multicolumn{9}{l}{Model:~~$2a_{\rm b}= 20.0$~AU, CHZ} \\
\noalign{\smallskip}
\hline
\noalign{\smallskip}
 0.00  &  1.0513  &  1.5027  &  1.0513  &  1.5027   & 1.05 & 1.50 & 2.74   & 100.0 & 100.0 & {\it S}   \\
 0.10  &  1.0517  &  1.5035  &  1.0510  &  1.5021   & 1.05 & 1.50 & 2.41   &  99.8 &  99.8 & {\it S}   \\
 0.20  &  1.0523  &  1.5046  &  1.0508  &  1.5017   & 1.05 & 1.50 & 2.08   &  99.5 &  99.5 & {\it S}   \\
 0.30  &  1.0531  &  1.5061  &  1.0506  &  1.5013   & 1.05 & 1.50 & 1.77   &  99.3 &  99.3 & {\it S}   \\
 0.40  &  1.0545  &  1.5084  &  1.0505  &  1.5010   & 1.05 & 1.50 & 1.47   &  98.9 &  91.9 & {\it ST}  \\
 0.50  &  1.0570  &  1.5120  &  1.0504  &  1.5008   & 1.06 & 1.50 & 1.18   &  98.3 &  26.7 & {\it ST}  \\
 0.60  &  1.0619  &  1.5180  &  1.0503  &  1.5006   & 1.06 & 1.50 & 0.90   &  97.2 &   0.0 & ...       \\
 0.70  &  1.0743  &  1.5297  &  1.0502  &  1.5004   & 1.07 & 1.50 & 0.62   &  94.4 &   0.0 & ...       \\
 0.80  &  1.1277  &  1.5568  &  1.0501  &  1.5003   & 1.13 & 1.50 & 0.36   &  82.5 &   0.0 & ...       \\
\noalign{\smallskip}
\hline
\noalign{\smallskip}
\multicolumn{9}{l}{Model:~~$2a_{\rm b}= 20.0$~AU, GHZ} \\
\noalign{\smallskip}
\hline
\noalign{\smallskip}
 0.00  &  0.9287  &  1.8384  &  0.9287  &  1.8384   &  0.93 & 1.84 & 2.74  & 100.0 & 100.0 & {\it S}  \\
 0.10  &  0.9290  &  1.8398  &  0.9285  &  1.8373   &  0.93 & 1.84 & 2.41  &  99.9 &  99.9 & {\it S}  \\
 0.20  &  0.9294  &  1.8416  &  0.9284  &  1.8365   &  0.93 & 1.84 & 2.08  &  99.7 &  99.7 & {\it S}  \\
 0.30  &  0.9300  &  1.8443  &  0.9283  &  1.8359   &  0.93 & 1.84 & 1.77  &  99.6 &  92.6 & {\it ST} \\
 0.40  &  0.9309  &  1.8481  &  0.9282  &  1.8354   &  0.93 & 1.84 & 1.47  &  99.4 &  59.2 & {\it ST} \\
 0.50  &  0.9325  &  1.8542  &  0.9281  &  1.8349   &  0.93 & 1.84 & 1.18  &  99.2 &  26.9 & {\it ST} \\
 0.60  &  0.9357  &  1.8643  &  0.9281  &  1.8346   &  0.94 & 1.83 & 0.90  &  98.8 &   0.0 & ...      \\
 0.70  &  0.9437  &  1.8834  &  0.9280  &  1.8343   &  0.94 & 1.83 & 0.62  &  97.9 &   0.0 & ...      \\
 0.80  &  0.9746  &  1.9262  &  0.9280  &  1.8340   &  0.97 & 1.83 & 0.36  &  94.5 &   0.0 & ...      \\
\enddata
\end{deluxetable}

\clearpage

%+++++++++++++++++++++++++++++++++++++++++++++++++++++++++++++++++++++++++

%%% *** Table 10
%%%%%%%%%%%%%%%%%%%%%%%%%%%%%%%%%%%%%%%%%%%%%%%%%%%%%%%%%%%%%%%%%%%%%%%%%
\begin{deluxetable}{lcccccccccc}
\tabletypesize{\scriptsize}
\tablecaption{RHZs of $S$-Type Orbits for $M_1 = 1.25~M_\odot$, $M_2 = 0.75~M_\odot$}
\tablewidth{0pt}
\tablehead{
$e_{\rm b}$ & \multicolumn{2}{c}{$a_{\rm per}$} & \multicolumn{2}{c}{$a_{\rm ap}$} & \multicolumn{2}{c}{Orbit} & $a_{\rm cr}$ & {\%}RHZ & {\%}HZ & Type \\
\noalign{\smallskip}
\hline
\noalign{\smallskip}
...    & RHZ$_{\rm in}$ & RHZ$_{\rm out}$ & RHZ$_{\rm in}$ & RHZ$_{\rm out}$ & RHZ$_{\rm in}$ & RHZ$_{\rm out}$ & ...  & ... & ... & ... \\
...    & (AU)           & (AU)            & (AU)           & (AU)            & (AU)           & (AU)            & (AU) & ... & ... & ...
}
\startdata
% \noalign{\smallskip}
% \hline
% \noalign{\smallskip}
\multicolumn{9}{l}{Model:~~$2a_{\rm b}= 20.0$~AU, CHZ} \\
\noalign{\smallskip}
\hline
\noalign{\smallskip}
 0.00  &  1.3703  &  1.8981  &  1.3703  &  1.8981    &  1.37 & 1.90 & 3.22  &  100.0 & 100.0 & {\it S}  \\
 0.10  &  1.3705  &  1.8984  &  1.3702  &  1.8979    &  1.37 & 1.90 & 2.77  &   99.9 &  99.9 & {\it S}  \\
 0.20  &  1.3707  &  1.8988  &  1.3701  &  1.8977    &  1.37 & 1.90 & 2.39  &   99.8 &  99.8 & {\it S}  \\
 0.30  &  1.3711  &  1.8994  &  1.3700  &  1.8975    &  1.37 & 1.90 & 2.02  &   99.7 &  99.7 & {\it S}  \\
 0.40  &  1.3717  &  1.9003  &  1.3699  &  1.8974    &  1.37 & 1.90 & 1.67  &   99.6 &  55.8 & {\it ST} \\
 0.50  &  1.3727  &  1.9017  &  1.3699  &  1.8973    &  1.37 & 1.90 & 1.33  &   99.4 &   0.0 & ...      \\
 0.60  &  1.3749  &  1.9039  &  1.3699  &  1.8972    &  1.37 & 1.90 & 1.01  &   99.0 &   0.0 & ...      \\
 0.70  &  1.3804  &  1.9081  &  1.3699  &  1.8972    &  1.38 & 1.90 & 0.70  &   97.9 &   0.0 & ...      \\
 0.80  &  1.4046  &  ...     &  1.3699  &  1.8971    &  1.40 & 1.90 & 0.40  &   93.3 &   0.0 & ...      \\
\noalign{\smallskip}
\hline
\noalign{\smallskip}
\multicolumn{9}{l}{Model:~~$2a_{\rm b}= 20.0$~AU, GHZ} \\
\noalign{\smallskip}
\hline
\noalign{\smallskip}
 0.00  &  1.2087  &  2.3394  &  1.2087  &  2.3394    &  1.21 & 2.34 & 3.22  &  100.00 & 100.00 & {\it S}  \\
 0.10  &  1.2088  &  2.3399  &  1.2086  &  2.3390    &  1.21 & 2.34 & 2.77  &   99.95 &  99.95 & {\it S}  \\
 0.20  &  1.2090  &  2.3407  &  1.2085  &  2.3387    &  1.21 & 2.34 & 2.39  &   99.91 &  99.91 & {\it S}  \\
 0.30  &  1.2092  &  2.3417  &  1.2085  &  2.3384    &  1.21 & 2.34 & 2.02  &   99.87 &  72.1  & {\it ST} \\
 0.40  &  1.2096  &  2.3431  &  1.2085  &  2.3382    &  1.21 & 2.34 & 1.67  &   99.82 &  40.8  & {\it ST} \\
 0.50  &  1.2103  &  2.3453  &  1.2084  &  2.3381    &  1.21 & 2.34 & 1.33  &   99.7  &  10.8  & {\it ST} \\
 0.60  &  1.2117  &  2.3489  &  1.2084  &  2.3379    &  1.21 & 2.34 & 1.01  &   99.6  &   0.0  & ...      \\
 0.70  &  1.2152  &  2.3554  &  1.2084  &  2.3378    &  1.22 & 2.34 & 0.70  &   99.3  &   0.0  & ...      \\
 0.80  &  1.2293  &  ...     &  1.2084  &  2.3377    &  1.23 & 2.34 & 0.42  &   98.0  &   0.0  & ...      \\
\enddata
\end{deluxetable}

\clearpage

%+++++++++++++++++++++++++++++++++++++++++++++++++++++++++++++++++++++++++

%%% *** Table 11
%%%%%%%%%%%%%%%%%%%%%%%%%%%%%%%%%%%%%%%%%%%%%%%%%%%%%%%%%%%%%%%%%%%%%%%%%
\begin{deluxetable}{lcccc}
\tablecaption{Critical Values of $e_{\rm b}$ for Models of {\it S}/{\it ST}-Type Habitability}
\tablewidth{0pt}
\tablehead{
Binary Major Axis $(2a_{\rm b})$ &  \multicolumn{2}{c}{15.0~AU} & \multicolumn{2}{c}{20.0~AU} \\
\noalign{\smallskip}
\hline
\noalign{\smallskip}
Model  & CHZ    &  GHZ   & CHZ    & GHZ
}
\startdata
$M_1=1.25~M_\odot$, $M_2=1.25~M_\odot$ &  0.28  &  0.35  &  0.43  &  0.49  \\
$M_1=1.25~M_\odot$, $M_2=1.00~M_\odot$ &  0.31  &  0.38  &  0.46  &  0.51  \\
$M_1=1.25~M_\odot$, $M_2=0.75~M_\odot$ &  0.35  &  0.42  &  0.49  &  0.54  \\
$M_1=1.25~M_\odot$, $M_2=0.50~M_\odot$ &  0.39  &  0.45  &  0.51  &  0.56  \\
$M_1=1.00~M_\odot$, $M_2=1.00~M_\odot$ &  0.42  &  0.48  &  0.54  &  0.59  \\
$M_1=1.00~M_\odot$, $M_2=0.75~M_\odot$ &  0.46  &  0.51  &  0.57  &  0.61  \\
$M_1=1.00~M_\odot$, $M_2=0.50~M_\odot$ &  0.49  &  0.54  &  0.59  &  0.63  \\
$M_1=0.75~M_\odot$, $M_2=0.75~M_\odot$ &  0.64  &  0.68  &  0.72  &  0.74  \\
$M_1=0.75~M_\odot$, $M_2=0.50~M_\odot$ &  0.67  &  0.70  &  0.73  &  0.76  \\
$M_1=0.50~M_\odot$, $M_2=0.50~M_\odot$ & $\dag$ & $\dag$ & $\dag$ & $\dag$ \\
\enddata
\tablecomments{
($\dag$) means that the critical value of $e_{\rm b}$ is larger than 0.80; it
could not be determined owing to the limitations of the work by HW99.
}
\end{deluxetable}

%+++++++++++++++++++++++++++++++++++++++++++++++++++++++++++++++++++++++++

\end{document}